\documentclass[twocolumn]{aastex631}

\usepackage{amsmath}

\accepted{in ApJ}

\begin{document}

\title{Star and Cluster Formation in the Sh2-112 Filamentary Cloud Complex}

\shorttitle{Sh2-112}
\shortauthors{Panja et al.}

\correspondingauthor{Alik Panja}
\email{alik.panja@gmail.com}

\author[0000-0002-4719-3706]{Alik Panja}
\affiliation{S. N. Bose National Centre for Basic Sciences, Kolkata 700106, India}

\author[0000-0002-3904-1622]{Yan Sun}
\affiliation{Purple Mountain Observatory, Chinese Academy of Sciences, 10 Yuanhua Road, Nanjing 210033, China}

\author[0000-0003-0262-272X]{Wen Ping Chen}
\affiliation{Institute of Astronomy, National Central University, 300 Zhongda Road, Zhongli, Taoyuan 32001, Taiwan}
\affiliation{Department of Physics, National Central University, 300 Zhongda Road, Zhongli, Taoyuan 32001, Taiwan}

\author[0000-0003-1457-0541]{Soumen Mondal}
\affiliation{S. N. Bose National Centre for Basic Sciences, Kolkata 700106, India}

\begin{abstract}
We present the star formation activity around the emission nebula Sh2-112.  At a distance of $\sim2.1$~kpc, this \ion{H}{2} complex, itself 3~pc in radius, is illuminated by the massive star (O8\,V) BD$+$45\,3216.  The associated molecular cloud extends in angular scales of $2\fdg0\times0\fdg83$, corresponding to linear sizes of 73~pc by 30~pc, along the Galactic longitude.  The high-resolution ($30\arcsec$) extinction map reveals a chain of dust clumps aligned with the filament-like structure with an average extinction of $A_{V} \sim 2.78$~mag, varying up to a maximum of $\sim17$~mag.  Our analysis led to identification of a rich population ($\sim 500$) of young (average age of $\sim 1$~Myr) stars, plus a numerous number ($\sim 350$) of H$\alpha$ emitters, spatially correlated with the filamentary clouds. Located near the edge of the cloud, the luminous star BD$+$45\,3216 has created an arc-like pattern as the ionizing radiation encounters the dense gas, forming a blister-shaped morphology.  We found three distinct young stellar groups, all coincident with relatively dense parts of the cloud complex, signifying ongoing star formation.  Moreover, the cloud filament (excitation temperature $\sim 10$~K) traced by the CO isotopologues and extending nearly $\sim 80$~pc is devoid of ionized gas except at the dense cores (excitation temperature $\sim$ 28--32~K) wherein significant ionized emission excited by OB stars (dynamical age $\sim$ 0.18--1.0~Myr) pertains.  The radial velocity is dynamic (median $\sim -3.65$~km~s$^{-1}$) along the main filament, increasing from Galactic east to west, features mass flow to form the massive stars/clusters at the central hubs.
\end{abstract}


\section{Introduction} \label{sec:intro}

Most and perhaps all stars form in clustered environments within molecular clouds.  Internal gravitational dynamics prompts cloud fragmentation, with each fragment then collapsing and leading to the onset of prestellar core formation \citep{andre16}.  Alternatively, propagation of the ionizing or explosive shocks from massive stars may compress neighboring clouds, hence triggering the next epoch of star formation \citep{elm98, deh05}.  Molecular clouds exhibit complex geometries, including substructures such as sheets and filaments to elongated networks \citep{evans91, fal91, elm93, mye09}.  The turbulence from expanding \ion{H}{2} regions near a filamentary molecular cloud can generate sequential waves of star-forming cores along the long axis of the filament on either side of an \ion{H}{2} region \citep{fuk00}.  As the ionization front passes through the cloud, it sweeps up neutral gas, potentially increasing the star formation rate in a dense shell \citep{elm77}.  The young protostars are preferentially aligned along the filamentary axis, bearing the imprint of fragmentation of the parental cloud.

The optically visible \ion{H}{2} region Sh2-112 (hereafter S112, $\ell = 83\fdg7589$; $b = +03\fdg2750$), located at a distance of $\sim2.1$~kpc \citep{bli82} toward the rich region of the Orion arm, is physically associated with one of the most active nebulous systems of Cygnus\,X.  Illuminated by the massive source BD$+$45\,3216 of a spectral type of O8\,V \citep{lah85}, probably a double or multiple system, the region shows a circular morphology.  The surrounding ionized region created by the massive star could be the possible outcome of a triggering effect and hence an efficient site for next-generation star formation, given its distinct blister shaped distribution \citep{isr78}.  The associated gas with the region has long been investigated, e.g., the radial velocity \citep[V$_{{\rm CO}}=-4.0\pm2.0$~km\,s$^{-1}$,][]{bli82}, $^{13}$CO cloud mass \citep[1880~M$_{\sun}$,][]{dob94}, H$\alpha$ and infrared luminosities \citep{hun90}, radio recombination lines \citep{gar83}, etc. So far, the studies have been focused within a diameter of $15\arcmin$, despite the fact that S112 is not an isolated region, as we shall demonstrate in this work, that there exist dust structures spanning  $\sim 2\degr$ parallel to the the Galactic plane, with a noticeable filamentary pattern connecting all the sub-structures with S112 near the center.  This paper aims to diagnose the global star-forming activity via a comprehensive sample of young stellar population plus detailed characterization of the associated molecular and ionized gases on a larger scale than previously reported in the literature.

The paper is organized as follows.  Section~\ref{sec:s112_data} summarizes the observations and data reduction techniques, followed by Section~\ref{sec:s112_res}, which derives the dust distribution in the region, and identifies the young stellar population from optical H$\alpha$ emitters, disk-bearing young stars, to embedded protostars.  Their spatial distribution in relevance to radio emission is presented.   Section~\ref{sec:s112_dis} discusses the interplay and possible feedback of massive stars to nearby molecular and ionized gas. Section~\ref{sec:s112_mol} presents a large-scale ($\sim 2\degr$) filamentary structure hosting multiple dense clumps, for which we investigate with CO emission lines the molecular gas parameters and kinematics, rendering an overall star formation activity in the cloud complex from parental clouds to protostellar formation.  Finally, we present a summary of main results in Section~\ref{sec:s112_sum}.

\section{Data Acquisition and Reduction}
 \label{sec:s112_data}

Diagnosing the star formation history in a young cloud complex is hampered partly by the initial series of events occurring within a relatively short time scale ($\sim5$~Myr).  Multi-wavelength observations are necessary to get a comprehensive young stellar sample.  For example, optical data allow us to select H$\alpha$ emitters and massive members, whereas infrared observations can reveal embedded or disk-bearing objects.  On the other hand, the distribution of dense gas, out of which stars form, is traced by molecular line emission, whereas the dust is detected either by their thermal radiation or inferred by the level of extinction of background stars.  Gas photoionized by massive stars is disclosed by radio continuum radiation.  Our work makes use of a variety of these tools, either collected by our own or with archival data sets, with each of which briefed in the following.

\subsection{Observational Data}
 \label{ssec:s112_obs_dat}

HCT:  Optical slit spectroscopic observations toward S112 were carried out using the Himalaya Faint Object Spectrograph and Camera mounted on the 2~m Himalayan Chandra Telescope (HCT). Grism 7 (380--684~nm) with a resolution of 1330 was chosen to cover the critical spectral features seen in massive stars.  After the bias and cosmic ray correction, the 1-d spectra were extracted using the {\rm \footnotesize {APALL}} task in the {\rm \footnotesize {IRAF}} software. The spectra were wavelength calibrated by using the Fe-Ar arc lamp, and then flux calibrated with standard star \citep{oke90} observations.  The data were also corrected for the atmospheric extinction and instrument sensitivity availing the standard star observations.

PMO: The molecular line data for the three CO ($^{12}$CO, $^{13}$CO, and C$^{18}$O) $J$=1--0 isotopologues are obtained as parts of the Milky Way Imaging Scroll Painting (MWISP) project \citep{su19}.   This ongoing project with an expected time span of more than ten years (2011--2022) provides large-scale CO maps of the northern Galactic plane ($-10\degr < \ell < +250\degr$ and $|b| \lesssim 5\fdg2$) with a planned sky coverage of $\sim 2600$~deg$^{2}$.  The observations are carried out by a 13.7~m diameter single-dish millimetre-wavelength telescope, located in Delingha, China, and is managed by the Purple Mountain Observatory (PMO).  The MWISP survey delivers high-quality mapping with uniform sensitivity and moderate resolution ($\sim 50\arcsec$), and features a high spatially dynamic range.   A multibeam sideband-separating Superconducting Spectroscopic Array Receiver system with an instantaneous bandwidth of 1~GHz is employed for simultaneous observations.  Typical system temperatures are $\sim 250$~K for $^{12}$CO at the upper sideband, and $\sim 140$~K for $^{13}$CO and C$^{18}$O at the lower sideband.  The observations are made in position-switch On-The-Fly mode with a sampling interval of $10\arcsec$--$15\arcsec$.  Typical rms noise levels are $\sim 0.5$~K for $^{12}$CO at the velocity resolution of 0.16~km~s$^{-1}$, $\sim 0.3$~K for $^{13}$CO and C$^{18}$O at 0.17~km~s$^{-1}$.  The noise suppression and signal identification methods are described in detail by \citet{sun21}.  Finally the raw data are resampled with a grid spacing of $30\arcsec$ and mosaicked into FITS cubes using the GILDAS \citep{gil13} software.

\subsection{Archival Data} 
 \label{ssec:s112_arc_dat}

Gaia DR2: The Gaia Data Release~2 \citep[Gaia DR2;][]{gai18} contains homogeneous astrometry on five parameters (celestial coordinates, trigonometric parallaxes, and proper motions) for more than 1.3 billion objects, supplemented with photometry of three broad-band magnitudes in $G$ (330--1050~nm), $G_\mathrm{BP}$ (330--680~nm), and $G_\mathrm{RP}$ (630--1050~nm) with unprecedented accuracy.  We adopted the distances computed by \citet{bai18}, which provides estimated distances using Gaia parallaxes, with a probabilistic inference approach, by taking into account for the nonlinearity of the transformation and the positivity constraint of a distance value.  For this work, the Gaia DR2 measurements are found consistent with those in Gaia Early Data Release~3 \citep[Gaia EDR3;][]{gai21a1}. 
IPHAS: The Isaac Newton Telescope (INT)/Wide Field Camera (WFC) Photometric H$\alpha$ Survey of the Northern Galactic Plane \citep[IPHAS;][]{dre05} is an imaging survey covering an 1800~deg$^2$ sky in broadband Sloan $r$ (624~nm), $i$ (774.3~nm), and narrowband H$\alpha$ (656.8~nm) filters.  The WFC generates a mosaic of four CCD images at a pixel scale of $0\farcs33$~pixel$^{-1}$, offering the capability to detect H$\alpha$ emission-line candidates by on-off (H$\alpha$ and continuum) photometry. 

UKIDSS and 2MASS: The near-infrared $J$ (1.25~$\mu$m), $H$ (1.65~$\mu$m), and $K$ (2.16~$\mu$m) bands photometric data are obtained from the UKIDSS DR10PLUS Galactic Plane Survey (GPS; \citealt{law07}) and the 2MASS Point Source Catalog (PSC; \citealt{skr06}).  UKIDSS has a finer angular resolution ($0\farcs8$ pixels) compared to 2MASS ($2\farcs0$ pixels), as well as a deeper sensitivity.  The reliable UKIDSS sources are accessed using the Structured Query Language (SQL\footnote{\url{http://wsa.roe.ac.uk/sqlcookbook.html}}) interface \citep{luc08}.  To avoid the inferior photometry, UKIDSS brighter ($J < 13.75$~mag, $H < 13.25$~mag, and $K < 12.50$~mag) sources are supplemented with 2MASS \citep{ale13}.  
For our work the photometric error for each of the three bands is restricted within 0.1~mag as a quality criterion so as to get a signal-to-noise ratio $\gtrsim10$.

WISE: The Wide-field Infrared Survey Explorer \citep[WISE;][]{wri10} has scanned the entire sky in four wavebands (3.4, 4.6, 12, and 22~$\mu$m) with an angular resolution of $6\farcs1$, $6\farcs4$, $6\farcs5$, and $12\farcs0$, respectively.
The 3.4~$\mu$m and 12~$\mu$m filters encompass prominent polycyclic aromatic hydrocarbon features, whereas the 4.6~$\mu$m filter measures the continuum emission from small grains, and the 22~$\mu$m filter detects stochastic emission from small grains or the Wien's tail of thermal emission from large grains \citep{wri10}. To ensure good quality photometry, we considered only sources with magnitude uncertainties $\lesssim0.2$~mag.

AKARI: The AKARI survey \citep{mur07} covers about 90\% of the sky in four far-infrared  bands centring at 65, 90, 140, and 160~$\mu$m, with spatial resolutions ranging from $1\arcmin$ to $1\farcm5$. The detection limit of the four bands reaches 2.5--16~MJy~sr$^{-1}$ with a relative accuracy $<20$\%. The AKARI data provide information on the properties of dusty material in the interstellar medium that emits primarily between $\sim50$ and 200~$\mu$m \citep{doi15}.

NVSS: The National Radio Astronomy Observatory (NRAO) Very Large Array (VLA) Sky Survey \citep[NVSS;][]{con98} covers the northern sky (82\% of the celestial sphere) at 1.4~GHz (21~cm) with nearly uniform sensitivity and a $\sim45\arcsec$ (FWHM) angular resolution. The radio continuum (Stokes $I$) maps are extracted from the NVSS archive for our study to trace the ionized gas.

Planck: The Planck space mission measured the anisotropy of the cosmic microwave background in nine frequency bands covering 30--857~GHz with angular resolutions ranging from $31\arcmin$ to $5\arcmin$ \citep{pla16}.
With its high sensitivity and wide wavelength coverage, Planck provides all-sky maps of the thermal dust emission and, in particular from cold dust mainly associated with dense regions within molecular clouds, relevant for studies of the early phases of star formation.

\section{Dust Distribution and Young Stars}
 \label{sec:s112_res}

The YSOs in the region are identified and characterized by their infrared colors.  The H$\alpha$ stars are recognized by their excessive flux in the H$\alpha$ filter relative to that in the short-red (as continuum) filter.  Notwithstanding the possibility of red dwarfs with active chromospheric activity, an H$\alpha$ sample seen against a star-forming region is dominated by PMS stars.  The spatial distribution of YSO population at different evolutionary stages is then correlated with the dust distribution, estimated by the level of extinction of background stars, to infer the starbirth sequence, as discussed below.

\subsection{The Extinction Map}
 \label{ssec:s112_ext_map}

The dust distribution is traced by the extinction of background starlight.  We utilized the combined UKIDSS and 2MASS $H$- and $K$-band photometry, and constructed a stellar number density count by defining a spatial grid over the target area \citep{gut05}.  First, the region of our interest is subdivided into rectilinear grids, each of a size of $30\arcsec\times30\arcsec$.  The 20 nearest-neighbor sources from the center of each grid are selected to calculate the mean and standard deviation of the ($H-K$) color for each grid, excluding the sources for which the ($H-K$) values deviate $\gtrsim 3\sigma$ from the mean value \citep{pan21}.  The mean ($H-K$) color for each grid is then converted to $A_{K}$, using the reddening law $A_{K} = 1.82 \times [(H-K)_{\rm obs}-(H-K)_{\rm int}]$, the difference between the observed and the intrinsic color \citep{fla07}.

An extinction map thus produced is somewhat limited in angular resolution by the detection of a fair number of background stars.  Moreover, the color excess for any particular grid is derived in a statistical manner \citep{lad94}. Empirically, after a series of trials, we found a $\sim30\arcsec$ grid size, and $\sim20$ nearest neighbor stars to be optimal  choices, as a compromise between sensitivity and resolution \citep{pan20}.

\begin{figure*}
\centering
        \includegraphics[width=\textwidth]{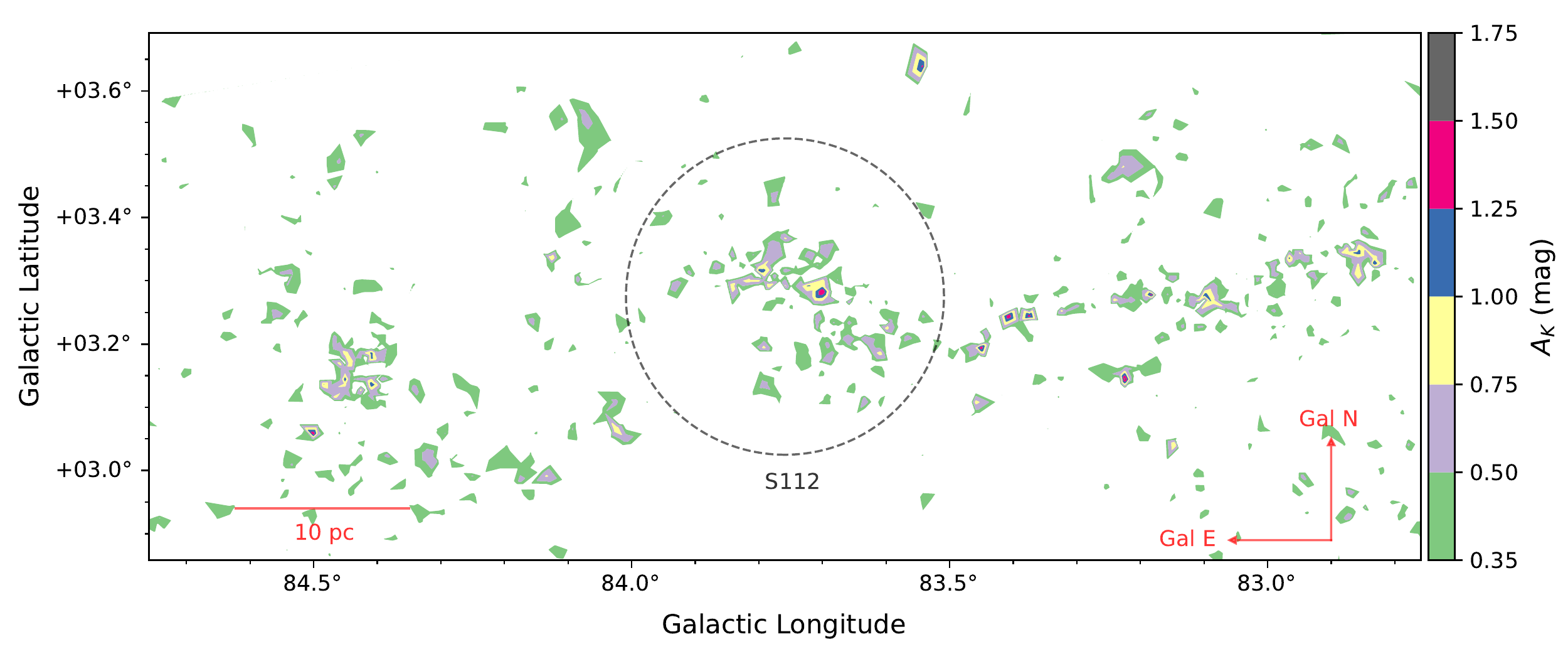}
  \caption{The extinction map around S112 computed by the $H$- and $K$-band photometry with a grid size of $30\arcsec\times30\arcsec$.  The dashed circle marks the 15$\arcmin$ radius around S112.  Enhanced extinction is seen generally in a zone in the Galactic east-west direction, with a multitude of dense clumps where extinction exceeds $A_K\sim1.55$~mag, or $A_V \ga 17$~mag.  A 10~pc linear scale is displayed assuming a heliocentric distance of 2.1~kpc to the complex. }
  \label{fig:s112_extinction_map}
\end{figure*}

The average intrinsic color $(H-K)_{\rm int}$ of the background population is measured to be $\sim 0.2$~mag, by using a nearby control field with a nominal extinction of $A_V=1.3$~mag.  
The resulting extinction map is displayed in Figure~\ref{fig:s112_extinction_map}.  The derived extinction values range from $A_{V} \simeq 1.33$--$17.20$~mag, or $A_{K} \simeq 0.12$--$1.55$~mag.  The extinction is relatively low for most of the region, with an average of $A_{V} \sim 2.78$~mag, despite the young (age $\sim 1$~Myr) nature of the complex and nebulous appearance, plausibly as the consequence of dispersal of parental cloud.

While the extinction is nonuniform over the region, a pattern stands out connecting chains of clumps and extending in the Galactic east-west direction, along which the average extinction varies in the order of $A_{V} \sim 3.5$~mag.  
The maximum extinction of $A_{V} \sim 17.20$~mag is observed around ($\ell, b) = (083\fdg70, +03\fdg28$) within the S112 region, which is located roughly at the center of the pattern.  Whereas toward the Galactic east (maximum $A_V\sim 16.48$~mag) or west (maximum $A_V\sim 15.78$~mag), the extinction marginally decreases from the peak value.

Our extinction map with an angular resolution of $30\arcsec$ and sensitivity down to $A_V\sim 20$~mag serves to guide the identification of heavily embedded sources such as protostellar objects, as will be discussed in the next section.

\subsection{The Young Stellar Population}
 \label{ssec:s112_yso_ide}

The infrared color excess is an indicator of the amount of retaining circumstellar dust, therefore the evolutionary status, of a YSO.  The level of the infrared excess dictates the location of a YSO in an infrared color-color diagram.  Initially, we have adopted the three-phase classification scheme from the \citet{koe12}, by using a set of UKIDSS and WISE color criteria.  Following which, we have used the UKIDSS data alone to identify additional young sources having excess in the near-infrared bands.  In addition to WISE data \citep[angular resolution $\sim6\arcsec$--$12\arcsec$;][]{wri10}, we also make use of the images where applicable taken by the Spitzer Infrared Array Camera \citep{faz04} which has a higher resolution (pixel scale $\sim 1\farcs2$~pixel$^{-1}$) but does not cover the whole region of our study.  

The infrared sample is ``sanitized'' by removing possible contaminants.  Galaxies with elevated star formation activity exhibiting increased polycyclic aromatic hydrocarbon (PAH) emission may mimic a YSO color.  Unresolved broad-line active galactic nuclei (AGNs) possess mid-infrared colors very similar to those of young stars \citep{gut09}.  These PAH/star-forming galaxies, AGNs, the shock-excited extended sources are winnowed out from the YSO sample using a combination of WISE colors.

\begin{figure*}
 \centering
        \includegraphics[width=\textwidth]{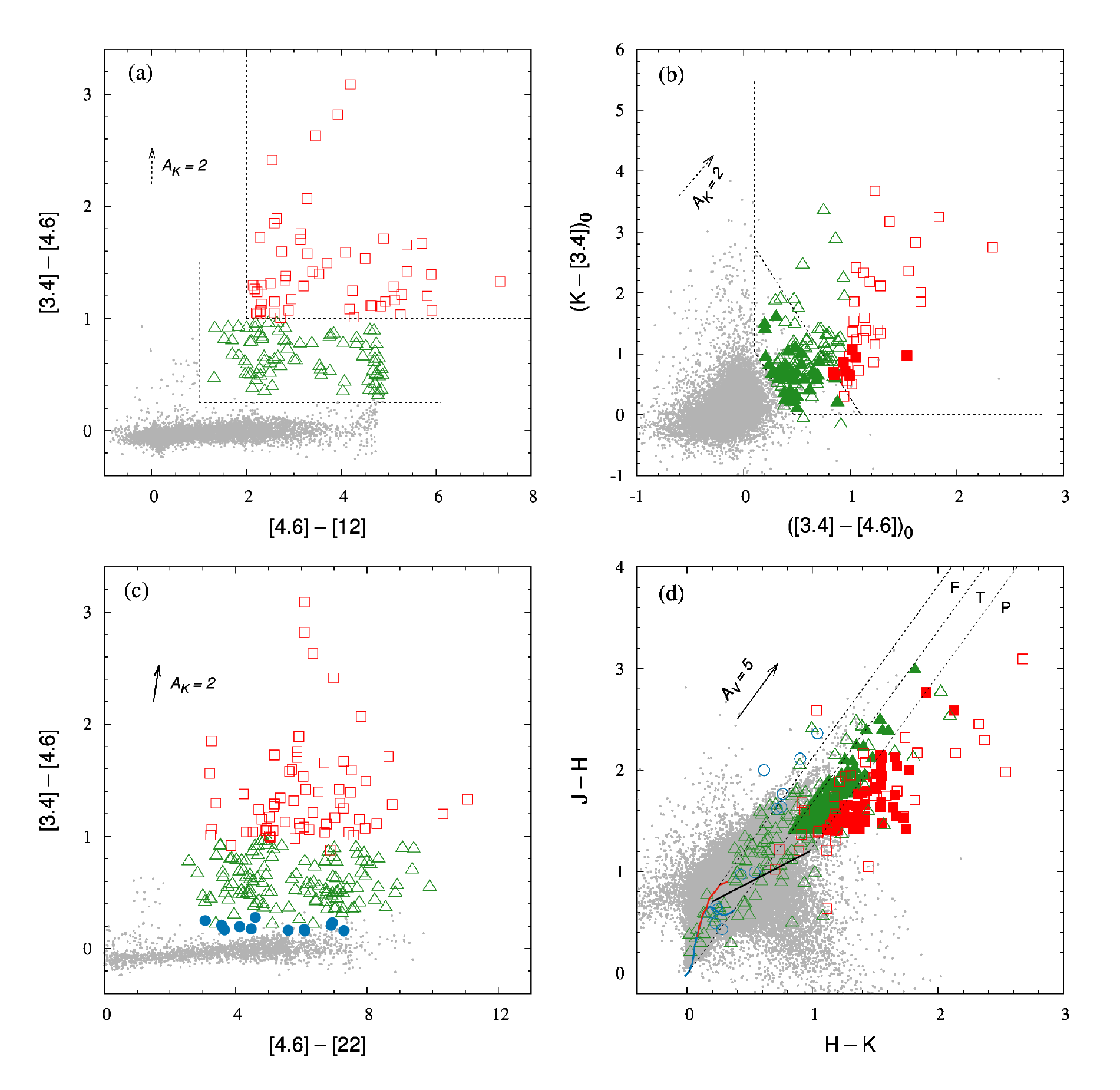}
  \caption{Color-color diagrams of Class~I (red squares), Class~II (green triangles), transition-disk (blue circles), and non-excess (gray dots) sources, identified by complementary classification schemes \citep{koe12}.  In each scheme, the dotted lines denote the adopted color criteria.
  (a)~Class~I and Class~II objects are selected by the WISE 3.4, 4.6, and 12~$\mu$m colors.  (b)~Heavily embedded sources are recognized by a combination of WISE 3.4 and 4.6~$\mu$m, and UKIDSS data. Sources are dereddened in this analysis.  
  Sources that are already found in the previous scheme are shown in open symbols, whereas the additional candidates selected in this scheme are in solid symbols, and a similar sequence is followed for the consecutive plots.  
  (c)~This WISE color-color diagram plots all the YSOs identified  consistently with three analysis schemes, including the transition-disk objects recognized by WISE 3.4, 4.6, and 22~$\mu$m data.  
  (d) Distribution of all the YSOs identified in the previous schemes along with the additional Class I and Class II sources identified only using the excess emission in the UKIDSS bands.  The solid blue and red lines mark the locus of dwarfs and giants, both taken from \citet{bes88}, and the solid black line indicates the CTTs locus, taken from \citet{mey97}.  The three parallel dashed lines represent the reddening vectors.  All the photometric data points in this figure are converted to CIT \citep{eli82} system.  }
  \label{fig:s112_wise_multi}
 \end{figure*}

Next, we identified YSO candidates by using the WISE and other color criteria \citep{koe12}.  Such a diagnostic color-color diagram using $W1$, $W2$, and $W3$, shown in Figure~\ref{fig:s112_wise_multi}(a), immediately resulted in 55 Class~I and 83 Class~II objects. To affirm the near-infrared sample, we added the UKIDSS photometry in addition to WISE 3.4, and 4.6~$\mu$m data for heavily embedded protostellar candidates.  This method requires dereddening of an object by removing the extinction of the nearest grid in the extinction map discussed in the previous section.  Additional eight Class~I and 39 Class II objects were rectified with this analysis; the dereddened color-color diagram of this added set of YSOs is shown in Figure~\ref{fig:s112_wise_multi}(b).  

The analysis discussed thus far does not include those sources visible in near-infrared, but lacking reliable detection in 12 or 22~$\mu$m due to the reduced instrument sensitivity and the bright background emission present at these longer wavelengths.  The final scrutiny therefore  utilized the WISE 3.4, 4.6, and 22~$\mu$m bands to select evolved transition-disk sources, i.e., those with little excess between 3.4 and 12~$\mu$m, but being bright at 22~$\mu$m, presented in Figure~\ref{fig:s112_wise_multi}(c)

The three-pronged analysis scheme outlined above must be self-consistent; namely a YSO that has been classified as a candidate by one scheme would still be reaffirmed with the others.  For example if an object mimicking a Class~I behavior in the near-infrared turns out not to have a rising spectral energy distribution at 22~$\mu$m, it would be reclassified as a reddened Class~II.  Likewise, a Class~II object with unusually blue colors \citep{koe12} would be placed back to the unclassified sample.  At the end, with a combination of all three analysis schemes, a total of 63 Class~I, 122 Class~II, and 13 transition-disk objects are identified, whose color-color plot is presented in Figure~\ref{fig:s112_wise_multi}(c).  

Furthermore, we have used the UKIDSS colors (Figure~\ref{fig:s112_wise_multi}(d)) to diagnose young objects with excess near-infrared emission, but lacking higher wavelength detection in the previous scheme.  The detail methodology is prescribed in \citet{ojh11}, \citet{pan20}, and references therein.  We have set an additional ($J-H$) color cut of 1.4~mag in order to remove any possible effects of contamination, based on the analysis of a nearby control field ($\ell = 083\fdg1441$; $b = +04\fdg2823$).  However, we note that below this limit there would still be a significant number of disk-bearing young sources, but are not included to keep our YSO sample credible.  Hereafter, from this method, we found 92 Class~I and 252 Class~II sources, whereas 41 of them were already detected in any of the above phases.  In case a source is selected with multiple methods, priority is given to WISE colors, because the higher wavelength data provide more robust information about the disk and envelope properties.  Thus, finally combining all the methods, we have detected a total of 135 Class~I, 353 Class~ II, and 13 transition-disk objects and their distribution is shown in Figure~\ref{fig:s112_wise_multi}(d).  Then again, any of these numbers should be considered as a conservative lower limit, as there are sources detected only in the $H$ and $K$ bands, or only in the $K$ band, but not in the $J$ band.
These young stars are listed in Table~\ref{tab:s112_yso_par}, including relevant infrared photometry, and Gaia data when available; only about 12\% of the sources in the table have Gaia DR2 parallax measurements.

\begin{longrotatetable}
\centerwidetable
\movetabledown=3mm
\begin{deluxetable*}{cccccccccccccc}
\tablecaption{Photometric catalog of the YSOs toward S112. The table is available in its entirety in machine-readable form}
\tabletypesize{\tiny}
\label{tab:s112_yso_par}
\tablehead{
\colhead{Sl.} & \colhead{Glon.} & \colhead{Glat.} & \colhead{[3.4]~$\mu$m} & \colhead{[4.6]~$\mu$m} & \colhead{[12]~$\mu$m} & \colhead{[22]~$\mu$m} & \colhead{$J$} & \colhead{$H$} & \colhead{$K$} & \colhead{$G$} & \colhead{$G_\mathrm{BP}$}  & \colhead{$G_\mathrm{RP}$}  & \colhead{Distance}  \\
  \colhead{No.} & \colhead{(deg)} & \colhead{(deg)} & \colhead{(mag)} &  \colhead{(mag)} & \colhead{(mag)} & \colhead{(mag)} & \colhead{(mag)} &  \colhead{(mag)} & \colhead{(mag)} & \colhead{(mag)} &  \colhead{(mag)} & \colhead{(mag)} &  \colhead{(kpc)}           
		}
\startdata
\multicolumn{14}{c}{Class I} \\
\hline 
  1  &  83.403456  &  3.244317  &  12.307 $\pm$ 0.031  &  11.043 $\pm$ 0.022  &  8.872 $\pm$ 0.033  &  5.878 $\pm$ 0.041  &  17.522 $\pm$ 0.020  &  15.682 $\pm$ 0.007  &  14.125 $\pm$ 0.005  &  \nodata  &  \nodata  &  \nodata  &  \nodata  \\
  2  &  83.395593  &  3.142038  &  16.110 $\pm$ 0.112  &  15.036 $\pm$ 0.125  &  9.128 $\pm$ 0.056  &  7.093 $\pm$ 0.120  &  \nodata  &  \nodata  &  \nodata  &  \nodata  &  \nodata  &  \nodata  &  \nodata  \\
  3  &  83.363773  &  3.247085  &  14.877 $\pm$ 0.074  &  12.807 $\pm$ 0.031  &  9.529 $\pm$ 0.058  &  4.985 $\pm$ 0.029  &  \nodata  &  \nodata  &  \nodata  &  \nodata  &  \nodata  &  \nodata  &  \nodata  \\
  4  &  83.612797  &  3.186152  &  11.288 $\pm$ 0.023  &  9.946 $\pm$ 0.019  &  7.136 $\pm$ 0.019  &  4.589 $\pm$ 0.026  &  15.410 $\pm$ 0.056  &  13.982 $\pm$ 0.043  &  13.018 $\pm$ 0.035  &  18.828 $\pm$ 0.014  &  20.130 $\pm$ 0.082  &  17.300 $\pm$ 0.032  &  2.375$_{-0.77}^{+1.54}$  \\
  5  &  83.330033  &  3.268075  &  13.229 $\pm$ 0.031  &  12.219 $\pm$ 0.026  &  12.307 $\pm$ \nodata  &  8.982 $\pm$ \nodata  &  18.446 $\pm$ \nodata  &  16.154 $\pm$ \nodata  &  13.930 $\pm$ 0.057  &  \nodata  &  \nodata  &  \nodata  &  \nodata  \\
  6  &  83.360964  &  2.990175  &  9.620 $\pm$ 0.022  &  8.468 $\pm$ 0.021  &  3.544 $\pm$ 0.015  &  0.698 $\pm$ 0.005  &  14.651 $\pm$ 0.057  &  13.565 $\pm$ 0.061  &  12.056 $\pm$ 0.031  &  \nodata  &  \nodata  &  \nodata  &  \nodata  \\
  7  &  83.457345  &  3.188119  &  10.501 $\pm$ 0.024  &  8.922 $\pm$ 0.019  &  5.642 $\pm$ 0.015  &  3.287 $\pm$ 0.019  &  16.406 $\pm$ 0.008  &  14.789 $\pm$ 0.004  &  12.988 $\pm$ 0.002  &  19.767 $\pm$ 0.018  &  21.021 $\pm$ 0.158  &  18.263 $\pm$ 0.048  &  2.203$_{-1.00}^{+1.98}$  \\
  8  &  83.251595  &  3.255542  &  13.163 $\pm$ 0.029  &  12.099 $\pm$ 0.024  &  12.126 $\pm$ \nodata  &  8.822 $\pm$ 0.372  &  17.739 $\pm$ \nodata  &  15.708 $\pm$ 0.152  &  14.152 $\pm$ 0.063  &  \nodata  &  \nodata  & \nodata  &  \nodata  \\
  9  &  83.135417  &  3.271612  &  13.042 $\pm$ 0.031  &  12.164 $\pm$ 0.028  &  9.833 $\pm$ 0.211  &  5.287 $\pm$ 0.042  &  16.859 $\pm$ 0.012  &  15.033 $\pm$ 0.004  &  13.798 $\pm$ 0.004  &  \nodata  &  \nodata  &  \nodata  &  \nodata  \\
  10  &  83.135074  &  3.278172  &  13.408 $\pm$ 0.031  &  12.373 $\pm$ 0.027  &  10.353 $\pm$ \nodata  &  5.663 $\pm$ 0.056  &  18.140 $\pm$ 0.036  &  16.119 $\pm$ 0.011  &  14.547 $\pm$ 0.007  &  \nodata  &  \nodata  &  \nodata  &  \nodata  \\
\hline
\multicolumn{14}{c}{Class II} \\
\hline
  1  &  83.131848  &  3.232247  &  13.443 $\pm$ 0.040  &  12.669 $\pm$ 0.041  &  9.545 $\pm$ 0.194  &  5.951 $\pm$ 0.065  &  16.601 $\pm$ 0.010  &  15.451 $\pm$ 0.006  &  14.611 $\pm$ 0.007  &  20.039 $\pm$ 0.012  &  21.347 $\pm$ 0.230  &  18.737 $\pm$ 0.044  &  2.379$_{-1.11}^{+2.04}$  \\
  2  &  83.068837  &  3.268861  &  10.922 $\pm$ 0.025  &  10.422 $\pm$ 0.021  &  8.455 $\pm$ 0.063  &  3.659 $\pm$ 0.035  &  13.667 $\pm$ 0.025  &  12.675 $\pm$ 0.026  &  11.864 $\pm$ 0.024  &  17.003 $\pm$ 0.002  &  18.281 $\pm$ 0.011  &  15.779 $\pm$ 0.005  &  2.457$_{-0.47}^{+0.72}$  \\
  3  &  83.327383  &  3.278623  &  13.418 $\pm$ 0.029  &  12.983 $\pm$ 0.031  &  11.613 $\pm$ \nodata  &  9.203 $\pm$ \nodata  &  16.418 $\pm$ \nodata  &  15.215 $\pm$ 0.093  &  14.364 $\pm$ 0.076  &  \nodata  &  \nodata  &  \nodata  &  \nodata  \\
  4  &  83.001149  &  3.390222  &  14.675 $\pm$ 0.035  &  14.181 $\pm$ 0.046  &  9.462 $\pm$ 0.052  &  7.909 $\pm$ 0.160  &  16.601 $\pm$ 0.009  &  15.674 $\pm$ 0.007  &  15.245 $\pm$ 0.012  &  20.728 $\pm$ 0.013  &  21.369 $\pm$ 0.291  &  19.230 $\pm$ 0.055  &  2.239$_{-1.30}^{+2.14}$  \\
  5  &  83.323391  &  3.250221  &  12.478 $\pm$ 0.026  &  11.703 $\pm$ 0.022  &  11.325 $\pm$ \nodata  &  9.131 $\pm$ \nodata  &  18.297 $\pm$ \nodata  &  15.615 $\pm$ 0.137  &  13.437 $\pm$ 0.039  &  \nodata  &  \nodata  &  \nodata  &  \nodata  \\
  6  &  83.057411  &  3.271772  &  11.963 $\pm$ 0.026  &  11.436 $\pm$ 0.023  &  7.677 $\pm$ 0.023  &  4.689 $\pm$ 0.048  &  15.236 $\pm$ 0.056  &  13.938 $\pm$ 0.052  &  12.891 $\pm$ 0.033  &  19.469 $\pm$ 0.007  &  20.630 $\pm$ 0.207  &  18.044 $\pm$ 0.019  &  2.691$_{-1.19}^{+2.08}$  \\
  7  &  83.107832  &  3.285580  &  9.800 $\pm$ 0.021  &  8.984 $\pm$ 0.020  &  7.457 $\pm$ 0.024  &  1.995 $\pm$ 0.012  &  14.068 $\pm$ 0.032  &  12.206 $\pm$ 0.024  &  10.964 $\pm$ 0.018  &  19.512 $\pm$ 0.008  &  20.885 $\pm$ 0.528  &  17.819 $\pm$ 0.029  &  2.530$_{-1.11}^{+2.02}$  \\
  8  &  83.324920  &  3.137974  &  13.559 $\pm$ 0.028  &  12.906 $\pm$ 0.030  &  10.537 $\pm$ 0.089  &  8.468 $\pm$ 0.324  &  16.116 $\pm$ 0.098  &  14.843 $\pm$ 0.069  &  14.426 $\pm$ 0.080  &  19.535 $\pm$ 0.016  &  20.510 $\pm$ 0.103  &  18.263 $\pm$ 0.041  &  1.700$_{-0.71}^{+1.76}$  \\
  9  &  83.325522  &  3.131974  &  11.872 $\pm$ 0.024  &  11.374 $\pm$ 0.021  &  9.287 $\pm$ 0.052  &  6.974 $\pm$ 0.075  &  14.320 $\pm$ 0.028  &  13.273 $\pm$ 0.027  &  12.652 $\pm$ 0.023  &  17.538 $\pm$ 0.005  &  18.682 $\pm$ 0.043  &  16.398 $\pm$ 0.013  &  2.601$_{-0.55}^{+0.89}$  \\
  10  &  83.284975  &  3.249319  &  10.168 $\pm$ 0.024  &  9.541 $\pm$ 0.019  &  6.542 $\pm$ 0.015  &  4.878 $\pm$ 0.031  &  13.555 $\pm$ 0.061  &  12.399 $\pm$ 0.051  &  11.518 $\pm$ 0.034  &  16.892 $\pm$ 0.002  &  18.171 $\pm$ 0.020  &  15.708 $\pm$ 0.007  &  2.289$_{-0.38}^{+0.56}$  \\
\hline
\multicolumn{14}{c}{Transition Disk} \\
\hline
  1  &  83.281329  &  3.251314  &  11.473 $\pm$ 0.023  &  11.299 $\pm$ 0.022  &  9.117 $\pm$ 0.047  &  6.826 $\pm$ 0.083  &  15.017 $\pm$ 0.054  &  13.159 $\pm$ 0.053  &  12.348 $\pm$ 0.032  &  20.787 $\pm$ 0.011  &  21.095 $\pm$ 0.240  &  18.839 $\pm$ 0.043  &  1.296$_{-0.90}^{+2.17}$  \\
  2  &  83.216944  &  3.055202  &  12.743 $\pm$ 0.028  &  12.466 $\pm$ 0.030  &  10.450 $\pm$ 0.108  &  7.866 $\pm$ 0.193  &  14.936 $\pm$ 0.045  &  13.935 $\pm$ 0.041  &  13.468 $\pm$ 0.043  &  17.722 $\pm$ 0.006  &  18.719 $\pm$ 0.035  &  16.633 $\pm$ 0.016  &  2.218$_{-0.51}^{+0.87}$  \\
  3  &  83.214491  &  3.055450  &  11.249 $\pm$ 0.022  &  11.083 $\pm$ 0.022  &  9.902 $\pm$ 0.072  &  7.421 $\pm$ 0.126  &  12.532 $\pm$ 0.023  &  11.897 $\pm$ 0.024  &  11.618 $\pm$ 0.022  &  14.844 $\pm$ 0.002  &  15.656 $\pm$ 0.007  &  13.920 $\pm$ 0.010  &  3.540$_{-0.33}^{+0.40}$  \\
  4  &  83.094967  &  3.295275  &  12.438 $\pm$ 0.030  &  12.208 $\pm$ 0.029  &  10.077 $\pm$ 0.116  &  5.262 $\pm$ 0.083  &  14.033 $\pm$ 0.032  &  13.371 $\pm$ 0.037  &  13.106 $\pm$ 0.031  &  16.724 $\pm$ 0.001  &  17.819 $\pm$ 0.019  &  15.650 $\pm$ 0.006  &  2.002$_{-0.29}^{+0.41}$  \\
  5  &  83.087081  &  3.238889  &  12.171 $\pm$ 0.032  &  12.012 $\pm$ 0.029  &  9.819 $\pm$ 0.109  &  4.717 $\pm$ 0.032  &  15.637 $\pm$ 0.064  &  13.526 $\pm$ 0.041  &  12.869 $\pm$ 0.030  &  \nodata  &  \nodata  &  \nodata  &  \nodata  \\
  6  &  83.083953  &  3.286479  &  9.087 $\pm$ 0.023  &  8.880 $\pm$ 0.021  &  6.689 $\pm$ 0.027  &  1.971 $\pm$ 0.026  &  13.717 $\pm$ 0.038  &  11.218 $\pm$ 0.030  &  10.124 $\pm$ 0.041  &  \nodata  &  \nodata  &  \nodata  &  \nodata  \\
  7  &  82.923383  &  3.154172  &  11.477 $\pm$ 0.023  &  11.285 $\pm$ 0.023  &  9.932 $\pm$ 0.063  &  7.688 $\pm$ 0.165  &  13.263 $\pm$ 0.030  &  12.237 $\pm$ 0.034  &  11.656 $\pm$ 0.027  &  17.475 $\pm$ 0.003  &  19.631 $\pm$ 0.052  &  16.034 $\pm$ 0.014  &  3.704$_{-0.96}^{+1.58}$  \\
  8  &  83.735402  &  3.161190  &  12.275 $\pm$ 0.028  &  12.117 $\pm$ 0.028  &  10.436 $\pm$ 0.147  &  6.011 $\pm$ 0.066  &  15.131 $\pm$ 0.048  &  13.431 $\pm$ 0.030  &  12.667 $\pm$ 0.023  &  20.826 $\pm$ 0.011  &  21.409 $\pm$ 0.235  &  19.119 $\pm$ 0.060  &  1.996$_{-1.17}^{+2.09}$  \\
  9  &  83.818115  &  3.134543  &  11.068 $\pm$ 0.023  &  10.861 $\pm$ 0.020  &  9.394 $\pm$ 0.051  &  7.287 $\pm$ 0.126  &  14.068 $\pm$ 0.025  &  12.351 $\pm$ 0.026  &  11.552 $\pm$ 0.021  &  19.622 $\pm$ 0.005  &  21.154 $\pm$ 0.215  &  17.906 $\pm$ 0.023  &  2.415$_{-1.03}^{+2.00}$  \\
  10  &  83.968444  &  3.647253  &  11.612 $\pm$ 0.024  &  11.363 $\pm$ 0.022  &  10.491 $\pm$ 0.107  &  8.291 $\pm$ 0.182  &  12.440 $\pm$ 0.021  &  12.018 $\pm$ 0.026  &  11.703 $\pm$ 0.021  &  14.758 $\pm$ 0.000  &  15.577 $\pm$ 0.003  &  13.841 $\pm$ 0.002  &  2.751$_{-0.20}^{+0.23}$  \\
\hline 
\enddata
\end{deluxetable*}
\end{longrotatetable}

\subsection{The H$\alpha$ Emitters}
 \label{ssec:s112_iph_hal}

\begin{figure}
        \includegraphics[width=\columnwidth]{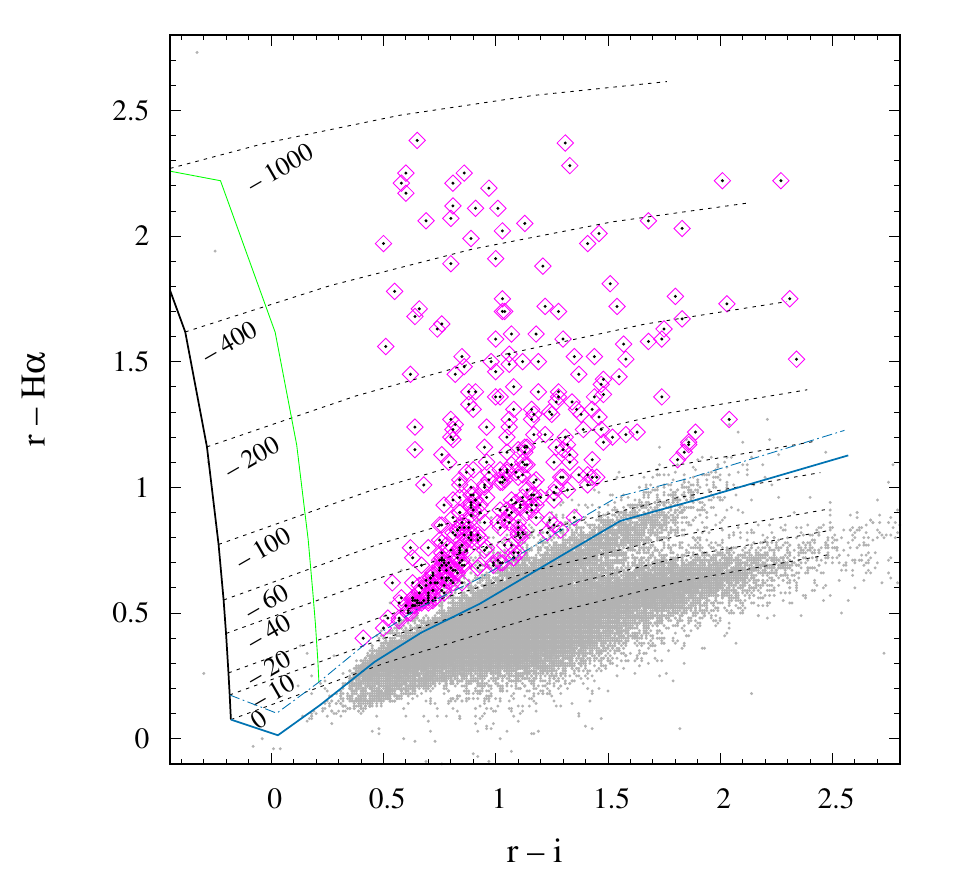}
  \caption{The IPHAS $r-$H$\alpha$ versus $r-i$ color-color diagram to select H$\alpha$ emission stars.  Candidate H$\alpha$ emitters are represented by magenta diamonds, whereas the rest field interlopers are depicted as gray dots. The solid and dashed blue lines represent the unreddened main sequence, which normally serves as the upper bound to the main stellar locus, and the expected position of unreddened main-sequence stars with H$\alpha$ emission strengths of an equivalent width of $-10$~\AA.
  The nearly vertical solid black and green lines show the trend for an unreddened Rayleigh-Jeans continuum and the unreddened continuum of an F0 spectral type, respectively. The black broken lines are the predicted lines of constant net emission.}
  \label{fig:s112_iphas_ccd}
\end{figure}

Many young low-mass stars show H$\alpha$ in emission, from either chromospheric activity or circumstellar accretion.  The IPHAS survey avails the detection of H$\alpha$ emission-line stars, reaching a 10$\sigma$ sensitivity down to $r \simeq 20$~mag.  Figure~\ref{fig:s112_iphas_ccd} shows the distribution of sources detected in IPHAS imaging, by considering a reliability criteria of $r <$ 20 mag and photometric uncertainty $< 0.1$~mag in all three bands.  The benefit of forming ($r-i$) as abscissa and ($r-$H$\alpha$) as ordinate in the color-color plot is that objects with H$\alpha$-band excesses appear higher within the diagram, while intrinsically redder or more highly reddened objects are over to the right \citep{dre05}, and because the ($r-$H$\alpha$) color tends to act as a coarse proxy for the spectral type and is less sensitive to the reddening than ($r-i$).  A star is considered as a H$\alpha$ star if it is located above the track of unreddened main-sequence stars (1)~with H$\alpha$ emission strengths more than equivalent width of $-10$~\AA\ \citep{bar14}, and (2)~its ($r-$H$\alpha$) color is deviated from the main locus of non-emission-line objects by more than three times the average uncertainty of its ($r-$H$\alpha$) value.   

With this, we found a total of 357 H$\alpha$ emitters toward the S112 region.  In Figure~\ref{fig:s112_iphas_ccd}, a large number of sources have excessive ($r-$H$\alpha$) color, reflecting strong H$\alpha$ line strength, while a few of the sources are located toward the higher reddening zone.  In contrast, the field star distribution shows two well-defined loci, revealing two distinct stellar populations, one for the unreddened main-sequence and other for the giant stars.  An estimation of the reddening from such a color-color diagram would be inappropriate, as the reddening tracks are curved in a way that depends on the spectral energy distribution and the amount of reddening.  Among the 357 H$\alpha$ emitters, 14 are found to have infrared counterparts and are previously (Section~\ref{ssec:s112_yso_ide}) classified as either Class I, Class II, or transition-disk objects.  The photometric parameters of the H$\alpha$ emitting sources are detailed in Table~\ref{tab:s112_hal_cat}.  More than 97\% of the H$\alpha$ emitters have Gaia DR2 parallax measurements.

\begin{deluxetable*}{CCCCC CCc}
\tablecaption{Photometric catalog of the H$\alpha$ emitters toward S112. The entire catalog is available in the electronic version}
\tabletypesize{\footnotesize}
\label{tab:s112_hal_cat}
\tablehead{
\colhead{} & \colhead{Glon.} & \colhead{Glat.} & \colhead{$r$} & \colhead{$i$} & \colhead{H$\alpha$} & \colhead{Distance} & \colhead{YSO} \\
\colhead{No.} & \colhead{(deg)} & \colhead{(deg)} & \colhead{(mag)} & \colhead{(mag)} & \colhead{(mag)} & \colhead{(kpc)} & \colhead{Class}
           }
\startdata
  1  &  83.325522  &  3.131974  &  17.44 $\pm$ 0.01  &  16.33 $\pm$ 0.01  &  16.64 $\pm$ 0.01  &  2.601$_{-0.55}^{+0.89}$  &  II  \\
  2  &  83.216944  &  3.055202  &  17.92 $\pm$ 0.02  &  16.87 $\pm$ 0.01  &  16.86 $\pm$ 0.01  &  2.218$_{-0.51}^{+0.87}$  &  TD  \\
  3  &  83.131848  &  3.232247  &  19.60 $\pm$ 0.05  &  18.41 $\pm$ 0.03  &  18.10 $\pm$ 0.03  &  2.379$_{-1.11}^{+2.04}$  &  II  \\
  4  &  83.771466  &  3.278478  &  18.34 $\pm$ 0.01  &  17.31 $\pm$ 0.02  &  16.59 $\pm$ 0.01  &  3.316$_{-0.92}^{+1.59}$  &  II  \\
  5  &  83.767700  &  3.295630  &  18.04 $\pm$ 0.01  &  16.97 $\pm$ 0.01  &  16.43 $\pm$ 0.01  &  2.109$_{-0.50}^{+0.88}$  &  II  \\
  6  &  83.799645  &  3.291358  &  13.66 $\pm$ 0.00  &  13.02 $\pm$ 0.00  &  12.51 $\pm$ 0.00  &  $...$  &  II  \\
  7  &  83.759443  &  3.319530  &  19.92 $\pm$ 0.04  &  18.38 $\pm$ 0.04  &  18.20 $\pm$ 0.03  &  1.409$_{-0.56}^{+1.62}$  &  II  \\
  8  &  83.680314  &  3.232444  &  19.82 $\pm$ 0.04  &  18.52 $\pm$ 0.04  &  18.78 $\pm$ 0.04  &  $...$  &  II  \\
  9  &  83.612797  &  3.186152  &  19.51 $\pm$ 0.03  &  17.88 $\pm$ 0.03  &  18.29 $\pm$ 0.03  &  2.375$_{-0.77}^{+1.54}$  &  I  \\
  10  &  83.968444  &  3.647253  &  14.72 $\pm$ 0.00  &  13.86 $\pm$ 0.00  &  14.03 $\pm$ 0.00  &  2.751$_{-0.20}^{+0.23}$  &  TD  \\
\hline 
\enddata
\end{deluxetable*}

\subsection{Average Age of the YSOs}
 \label{ssec:s112_yso_age}

\begin{figure}
        \includegraphics[width=\columnwidth]{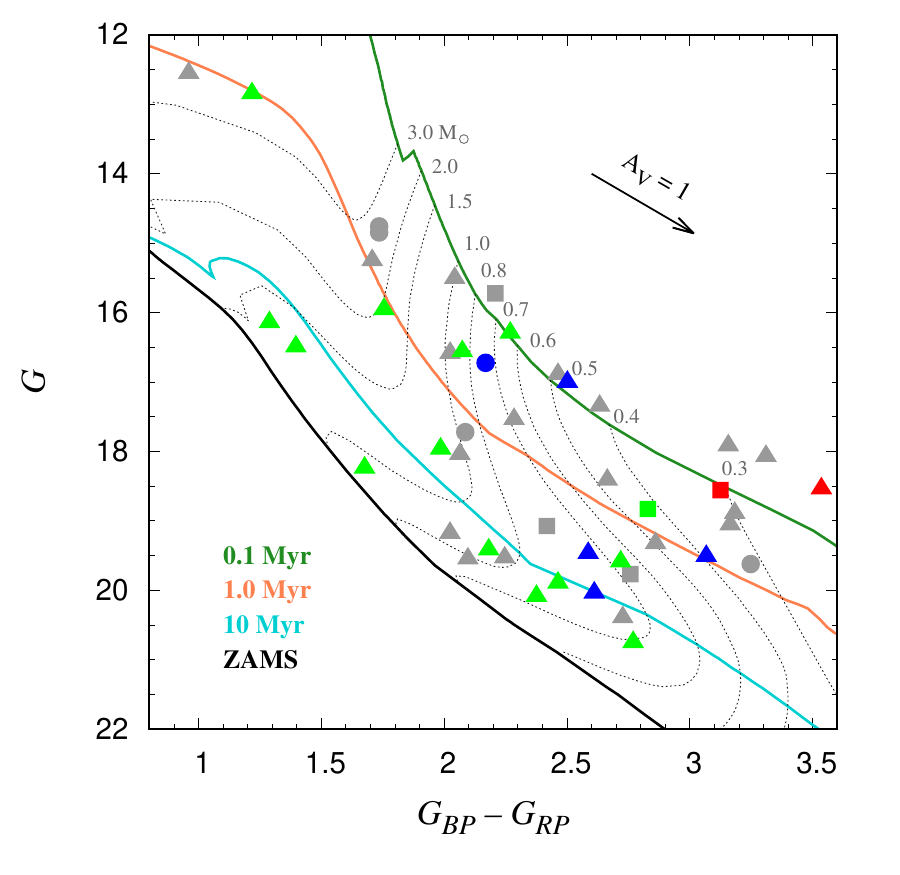}
  \caption{Gaia DR2 color-magnitude diagram of the YSOs (Class~I: squares; Class~II: triangles; transition-disk: circles) in the subregions Clump~A (red), S112 (green), Clump~B (blue), and in rest of the region (grey).  
  Overplotted as solid lines are the theoretical isochrones for ages of 0.1, 1.0, and 10~Myr along with the zero-age-main-sequence (ZAMS) taken from the PAdova and TRieste Stellar Evolution Code \citep{bre12}.
  The evolutionary tracks for different masses (as dotted lines) are also shown. All the isochrones are adjusted for a cluster distance of 2.1~kpc \citep{bli82} and a universal reddening $A_{V} \sim 2.78$~mag (Section~\ref{ssec:s112_ext_map}).}
  \label{fig:s112_gaia_cmd}
\end{figure}

The average age of the stellar sources, which are the most profoundly traceable entity of a cluster, hints on the timescale and dynamics of the cluster evolution.  We have estimated the ages of the YSOs by comparing their (Class~I, Class~II, and transition-disk) $G_\mathrm{BP}-G_\mathrm{RP}$ color vs $G$ magnitude with theoretical isochrones, presented in Figure~\ref{fig:s112_gaia_cmd}.  With the Gaia  sensitivity of $G \sim 20$~mag, plus generally a low extinction across the region, Gaia data afford the detection of a handful of YSOs with reliable parallax and proper motion.   Going even fainter ($G \gtrsim 20.5$~mag) limit, the detection of lower-mass (M$\lesssim 0.4$~$M_{\sun}$) membership becomes unreliable.  
The objects in different subregions are denoted separately to check if they are attributed to any specific age range, but we did not find any systematic trend in the age distribution for different clusters.  
We deliberately excluded the H$\alpha$ emitters in the age determination because other than being  pre-main-sequence objects this sample could incorporate post-main-sequence stars, interacting binaries, supergiants, luminous blue variables, or post-asymptotic giant branch stars among others \citep{dre05}. The exclusion of H$\alpha$ stars also applies to distance analysis, to avoid foreground or background contamination. 

The stellar evolutionary tracks are taken from the PAdova and TRieste Stellar Evolution Code \citep[{\texttt{PARSEC}};][]{bre12} interactive tool with version release 1.2S.  For Gaia sources, we adopted the photometric sensitivity of \citet{eva18}, as they give empirically the best fit to our data.
In the fit of isochrones to the data, displayed as Figure~\ref{fig:s112_gaia_cmd}, while there is a spread in the inferred age, the majority of the YSOs are scattered between 0.1--10~Myr, except for a few older than 10~Myr near the zero-age main sequence.  The reason for the spread, other than the uncertainties in the isochrones at very young ages, could likely arise from variable (interstellar and circumstellar) extinction of individual sources.  A compromising age of the YSO population therefore is $\sim1$~Myr.

\subsection{Spatial Distribution of the PMS Population}
 \label{ssec:s112_spa_dis}

\begin{figure*}
\centering
        \includegraphics[width=0.9\textwidth]{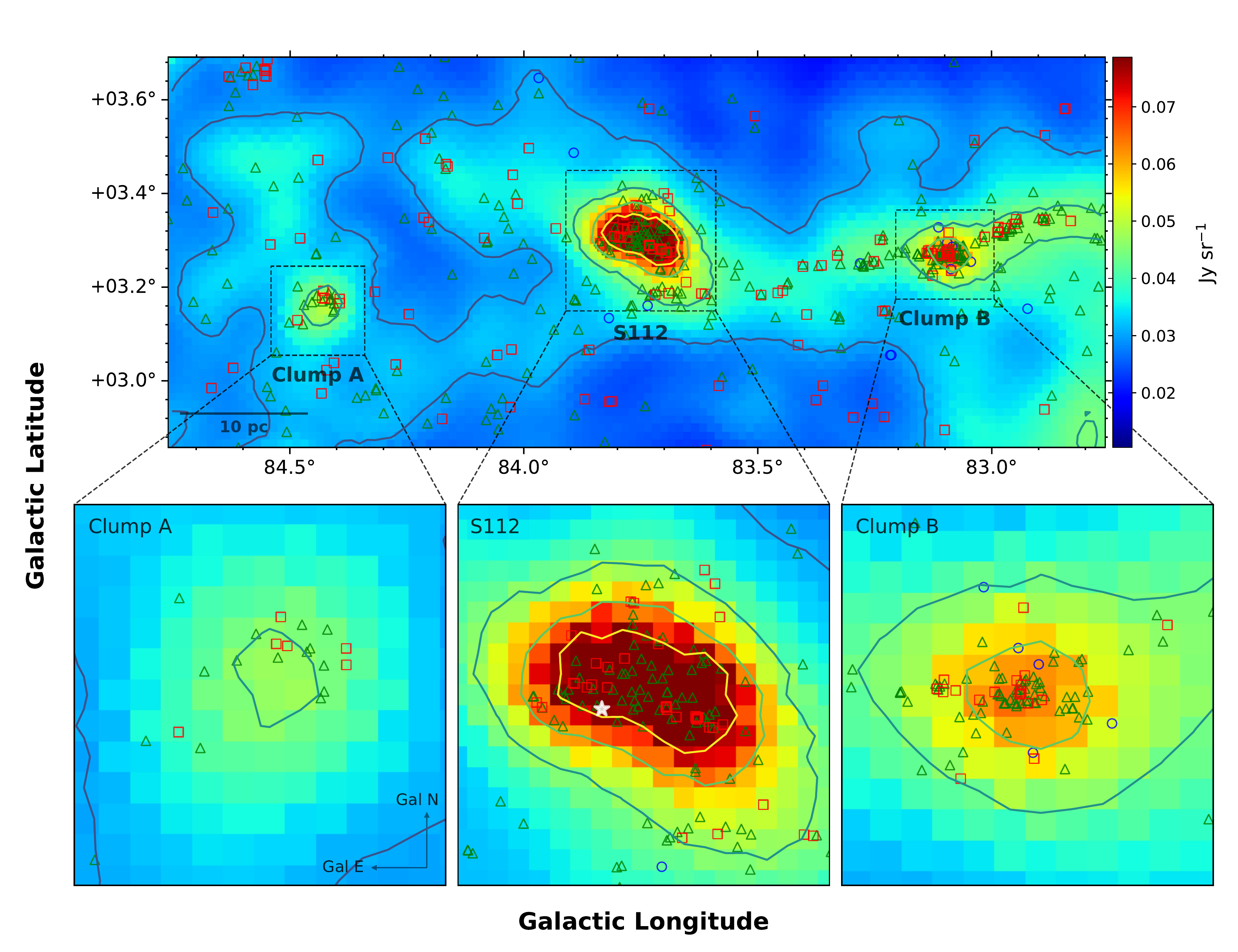}
    \caption{(Top) Spatial distribution of the YSOs (Class~I: red square; Class~II: green triangle; transition disk: blue~circle) overlaid on the Planck 353~GHz map.  The stellar density peaks at three subregions (Clump A, S112, and Clump B), and coincide well with the high intensity regions.  The contour levels are at 0.028, 0.040, 0.052, and 0.064~Jy~sr$^{-1}$.
    (Bottom) Zoomed-in maps of the three subregions.  The main ionizing source BD+45\,3216 is represented by a white asterisk symbol.}
    \label{fig:s112_pla_yso}
\end{figure*}

The spatial distribution of the young population reveals essential information regarding the length- and time-scale of the star formation activity in association with a cloud complex.  Figure~\ref{fig:s112_pla_yso} manifests the distribution of Class~I, Class~II, and transition-disk objects, superimposed on the Planck 353~GHz map, in which existence of a filamentary-like pattern aligned from the Galactic east to west is evident.  The location of three high intensity subregions (Clump A, S112, and Clump B) is revealed, with all of them being interconnected via the filamentary structure.  Not only the positions of the YSOs are traced along the higher intensity clouds, but also the YSO clustering nicely coincides with the subregions.  Moreover, a good degree of similarity is observed between this spatial map with that of the extinction map  (Figure~\ref{fig:s112_extinction_map}) derived from the infrared star counts (Section~\ref{ssec:s112_ext_map}).  Altogether, a combination of spatial distribution of young objects, intensity morphology, and extinction map favours of an ongoing star formation activity in a broader and continuous scale ($\sim 2\degr$) spanned across the Galactic longitude. 

The intensity map unveils fluctuation in the cloud density over the region, with the peak near the S112 subregion, where an elevated concentration of YSOs is also found.  In accordance with the spatial map, the relative number of Class~I sources decreases noticeably from the Galactic east to west.  Since the Class~I sources normally represent an earlier phase compared to the Class~II sources, the ratio of Class I/Class II sources could be used to interpret the evolutionary stage of the subregions, with a higher ratio indicative of a younger population.  The source ratio for the subregions Clump A, S112, and Clump B is estimated to be 0.40, 0.36, and 0.25, respectively, varying systematically from east to west.  Literature studies of several young and nearby star-forming regions have shown this ratio to vary in a wide range, between $\sim0.1$--0.8, with a median of 0.27 \citep{gut09}.  This suggests that the sources associated with Clump A and S112 subregions consist of younger population and are evolving almost in a similar timescale, while the sources to the western periphery are comparatively older in nature.  There is a slight possibility of undercounting of embedded stellar sources toward the western side as the reddening marginally varies compared to the other subregions.  
In S112, the majority of the YSOs are found to be distributed toward the northern vicinity of  BD+45\,3216, while the Class~I sources have preferentially formed two separate groups both away from the central core.  Actually, the S112 subregion itself is fragmented into two dense cores which will be seen in the molecular maps presented later but not resolved here in the Planck data, and these Class~I sources nicely coincide with both the cores.  As Class~I sources have less evolved circumstellar environments, this indicates that star formation is still continuing at those cores.  Most probably, these Class~I sources are the next-generation stars of the region whose formation is induced by the expansion of the \ion{H}{2} region powered by the massive star.

 \section{Massive Stars and Ionized Gas}
 \label{sec:s112_dis}

To investigate the roles massive, ionizing sources play on their surrounding environs, whether to prompt or to restrain subsequent star formation, we present below spectroscopic observations supplemented with infrared imaging data.  Furthermore, radio continuum measurements are included to curb the dynamical evolution of the ionized gas associated with the subregions.

\subsection{Spectra Classification }
 \label{ssec:s112_spe_res}

\begin{figure*}
\centering
        \includegraphics[width=0.9\textwidth]{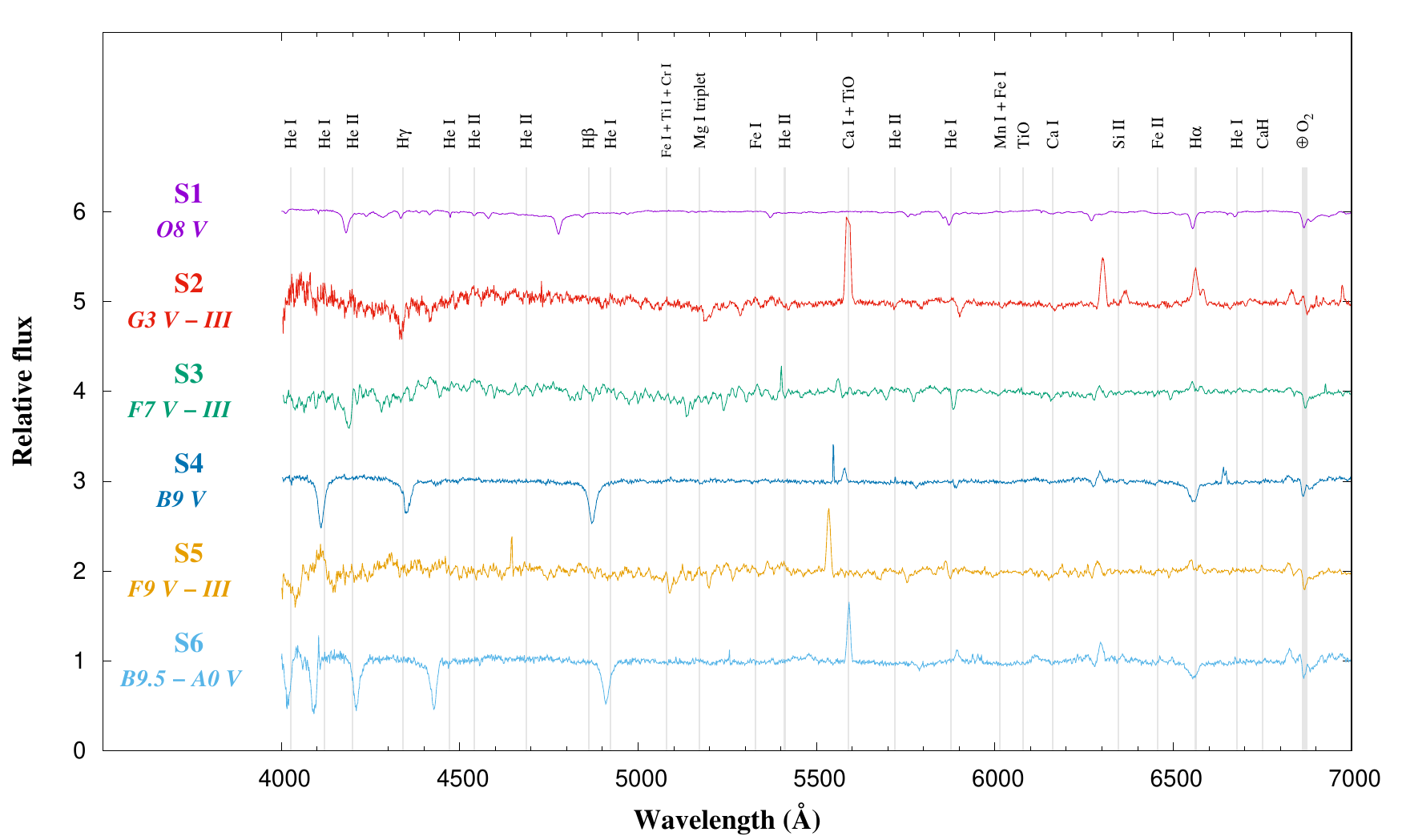}
  \caption{Flux-calibrated normalized spectra of the stars toward S112.  Prominent spectral line are marked.}
   \label{fig:s112_spe_gr7}
\end{figure*}

Optical spectroscopy for a few bright ($J < 13$~mag) stars toward S112 were conducted to gain knowledge of their stellar properties so as to identify the candidate massive stars as the ionizing sources to influence the surrounding star formation processes via any feedback mechanism.  The flux-calibrated normalized spectra of these stars are presented in Figure~\ref{fig:s112_spe_gr7}.

Spectral classification is done by comparison with standard spectral libraries \citep{jac84, wal90,tor93}, relying on the identification of conspicuous lines and their equivalent widths, following the procedure outlined in \citet{pan20}. Selected spectral indices are compared with the spectral atlas of \citet{dan94}, \citet{kob12}, and \citet{her04}.  Primarily, for early-type stars up to A, we judged by the line strengths of the Balmer series (H$\delta$~4102~\AA, H$\gamma$~4340~\AA, H$\beta$~4861~\AA, H$\alpha$~6563~\AA), \ion{He}{1} (5876, 6678, 7065\AA), and \ion{He}{2} (4200, 4541, 4686, 5411, 5720~\AA). For cooler stars of F or later, we relied on various metallic lines, such as \ion{Na}{1} (5890, 5896~\AA), \ion{Mg}{1} triplet (5167, 5172, 5183~\AA), \ion{Mg}{2} (4481, 6347~\AA), \ion{Ca}{1} (6122, 6162~\AA), \ion{Ca}{2} triplet (8498, 8542, 8662~\AA), and \ion{Fe}{1} (6495, 7749, 7834~\AA).  A brief description of the spectral features for each observed star is given as follows.

S1:  The star S1 shows prominent \ion{He}{1} (4144, 4387, 4471, 5876, 6678, 7065~\AA), \ion{He}{2} (4541, 4686, 5411 \r{A}), and hydrogen (4101, 4340, 4861, 6563 \r{A}) lines, with an evidence of strong absorption features at \ion{He}{1} (5876, 6678, 7065 \r{A}), \ion{He}{2} (4541, 5411 \r{A}), and H$\alpha$ (6563 \r{A}) lines, which restricts the spectral type to O as a dwarf.  In case of supergiants, the H$\alpha$ line is extremely weak from O5 to B8.  The presence of \ion{Si}{2} (4128 \r{A}), \ion{O}{3} (4415 \r{A}), and a moderate nitrogen enhancement are indicative of later O-type spectra.  The line ratio of \ion{He}{1} (4471 \r{A})/\ion{He}{2} (4541 \r{A}) is found to be marginally greater than 1, suggesting a spectral type later than O7.  Finally, by comparing the equivalent widths of \ion{He}{1} (5876 \r{A}), \ion{He}{2} (5411 \r{A}), and H$\alpha$ (6563 \r{A}) and a visual comparison of the spectrum with the spectral libraries, we assigned the spectral type of the star as O8~V.

S2:  The star S2 shows a peculiar feature of very strong H$\alpha$ (6563 \r{A}) emission, along with strong absorption at \ion{Na}{1} (5890, 5896 \r{A}) and \ion{Ca}{2} triplet (8498, 8542, 8662 \r{A}).  The presence of \ion{Ca}{1} (6162 \r{A}), \ion{Fe}{1} + \ion{Ti}{1} + \ion{Cr}{1} (6362 \r{A}), \ion{Fe}{1} (6495 \r{A}), and \ion{C}{2} (6580 \r{A}) suggests a spectral range between early- to mid-G type.  Moreover, the absorption at \ion{Mg}{1} triplet (5167, 5172, 5183 \r{A}) is an indication of a G-type spectrum.  Comparing the line ratios of H$\alpha$, \ion{Na}{1} doublet, and \ion{Ca}{2} triplet, we chose the spectral type to be G3~V/III, whereas for the same spectral type the \ion{Ca}{2} triplet line strength would be much stronger for a supergiant.

S3:  The declining strengths of \ion{Na}{1} (5890, 5896 \r{A}), H$\alpha$ (6563 \r{A}), and \ion{Ca}{2} triplet (8498, 8542, 8662 \r{A}) in the spectrum suggest a late-F star.  From the additional features of \ion{Ca}{2} (K) (3933 \r{A}), \ion{CH}{0} (G band) (4300 \r{A}), and \ion{Fe}{1} (4271, 5329, 7749 \r{A}), we determined a spectral type of F7~V/III.

S4:  The spectrum of S4 exhibits weak presence of \ion{He}{1} (5876 \r{A}) and very strong H$\alpha$ (6563 \r{A}) absorption, indicating a relatively hot star (late-B type).  The line ratio of H$\alpha$ (6563 \r{A})/\ion{He}{1} (5876 \r{A}) suggests a spectral type of B8 or later.  Considering the presence of \ion{C}{3} + \ion{O}{2} (4070 \r{A}), \ion{Si}{3} (4552 \r{A}), \ion{CN + Fe}{1} (4175 \r{A}), \ion{Fe}{1} (4383 \r{A}), and \ion{Na}{1} (5890, 5896 \r{A}) features, we assigned the spectral type of this star as B9~V.

S5:  This star shows \ion{CN + Fe}{1} (4175 \r{A}), \ion{Fe}{1} (4532, 6495, 7749 \r{A}), \ion{Ca}{1} (6162 \r{A}), and \ion{Fe}{2} (6242 \r{A}) blend, which are a signature of  late-F types.  From the additional features of \ion{Na}{1} (5890, 5896 \r{A}), H$\alpha$ (6563 \r{A}), and \ion{Ca}{2} triplet (8498, 8542, 8662 \r{A}), we estimated the spectral type to be F9~V/III.

S6:  The spectrum of S6 is similar to that of S4, with presence of weaker \ion{He}{1} (5876 \r{A}) and strong H$\alpha$ (6563 \r{A}) absorption, including \ion{C}{3} + \ion{O}{2} (4070 \r{A}), \ion{CN + Fe}{1} (4175 \r{A}), \ion{Si}{2} (4128 \r{A}), \ion{Si}{4} (4089 \r{A}), \ion{Mn}{1} + \ion{Fe}{1} (4458 \r{A}), and \ion{N}{3} (4511 \r{A}).  Considering the line strength of H$\alpha$ (6563 \r{A})/\ion{He}{1} (5876 \r{A}), we classified the star as a B9.5--A0~V type.

All the classified stars are primarily placed into the dwarfs/giants categories.  With our low-dispersion spectroscopy, the typical uncertainty for early-type stars up to the A type is $\pm1$ subtype, whereas for F-type and later the uncertainty is about $\pm3$ subtypes. However, the distribution of varying reddening and the contribution from circumstellar dust emission can have considerable effects on the nature of spectral class, as will be discussed in the next section.

\subsection{Reddening and Membership of the Observed Stars}
 \label{ssec:s112_red_mem}

Stellar parameters were derived from the estimated spectral types and the infrared photometry. To estimate the $A_{V}$ of each source, we utilized the $JHK$ color excess relations \citep{pan20}, adopting the reddening laws in \citet{coh81} and the intrinsic stellar colors from \citet{martins06} for the O-type star and from \citet{pec13} for other sources.    
Furthermore, the position of a particular source in a color-color diagram provides a consistency check of its reddening.

\begin{figure}
        \includegraphics[width=0.5\textwidth]{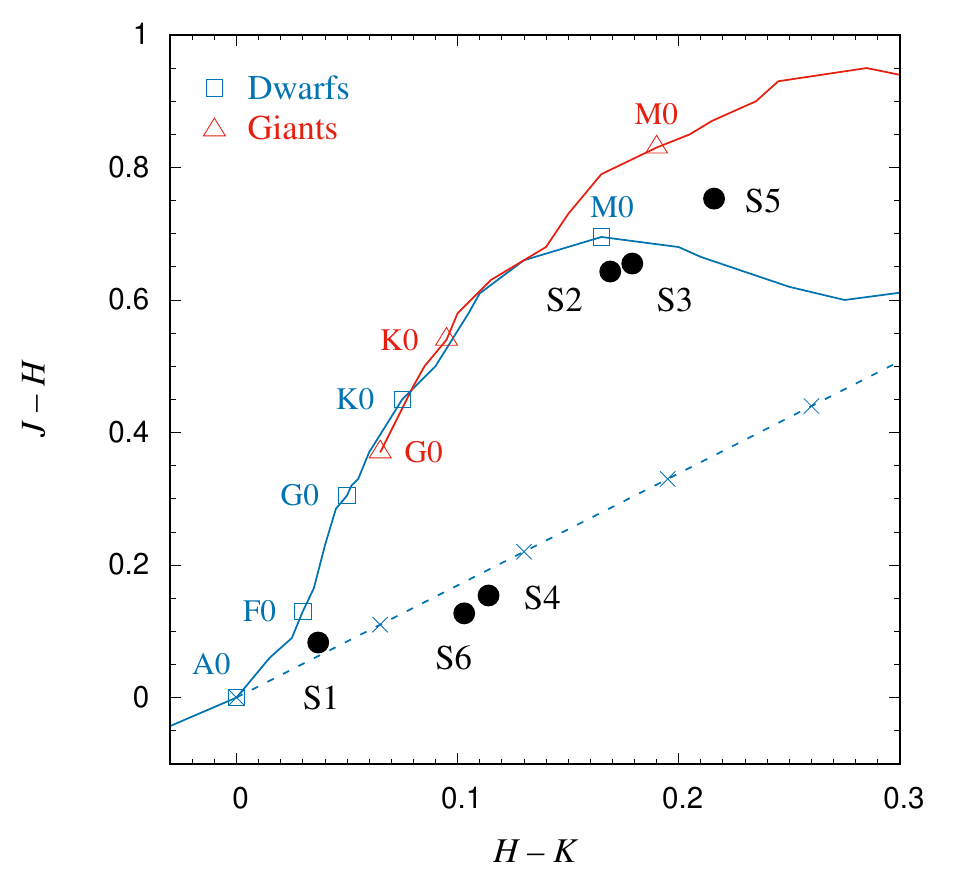}
        \includegraphics[width=0.5\textwidth]{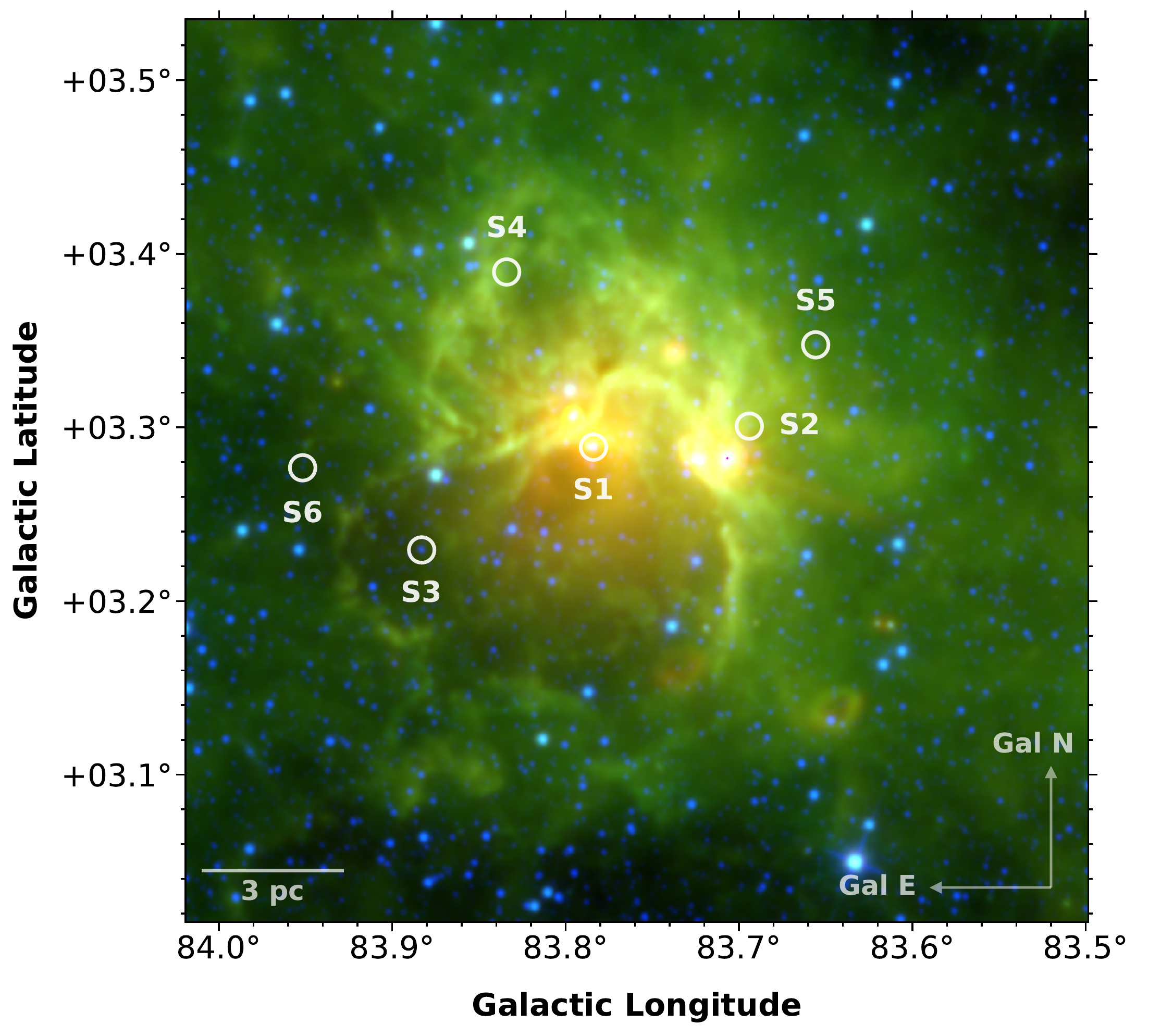}
  \caption{(left) The 2MASS color-color diagram for the spectroscopically observed stars, each marked with a black filled circle, overlaid with the unreddened main-sequence (blue solid line) and giant loci (red solid line), both taken from \citet{bes88}.  A reddening vector drawn from the tip of the main sequence, i.e., from A0 is represented by the blue dashed line, along which the consecutive cross symbols indicate an increment of reddening of $A_{V}=1$~mag. (right) Color composite image of S112 created with WISE $W4$ (22~$\mu$m in red), $W3$ (12~$\mu$m in green), and $W1$ (3.4~$\mu$m in blue) bands. Spectroscopically observed sources are marked (white circles) and labelled.}
  \label{fig:s112_jhk_spe}
\end{figure}

Table~\ref{tab:s112_spe_par} lists the stellar parameters of the spectroscopically observed targets.  The positions in the 2MASS color-color diagram, shown in Figure~\ref{fig:s112_jhk_spe} (left), can be accounted for either with extinction, i.e., $A_{V}$ of 3.2--3.8~mag for S3 and S5, and 2.5~mag for S2, or as excess emission for S1, S4, and S6 arising from thermal bremsstrahlung of these early-type stars.

The spectroscopic distance of each source, namely being derived from the spectroscopically determined extinction, apparent ($JHK$) and absolute magnitudes \citep{pec13} are by and large confirmed by those computed from the Gaia DR2 parallax using the estimates by \citet{bai18}.  As seen in Table~\ref{tab:s112_spe_par}, the distances are all consistent with being cluster members.  
Noteworthily the Gaia DR2 distance of BD$+$45\,3216 (S1) is $1.822_{-0.705}^{+1.704}$~kpc, whereas Gaia EDR3 has resolved both components of the source with relatively precise parallaxes ($0.4229\pm0.0477$~mas and $0.4094\pm0.0394$~mas), which correspond to geometric distances of $2.248_{-0.226}^{+0.275}$~kpc and $2.456_{-0.274}^{+0.271}$~kpc \citep{bai21}.  

Our spectroscopic observations yielded only a single massive (O8\,V) star capable of producing sufficient ultraviolet radiation, and responsible for creating and sustaining the ionized region. A color composite image of the S112 region is presented as Figure~\ref{fig:s112_jhk_spe} (right), generated with a combination of near- to mid-infrared wavebands, highlighting the locations of the spectroscopically observed stars.  A vivid rim-like structure, shaped semi-circularly, surrounding which trails of dust are prominent.  To the north, multiple filaments are seen interlaced with diffuse nebulosity, within which a significant amount of dust is contained.  The stellar feedback by the main illuminating source BD+45\,3216, i.e., S1 and the physical mechanism in formation of the bright arc-like structure are discussed in detail by \citet{panwar20}.  They have proposed a triggered star formation scenario as an outcome of the encounter between the ionized gas and molecular cloud for this blister-shaped \ion{H}{2} region.

\begin{rotatetable*}
\centerwidetable
\begin{deluxetable*}{cCCCC CCCcC CC}
\tablecaption{Parameters of the Spectroscopically Observed Stars}
\tabletypesize{\scriptsize}
\label{tab:s112_spe_par}
\tablehead{
\colhead{Star} & \colhead{2MASS}  & \colhead{Glon.} & \colhead{Glat.} & \colhead{$J$} & \colhead{$H$} & \colhead{$K$} & \colhead{Exposure} & \colhead{Spectral} & \colhead{Spectroscopic} & \colhead{Distance} & \colhead{Gaia DR2} \\
\colhead{ID} & \colhead{Designation}  & \colhead{(deg)} & \colhead{(deg)} & \colhead{(mag)} & \colhead{(mag)} & \colhead{(mag)} & \colhead{Time (s)} & \colhead{Type} & \colhead{$A_{V}$ (mag)} & \colhead{(kpc)} & \colhead{Distance (kpc)} 
           }
\startdata
S1  &  2MASS J20335033+4539412  &  083.783904  &  +03.288649  &  8.182  &  8.099  &  8.062  &  300  &  O8\,V  &  1.916  &  1.965  &  1.822  \\
  &  &  &  &  \pm 0.019  &  \pm 0.020  &  \pm 0.020  &  &  &  \pm 0.065  &  \pm 0.154  &  _{-0.705}^{+1.704}  \\
S2  &  2MASS J20332868+4535473  &  083.694004  &  +03.300749  &  12.870  &  12.227  &  12.058  &  1500  &  G3\,V/III  &  2.509  &  \nodata  &  1.971  \\
  &  &  &  &  \pm 0.021  &  \pm 0.021  &  \pm 0.019  &  &  &  \pm 0.128  &  &  _{-0.084}^{+0.092}  \\
S3  &  2MASS J20342691+4542206  &  083.883029  &  +03.229439  &  10.836  &  10.181  &  10.002  &  1200  &  F7\,V/III  &  3.206  &  \nodata  &  1.990  \\
  &  &  &  &  \pm 0.020  &  \pm 0.021  &  \pm 0.018  &  &  &  \pm 0.191  &  &  _{-0.060}^{+0.064}  \\
S4  &  2MASS J20333266+4545419  &  083.834016  &  +03.389542  &  12.920  &  12.766  &  12.652  &  1200  &  B9\,V  &  1.693  &  2.158  &  2.044  \\
  &  &  &  &  \pm 0.024  &  \pm 0.029  &  \pm 0.030  &  &  &  \pm 0.081  &  \pm 0.238  &  _{-0.071}^{+0.076}  \\
S5  &  2MASS J20330798+4535366  &  083.655728  &  +03.347573  &  11.079  &  10.326  &  10.110  &  1200  &  F9\,V/III  &  3.823  &  \nodata  &  2.049  \\
  &  &  &  &  \pm 0.020  &  \pm 0.019  &  \pm 0.016  &  &  &  \pm 0.204  &  &  _{-0.068}^{+0.073}  \\
S6  &  2MASS J20342791+4547203  &  083.951724  &  +03.276741  &  12.942  &  12.815  &  12.712  &  1200  &  B9.5--A0\,V  &  1.426  &  2.214  &  1.985  \\
  &  &  &  &  \pm 0.020  &  \pm 0.024  &  \pm 0.024  &  &  &  \pm 0.056  &  \pm 0.199  &  _{-0.067}^{+0.072}  \\
\hline 
\enddata
\end{deluxetable*}
\end{rotatetable*}

\subsection{Properties of the Ionized Gas}
 \label{ssec:s112_ionized}

\begin{figure*}
\centering
        \includegraphics[width=\textwidth]{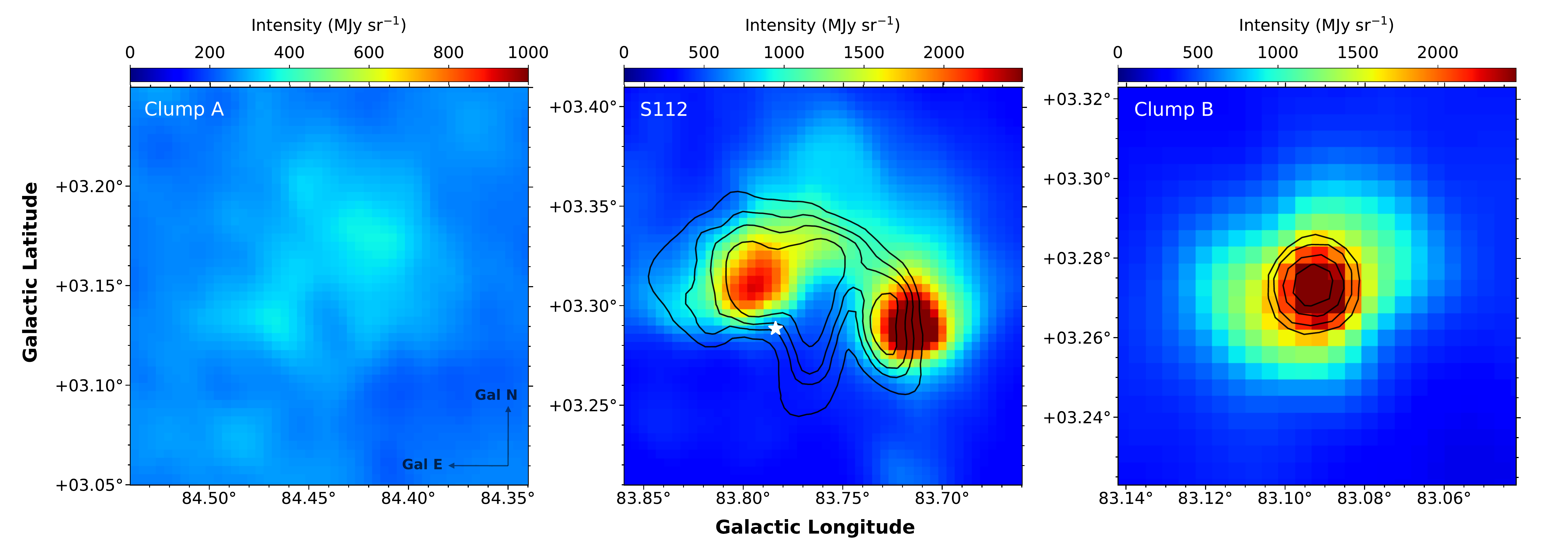}
    \caption{The AKARI 160~\micron\ emission (false color), superimposed with the NVSS 1.4~GHz radio continuum (contours) for Clump~A, S112, and Clump~B.  For Clump~A, no 1.4~GHz emission is detected.  The contour levels for S112 are at 0.003, 0.010, 0.020, and 0.040~Jy/beam, whereas the contour levels for Clump~B are at 0.003, 0.006, 0.012, and 0.020~Jy/beam. The main illuminating source BD+45\,3216 is indicated by a white asterisk symbol.  }
    \label{fig:s112_aka_nvs}
\end{figure*}

Compact \ion{H}{2} regions are distinct free-free radio sources.  The NVSS 1.4~GHz radio  continuum observations hence allow us to estimate the physical parameters of the ionized gas associated with the subregions.  The radio continuum distribution, overlaid on a far-infrared 160~\micron\ map for the three subregions, is shown in Figure~\ref{fig:s112_aka_nvs} and the parameters thus derived are given in Table~\ref{tab:s112_ion_gas}.  The number of Lyman continuum photons ($N_{UV}$), considering an optically thin, homogeneous and spherical \ion{H}{2} region, is given by \citep{mat76}

\begin{equation}
\label{eqn:s112_eq1}
\begin{split}
N_{UV} [{\rm s}^{-1}] = 7.5\, \times\, 10^{46}\, \left(\frac{S_{\nu}}{\rm Jy}\right)\,     
  \left(\frac{D}{\rm kpc}\right)^{2}\, \times\, \\
\left(\frac{T_{e}}{10^{4}~{\rm K}}\right)^{-0.45}\, \left(\frac{\nu}{\rm GHz}\right)^{0.1}
\end{split}
\end{equation}
where $S_{\nu}$ is the integrated 1.4~GHz flux density in Jy, $D$ is the distance in kpc, $T_{e}$ is the electron temperature, and $\nu$ is the frequency in GHz.  Adopted values are the measured radio fluxes listed in Table~\ref{tab:s112_ion_gas}, a distance of 2.1~kpc (Table~\ref{tab:s112_spe_par}), and an electron temperature of $10^{4}$~K \citep{mat76}.  

For Clump~A ($\ell = 084\fdg4325$; $b = +03\fdg1598$), little 1.4~GHz radio emission is detected therefore no parameters of ionized gas are available in Table~\ref{tab:s112_ion_gas}.  \citet{esi08} also have reported that this relatively cold core does not show any evidence of \ion{H}{2} emission, with no  presence of MSX \citep{ega03} or IRAS \citep{neu84} point sources.  
In lieu, Clump~A is designated as a cold core in the Planck Catalogue of Galactic Cold Clumps \citep[PGCC\,G084.43$+$03.16;][]{pla16}, which are good candidates for studies of the early evolutionary stages of star formation.  
In  S112, on the other hand two radio peaks are distinct, though the main exciting star (BD+45\,3216) coincides with neither of the peaks.  
Within this region two MSX sources (MSX\,G083.7071$+$03.2817 and MSX\,G083.7962$+$03.3058) are located and their interaction with the region is discussed in detail by \citet{panwar20}.  Studying both the ionized gas and the molecular cloud in association with the MSX sources near the massive star BD$+$45\,3216, they have proposed a sequential star formation scenario.  
Clump~B shows a single and definite maximum with a moderate ionization level.  
\citet{urq09} have presented the Red MSX Source (RMS) survey, programmed to detect the massive young stellar candidates in the northern hemisphere using the Very Large Array (VLA) high resolution radio continuum observations at 6~cm.  They have detected radio emission with an integrated flux of 101~mJy from the \ion{H}{2} region associated with Clump~B and catalogued the region as VLA\,G083.0934$+$03.2720 with a MSX counterpart as MSX\,G083.0936$+$03.2724.  Also \citet{and15} has mentioned this \ion{H}{2} region as G083.097$+$03.270 with an ionized gas velocity of $-7.1$~km~s$^{-1}$ traced by the radio recombination lines.  

Using the Lyman continuum flux to estimate the spectral class of a single ionizing star \citep{panagia73}, in S112 it would be O9.5--B0\,V,  which is in close agreement with our spectroscopic determinations (O8~V). 
Evidently from Figure~\ref{fig:s112_aka_nvs} the coincidence of dust emission with radio  continuum differs: in S112, both emissions near BD+45\,3216 are comparatively devoid, indicating the \ion{H}{2} region to be exclusively powered by BD+45\,3216.  
In contrary, in Clump~B a higher intensity plus the compact morphology points to the possibility of ionization caused by a group of early-B type stars still embedded in the cloud, rather than a single O-type star alone \citep{martin03}.  
Based on the infrared luminosity function, \citet{iva05} too suggested that Clump~B hosts a concentration of about half a dozen ionizing stars.  
This conforms to the notion that OB stars often exist in binaries or groups.

The size of the ionized gas as a function of the the total Lyman photons and the gas density is known as the Str\"{o}mgren radius ($R_{s}$), given by

\begin{equation}
\label{eqn:s112_eq2}
  R_{s} = (3 N_{UV} / 4 \pi n^{2}_{\rm{0}} \alpha_{B})^{1/3},
\end{equation}
where the radiative recombination coefficient $\alpha_{B}$ =  2.6 $\times$ 10$^{-13}$ (10$^{4}$ K/$T_{e}$)$^{0.7}$~cm$^{3}$~s$^{-1}$ \citep{kwa97}, and $n_{0}$ is the initial particle number density of the ambient neutral gas, typically $n_0\sim 10^{3}$~cm$^{-3}$. 

The dynamical age ($t_{dyn}$) of an \ion{H}{2} region is computed using the relation \citep{dys80}: 

\begin{equation}
\label{eqn:s112_eq3}
  t_{dyn} = \left(\frac{4 R_{s}}{7 c_{s}}\right) \,
   \left[ \left(\frac{R_{\rm H\,II}}{R_{s}}\right)^{7/4} - 1\right],
\end{equation}
where $c_{s}= 11$~km~s$^{-1}$ is the isothermal sound speed in the ionized gas   \citep{bis09}, and $R_{\rm H\,II}$ is the radius of the \ion{H}{2} region.  Adopting values in Table~\ref{tab:s112_ion_gas}, the dynamical age of the \ion{H}{2} regions is found to be 0.18--1.0~Myr, which is consistent with our earlier estimates (Section~\ref{ssec:s112_yso_age}).  The S112 region appears relatively evolved compared to the other two, though varying dust density may induce substantial changes in the dynamical age determination.

\begin{deluxetable*}{cccccccccc}
\tablecaption{Physical parameters of the ionized gas associated with the subregions}
\tabletypesize{\footnotesize}
\label{tab:s112_ion_gas}

\tablehead{
\colhead{Region} & \colhead{Glon.} & \colhead{Glat.} & \colhead{$R_{\rm H\,II}$} & \colhead{$S_{\nu}$} & \colhead{$N_{UV}$} & \colhead{$\log (N_{UV}$)} & \colhead{Star} & \colhead{$R_{s}$} & \colhead{$t_{dyn}$} \\
\colhead{ } & \colhead{(deg)} & \colhead{(deg)} & \colhead{(pc)} & \colhead{(mJy)} & \colhead{(s~$^{-1}$)} & \colhead{} & \colhead{type} & \colhead{(pc)} & \colhead{(Myr)}
           }
\startdata
%
%
S112  &  083.7589  &  +03.2750  &  3.038  &  1414  &  $4.836 \times 10^{47}$  &  47.684  &  O9.5--B0~V  &  0.2473  &  1.023  \\
Clump~B  &  083.0890  &  +03.2693  &  0.668  &  32.39  &  $1.108 \times 10^{46}$  &  46.044  &  B0.5--B1~V  &  0.0702  &  0.184  \\
\hline 
\enddata
\end{deluxetable*}

\section{Molecular Cloud Morphology}
 \label{sec:s112_mol}

We now present the molecular cloud parameters derived from the CO data.  The molecular gas traced by $^{12}$CO reveals structures of gas (density $\sim 10^2$~cm$^{-3}$), whereas the optically thinner $^{13}$CO or C$^{18}$O line traces denser ($\sim 10^{3}$--$10^{4}$~cm$^{-3}$) parts of the cloud.  A combination of $^{12}$CO, $^{13}$CO, and C$^{18}$O isotopologues thus provides complementary information of the molecular cloud morphology, from cloud envelopes to denser segments as giant molecular clouds, to cloud fragments to pre-stellar cores, and of cloud physical/chemical conditions. The detailed methodology to derive the cloud parameters from these observations are given in \citet{sun20}.

\subsection{Intensity Distribution}
 \label{ssec:s112_mol_intensity}

\begin{figure*}
\centering
        \includegraphics[width=1.0\textwidth]{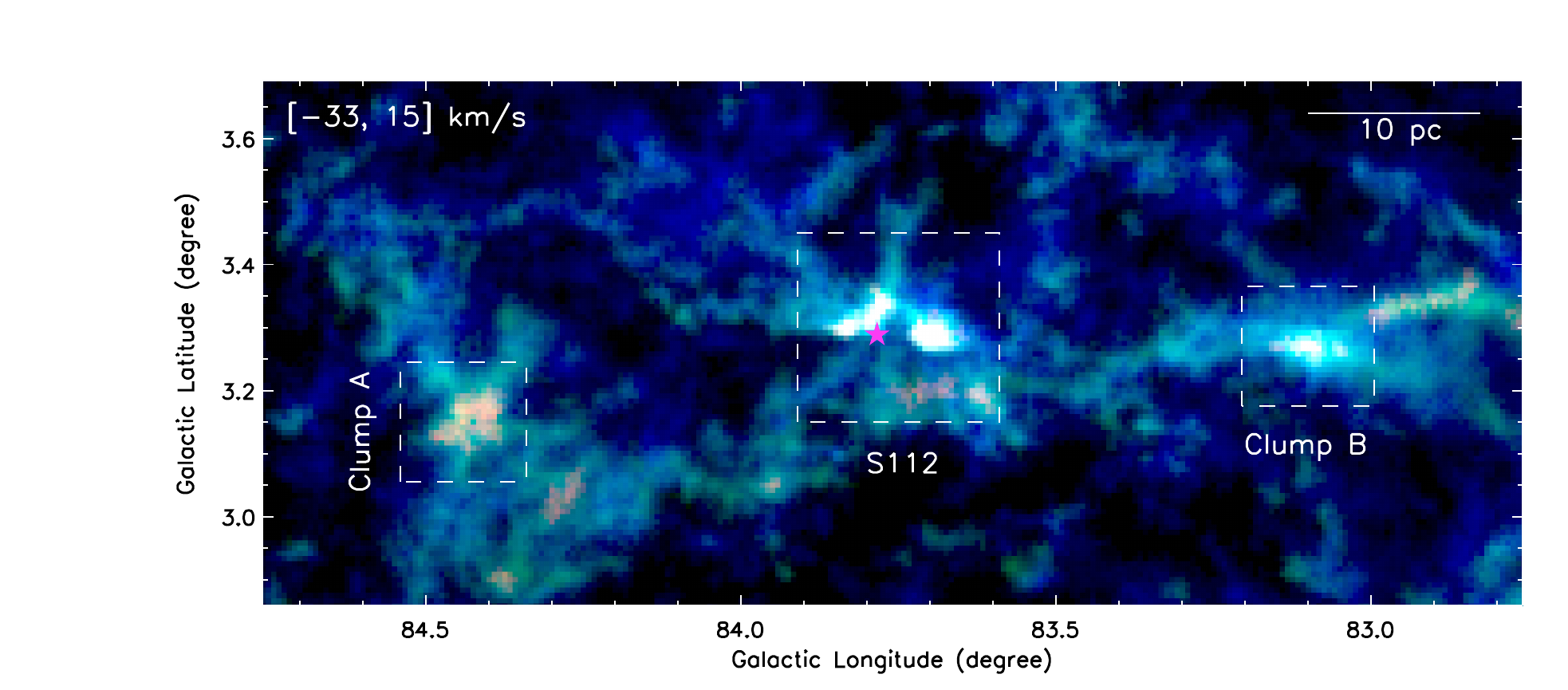}
    \caption{Integrated intensity map (moment~0) over the velocity range [$-33$, 15]~km~s$^{-1}$, created with a combination of $^{12}$CO (blue), $^{13}$CO (green), and C$^{18}$O (red).  Each of the three subregions is marked with a dashed rectangle.  The principal ionizing star BD+45\,3216, located within the S112 subregion, is indicated by an asterisk symbol.}
    \label{fig:s112_int_m0}
\end{figure*}

A color composite map of the integrated intensity in the velocity interval of $-33$ to 15~km~s$^{-1}$, constructed with a combination of three CO isotopologues is presented in Figure~\ref{fig:s112_int_m0}.  Large-scale extended structures are prominent in $^{12}$CO emission, whereas in the $^{13}$CO map, intricate filaments are seen connecting the subregions Clump~A and S112, along which Class~I objects are detected (see the YSO distribution in Figure~\ref{fig:s112_pla_yso}).  The complete filament extends over $\sim 80$~pc parallel to the Galactic plane, along which cloud fragments/cores in addition to the three subregions are located.  The C$^{18}$O emission is detected mostly at the location of the three subregions, where the compact structures are.  
A summary of the molecular cloud parameters for the three subregions is presented in Table~\ref{tab:s112_mol_par}.

\subsection{Velocity Distribution}
 \label{ssec:s112_mol_velocity}

\begin{figure*}
\centering
    \includegraphics[width=0.7\textwidth]{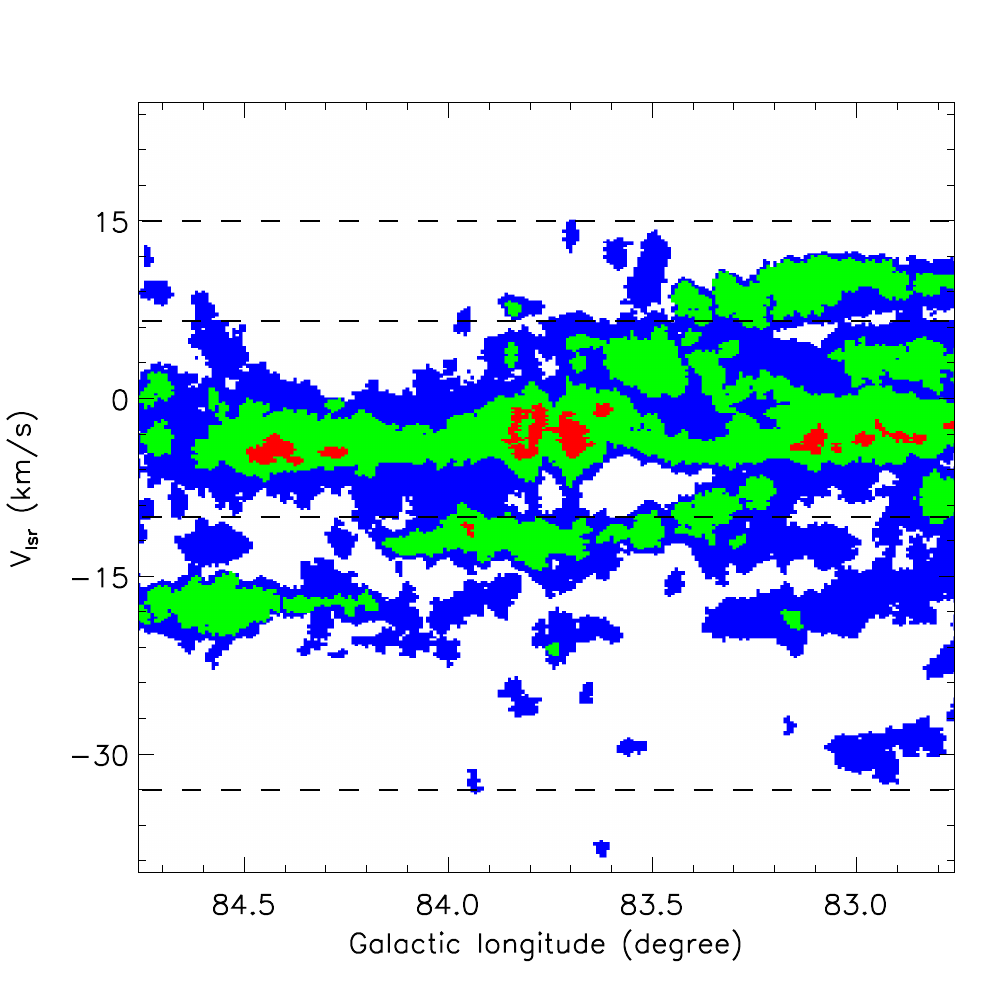}
    \caption{Longitude-velocity ($\ell-V$) map of the entire field toward S112, integrated over the Galactic latitude range of $b = [2\fdg8583, 3\fdg6917]$.  The blue, green, and red colors indicate $^{12}$CO, $^{13}$CO, and C$^{18}$O emission, respectively.  The four horizontal dashed lines denote $V_{lsr} = -33$, $-10$, 6.5, and 15~km~s$^{-1}$.  }
    \label{fig:s112_lv_rgb}
\end{figure*}

\begin{figure*}
\centering
        \includegraphics[width=1.0\textwidth]{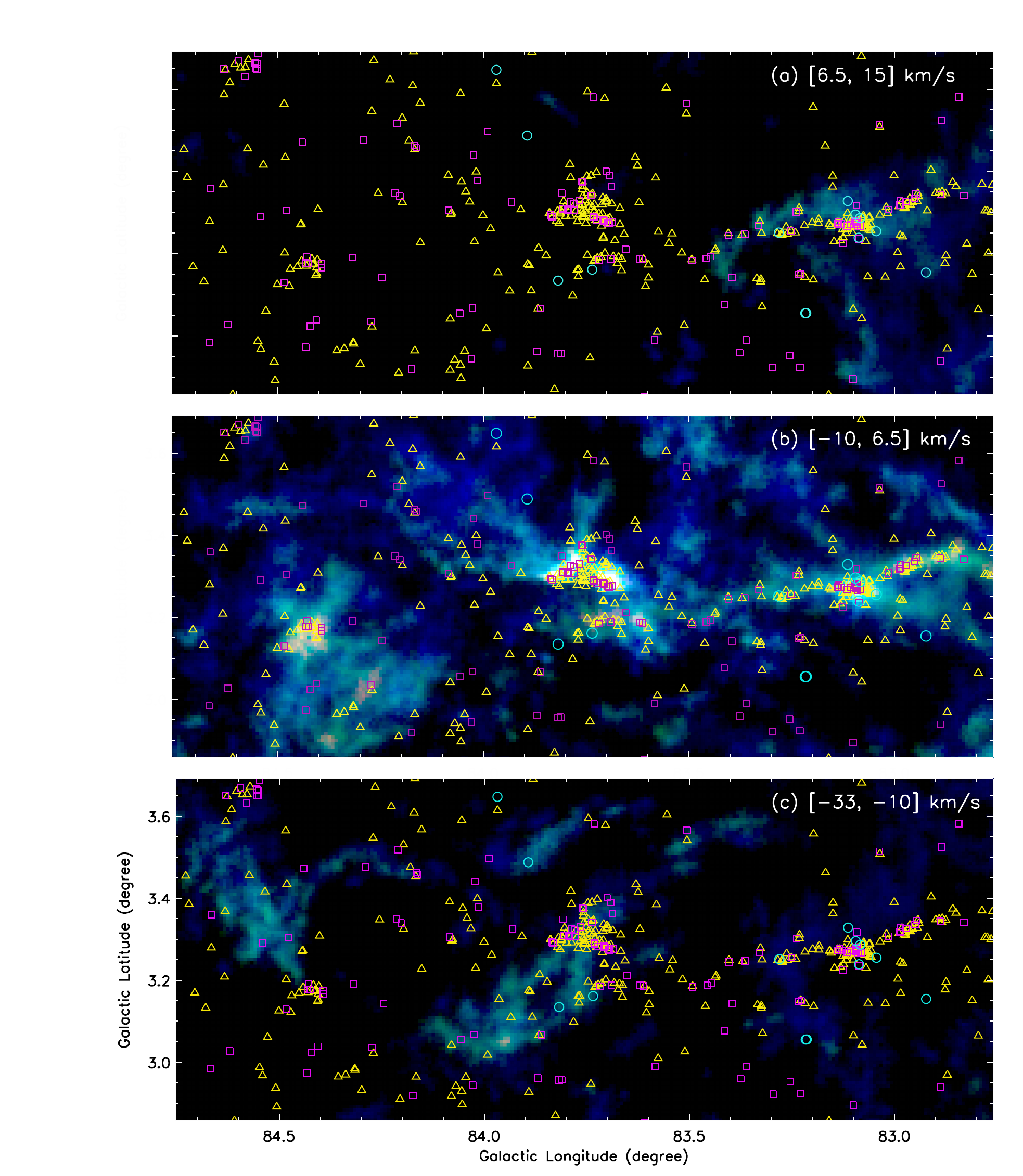}
    \caption{The channel maps for $^{12}$CO (blue), $^{13}$CO (green), and C$^{18}$O (red) with the velocity range depicted in each panel.  Overplotted are the YSOs (Class~I: squares; Class~II: triangles; transition disk objects: circles).}
    \label{fig:s112_channel_abc}
\end{figure*}

The overall velocity distribution (longitude-velocity map) of molecular gas within the mapped region is displayed as Figure~\ref{fig:s112_lv_rgb}.  The velocity range ([$-33$, 15]~km~s$^{-1}$) for the complete molecular structure is relatively wide.  Therefore, the cloud structures are studied in three velocity channels in [$-33$, $-10$]~km~s$^{-1}$, [$-10$, +6.5]~km~s$^{-1}$, and [6.5, 15]~km~s$^{-1}$, where prominent emission is detected.  The velocity structures for the three channels are shown in Figure~\ref{fig:s112_channel_abc}.  For C$^{18}$O, significant emission is detected only in the [$-10$, 6.5]~km~s$^{-1}$ channel.  Clearly the velocity channels of [$-33$, $-10$]~km~s$^{-1}$ and [6.5, 15]~km~s$^{-1}$ are not consistent with the cloud distribution, and hence the preferred velocity range of the molecular cloud would be [$-10$, 6.5]~km~s$^{-1}$.

\begin{figure*}
\centering
        \includegraphics[width=1.0\textwidth]{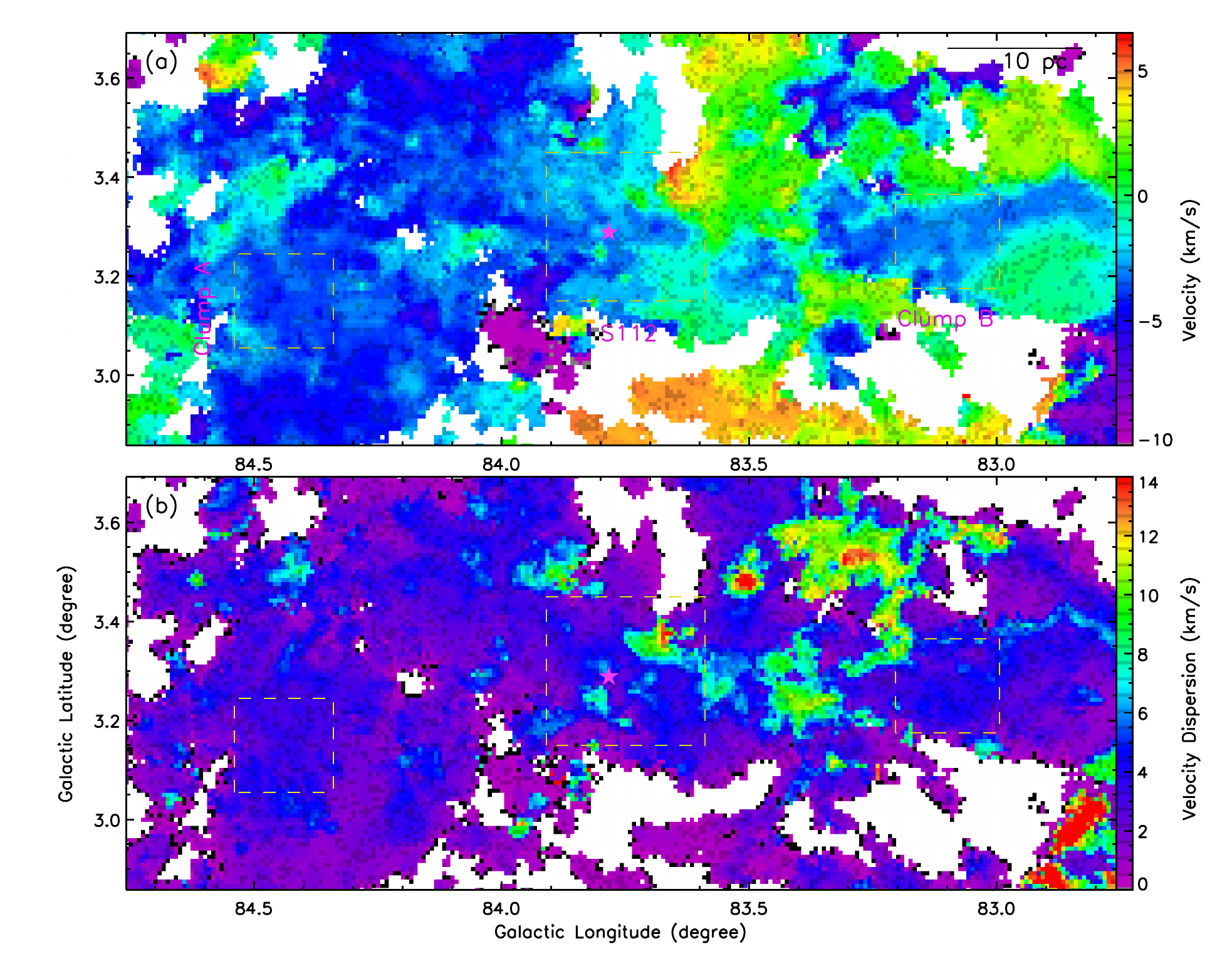}
    \caption{Velocity (moment~1) and velocity dispersion (moment~2) maps for $^{12}$CO in the velocity range [$-10$, 6.5]~km~s$^{-1}$.}
    \label{fig:s112_12co_vel}
\end{figure*}

\begin{figure*}
\centering
        \includegraphics[width=1.0\textwidth]{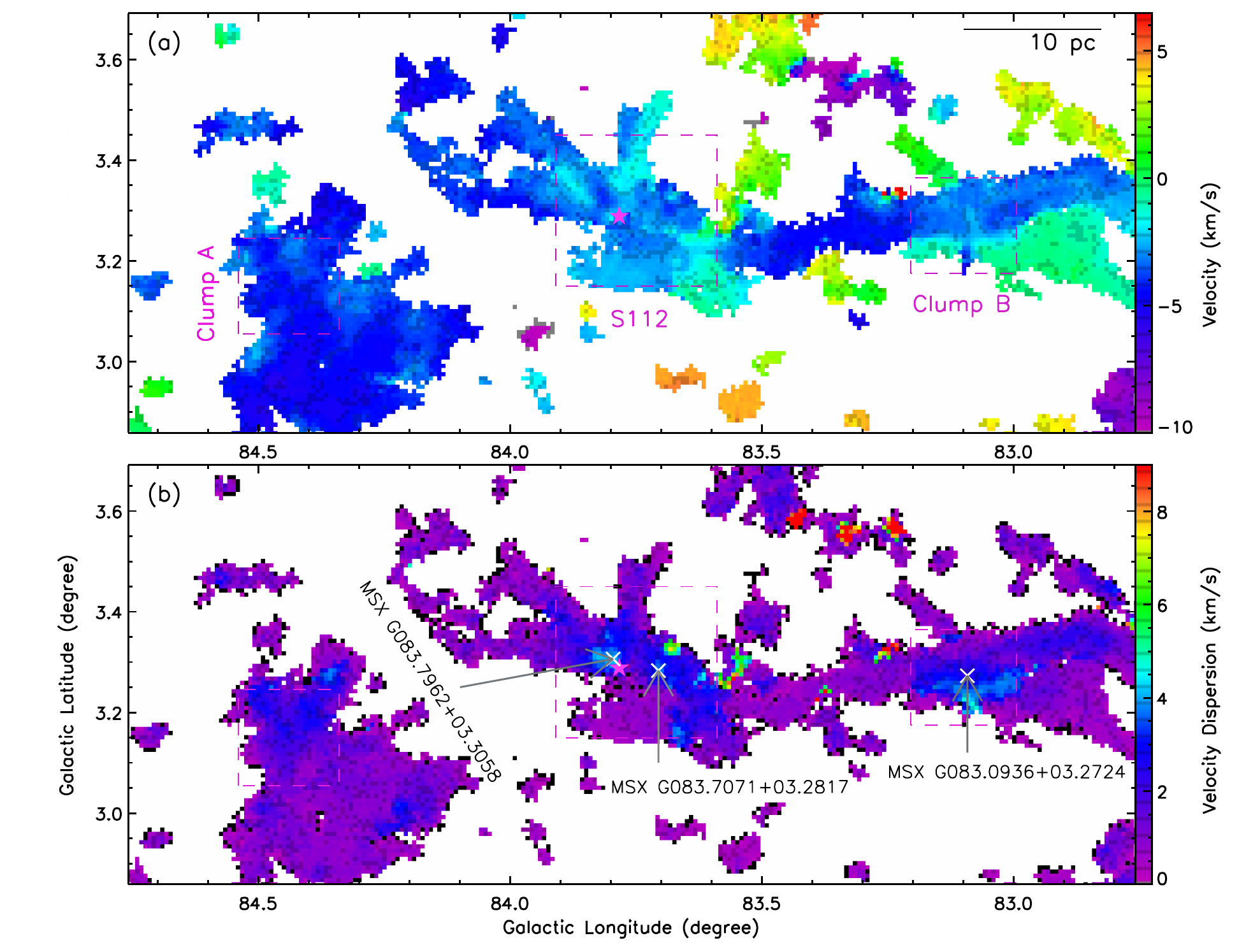}
    \caption{Moment~1 and moment~2 maps for $^{13}$CO.  The MSX sources, two in S112 and another in Clump~B are overlaid with cross symbols.}
    \label{fig:s112_13co_vel}
\end{figure*}

\begin{figure*}
\centering
        \includegraphics[width=1.0\textwidth]{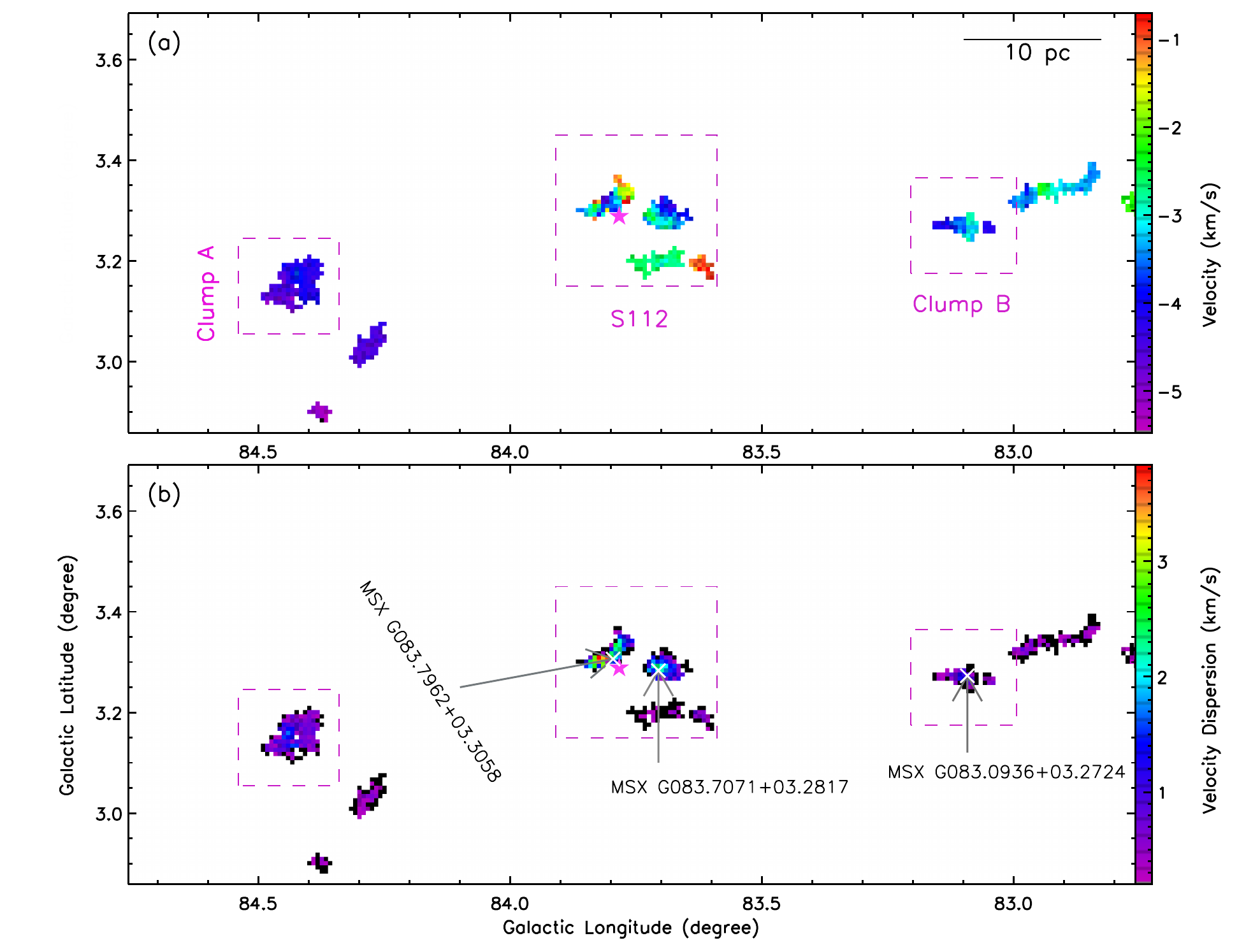}
    \caption{Moment~1 and moment~2 maps for C$^{18}$O.}
    \label{fig:s112_c18o_vel}
\end{figure*}

The velocity and its dispersion maps are displayed in Figure~\ref{fig:s112_12co_vel} for $^{12}$CO, Figure~\ref{fig:s112_13co_vel} for $^{13}$CO, and Figure~\ref{fig:s112_c18o_vel} for C$^{18}$O.  The moment~1 (velocity) map in $^{12}$CO shows considerable velocity gradient from Galactic east to west.  Toward the Galactic east, velocity varies almost uniformly between $-2.5$ and $-5.0$~km~s$^{-1}$ with an average of $-3.6$~km~s$^{-1}$,  whereas in the Galactic west, the velocity fluctuates more, except along the filamentary axis where the velocity remains $\sim-2.5$~km~s$^{-1}$.  The velocity dispersion (moment~2) is also lower (average $\sim 3.2$~km~s$^{-1}$) in the Galactic east and increases slightly (average $\sim 3.8$~km~s$^{-1}$) to the west.  

The moment~1 map of $^{13}$CO, being optically thinner, traces readily the molecular cloud distribution.  The velocity varies almost consistently along the entire filamentary structure from $-4.0$~km~s$^{-1}$ (east) to $-2.5$~km~s$^{-1}$ (west). The velocity dispersion is also much smaller, averaging $\sim 1.8$~km~s$^{-1}$, compared to that of $^{12}$CO.  Similar results apply to the C$^{18}$O map, where the velocity increases gradually from the Galactic east to west. 

In the velocity dispersion maps of $^{13}$CO (Figure~\ref{fig:s112_13co_vel}(b)) and C$^{18}$O (Figure~\ref{fig:s112_c18o_vel}(b)), the overlap of three MSX sources mentioned in Section~\ref{ssec:s112_ionized} are shown.  Interestingly the regions of elevated velocity dispersion nicely coincide with all the MSX source positions.  Toward S112, \citet{maud15b} reported massive (27.6~M$_{\sun}$) molecular outflow associated with MSX\,G083.7071$+$03.2817 having $V_{lsr} = -3.2$~km~s$^{-1}$.  Another source MSX\,G083.7962$+$03.3058 is less massive (5.0~M$_{\sun}$) and shows some evidence of outflow with $V_{lsr} = -3.7$~km~s$^{-1}$.  \citet{panwar20} also mentioned probable association of molecular gas with these outflow sources.  In the study of \citet{maud15b}, toward Clump~B, MSX\,G083.0936$+$03.2724 is highly luminous ($12 \times 10^{3}$~L$_{\sun}$) and is relatively massive (7.7~M$_{\sun}$) showing clear signature of molecular outflow with $V_{lsr} = -3.0$~km~s$^{-1}$.  These spatial and kinematic correlation between the outflows and the molecular gas imply that the high velocity dispersions are most probably related and caused by these outflows.  The outflow sources have dynamical timescales of 3--4$\times 10^{4}$~yr.

\subsection{Column Density Distribution}
 \label{ssec:s112_mol_column}

\begin{figure*}
\centering
        \includegraphics[width=1.0\textwidth]{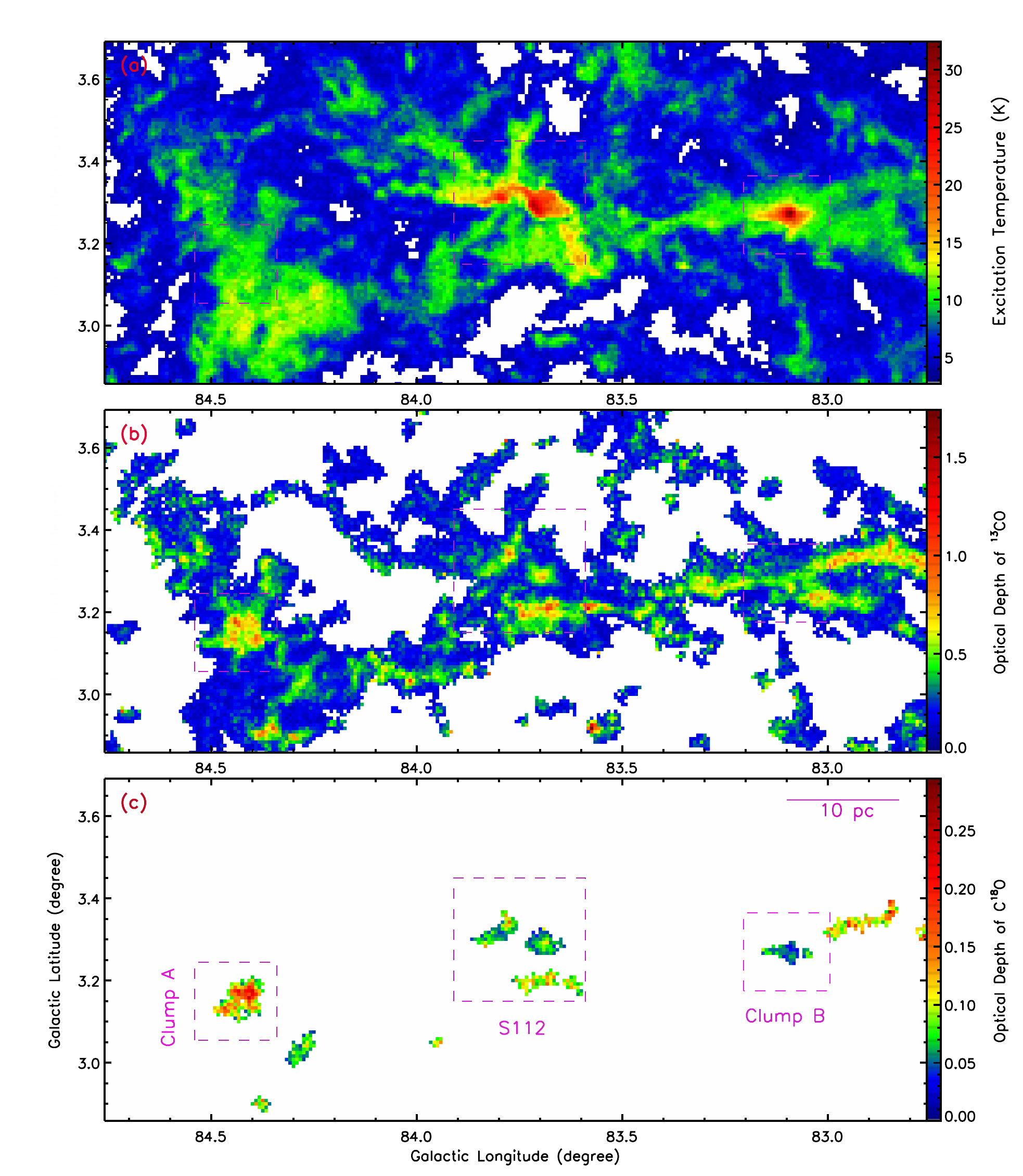}
    \caption{(a) Excitation temperature map for $^{12}$CO; optical depths for (b) $^{13}$CO and for (c) C$^{18}$O.}
    \label{fig:s112_tex_tau}
\end{figure*}

The distribution of the excitation temperature, derived from $^{12}$CO assuming an  optically thick condition, is presented in Figure~\ref{fig:s112_tex_tau}(a).  Along the filamentary cloud, the excitation temperature varies steadily with an average value of $\sim10$~K, which is similar to the typical kinetic temperature of molecular clouds, peaking near the three subregions, with the warmest for Clump~B (32.5~K), followed by S112 (28.6~K) and Clump~A (13.7~K).  The optical depths are presented in Figure~\ref{fig:s112_tex_tau}(b) for $^{13}$CO and in Figure~\ref{fig:s112_tex_tau}(c) for C$^{18}$O.  The optical depths are then used to make saturation corrections to the column densities.  

\begin{figure*}
\centering
        \includegraphics[width=1.0\textwidth]{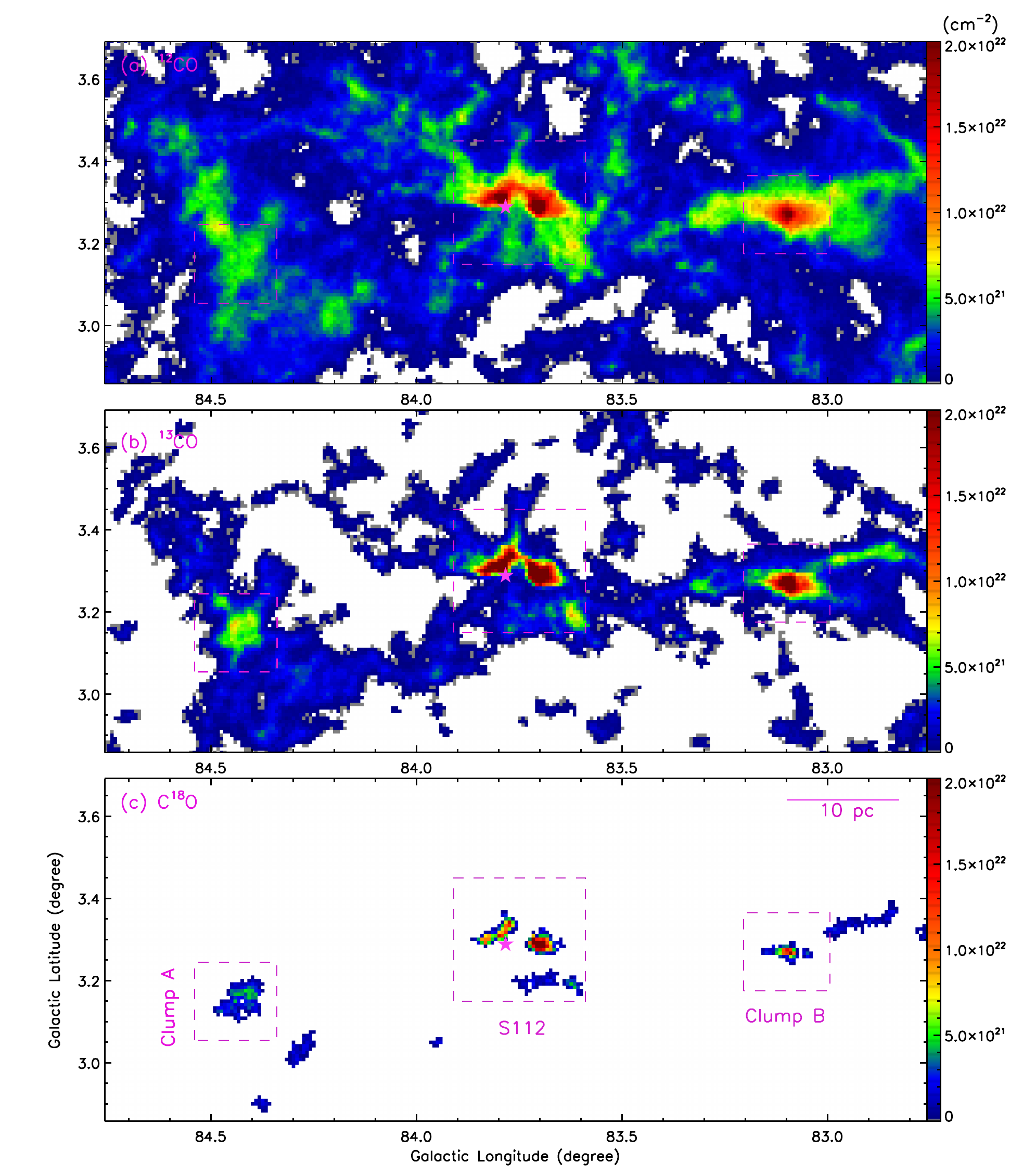}
    \caption{The H$_{2}$ column density distribution for $^{12}$CO, $^{13}$CO, and C$^{18}$O, respectively.}
    \label{fig:s112_col_den}
\end{figure*}

The H$_{2}$ column density is derived by adopting a CO-to-H$_{2}$ conversion factor, namely $X_{\mathrm{CO}} \equiv 2 \times 10^{20} {\rm cm}^{-2}~({\rm K~km~s}^{-1})^{-1}$ for $^{12}$CO and by assuming local thermodynamic equilibrium and abundance ratios of H$_{2}$ to CO for $^{13}$CO and for C$^{18}$O.  The H$_{2}$ column density map for the CO triplet is shown in Figure~\ref{fig:s112_col_den}.  
Within S112 and Clump~B, the column density varies on the order of 4--6 times  $10^{21}$~cm$^{-2}$, with the highest values ranging between 3--6 times  $10^{22}$~cm$^{-2}$.  

The corresponding values are relatively lower in Clump~A, averaging 2--3 times  $10^{21}$~cm$^{-2}$, peaking about $\sim 7.5 \times~10^{21}$~cm$^{-2}$.  For the results presented here, the C$^{18}$O emission is relatively discrete due to less velocity crowding and line blending \citep{su19}, in comparison with the extended $^{12}$CO and $^{13}$CO emission.  Therefore, the C$^{18}$O emission is useful in detecting the denser components, which are mainly associated with cloud cores.  

To combine the facts that we found prominent star formation activity mostly at the higher column density zones.  These results seem well consistent with the simulated analysis of \citet{clark14} according to which in the local interstellar medium the star formation is only possible when the area-averaged gas column density exceeds $10^{21}$~cm$^{-2}$.  Consequently similar results have been observationally obtained by \citet{das21} showing the correlation between star formation rate and gas column density in the Milky Way clouds.

\subsection{Molecular Mass}
 \label{ssec:s112_mol_mass}
Next we have derived the molecular cloud mass for the CO isotopologues from the integrated column density maps by adopting the velocity range [$-$33, 15]~km~s$^{-1}$ and a distance of 2.1~kpc.  The H$_{2}$ mass are obtained by assuming a constant CO-to-H$_{2}$ conversion factor of $2~\times~10^{20}~{\rm cm}^{-2}~({\rm K~km~s}^{-1})^{-1}$ \citep{bol13} for $^{12}$CO and local thermodynamic equilibrium (LTE) and abundance ratios of H$_{2}$ to $^{13}$CO for $^{13}$CO, respectively \citep{sun20}.  \\
We found the $^{13}$CO mass of Clump~A as $2.58 \times 10^{3}$~M$_\sun$.  
Whereas \citet{esi08} have computed the $^{13}$CO cloud mass as M$_{LTE}$ = $4.37 \times 10^{3}$~M$_\sun$ with a virial parameter of 0.25, which is well below the threshold value ($\lesssim 2$) for the onset of collapse of supercritical cloud fragments \citep{kau13}.  Note, however, that the low virial parameters for massive molecular clumps could be an artifact \citep{singh21}.  
In the PGCC catalogue \citep{pla16}, the mass of PGCC\,G084.43+03.16 associated with Clump~A was computed to be $4.30 \times 10^{3}$~M$_{\sun}$, based on an overestimated distance of 4.26~kpc.  But considering a proper clump distance of 2.1~kpc, the rescaled mass turns to be $\sim 1.03 \times 10^{3}$~M$_{\sun}$.  \citet{dob94} mentioned the molecular cloud as 084.4$+$03.2 and obtained similar kinematic parameters.  
In spite of significant deviations in the previously estimated masses, these results so far indicate that Clump~A is indeed a high-mass star forming region which is in the earliest of its evolutionary phase.  \\ 
S112 is a massive clump with a total $^{13}$CO mass of $9.02 \times 10^{3}$~M$_{\sun}$.  This clump merges with the molecular cloud 083.7$+$03.3 listed in \citet{dob94}.  \\ 
The molecular mass of Clump~B estimated from $^{13}$CO is $4.27 \times 10^{3}$~M$_{\sun}$.  By and large, Clump~B has the possibility to be associated with the molecular cloud 083.1$+$03.3 from \citet{dob94}, separating in angular scale of $\sim2.16\arcmin$.  Their cloud also depicts kinematic velocity ($V_{lsr} = -3.3$~km~s$^{-1}$) and distance (2.1~kpc) similar to those in Clump~B.  

\begin{rotatetable*}
\centerwidetable
\begin{deluxetable*}{cCCCC CCCCC C}
\tablecaption{Parameters of the molecular clouds associated with the subregions}
\tabletypesize{\scriptsize}
\label{tab:s112_mol_par}
\tablehead{
\colhead{CO} & \multicolumn{2}{c}{Intensity}  & \colhead{Velocity} & \colhead{Velocity Dispersion} & \multicolumn{2}{c}{Excitation Temperature} & \colhead{Optical Depth} & \multicolumn{2}{c}{$\rm H_{2}$ Column Density} & \colhead{$\rm H_{2}$ Mass} \\
\colhead{Survey} & \multicolumn{2}{c}{($\rm K~km~s^{-1}$)}  & \colhead{($\rm km~s^{-1}$)} & \colhead{($\rm km~s^{-1}$)} & \multicolumn{2}{c}{($\rm K$)} & \colhead{} & \multicolumn{2}{c}{($\rm 10^{21}$~cm$^{-2}$)} & \colhead{($\rm 10^{3}~M_{\sun}$)} \\
\cline{2-3} \cline{6-7} \cline{9-10} 
\colhead{} & \colhead{Peak}  & \colhead{Mean} & \colhead{} & \colhead{} & \colhead{Peak} & \colhead{Mean} & \colhead{} & \colhead{Peak}  & \colhead{Mean} & \colhead{} 
           }
\startdata
\hline
\multicolumn{11}{c}{Clump~A ($0\fdg20 \times 0\fdg19$)} \\
\hline
$^{12}$CO  &  35.68  &  18.03  &  -3.60  &  3.23  &  13.68  &  9.12  &  \nodata  &  7.14  &  3.61  &  4.13  \\
$^{13}$CO  &  9.82  &  3.08  &  -4.02  &  1.64  &  \nodata  &  \nodata  &  0.33  &  7.82  &  2.25  &  2.58  \\
\hline
\multicolumn{11}{c}{S112 ($0\fdg32 \times 0\fdg30$)} \\
\hline
$^{12}$CO  &  158.73  &  23.75  &  -2.36  &  3.88  &  28.61  &  10.06  &  \nodata  &  31.75  &  4.75  &  13.84  \\
$^{13}$CO  &  37.40  &  3.19  &  -2.61  &  1.72  &  \nodata  &  \nodata  &  0.25  &  64.90  &  3.10  &  9.02  \\
\hline
\multicolumn{11}{c}{Clump~B ($0\fdg21 \times 0\fdg19$)} \\
\hline
$^{12}$CO  &  105.11  &  29.48  &  -2.48  &  3.85  &  32.46  &  10.29  &  \nodata  &  21.02  &  5.90  &  6.73  \\
$^{13}$CO  &  27.38  &  4.04  &  -2.55  &  1.96  &  \nodata  &  \nodata  &  0.32  &  49.78  &  3.74  &  4.27  \\
\hline 
\enddata
\end{deluxetable*}
\end{rotatetable*}

\subsection{Star Formation Activity in the Filamentary Complex}
 \label{ssec:s112_mol_star}

We now discuss the global star formation processes in the region.  The YSOs are found exclusively at higher column density zones, which themselves are interconnected via filaments. The extinction pattern resembles the molecular cloud morphology.   This supports the notion of the ``hub-filament'' configuration where a hub, i.e., the conjunction of filaments along which, likely via magnetic fields, material channels to seed the formation of massive stars or star clusters \citep{kum22}.  The velocity measurements reveal the cloud geometry and kinematics.  Considering a median cloud velocity of $\sim -3.65$~km~s$^{-1}$ in $^{13}$CO, the axisymmetric Galactic rotation model \citep{rei19} produces a kinematic distance of $\sim 1.52 \pm 0.12$~kpc (local arm), marginally closer than that ($\sim 1.76 \pm 0.14$~kpc) of the main exciting star.  
In the velocity channel maps (Figure~\ref{fig:s112_channel_abc}), considerable amount of cloud structures were detected in the channels of $[-33, -10]$~km~s$^{-1}$ and $[6.5, 15]$~km~s$^{-1}$.  
The Galactic rotation model suggests these to be either background (distance $\gtrsim 3.0$~kpc) or foreground (distance $\lesssim 1.3$~kpc) clouds located along the same line-of-sight, although given the uncertainties in the distances the possibility of relation between these clouds cannot be ruled out.  A plausible explanation for the velocity gradient across the filamentary cloud is that the cloud distance varies, i.e., the eastern side cloud is tilted farther away compared to the western side.  

The higher fraction of Class~I objects toward the Galactic east indicates that the western side is relatively more evolved.  It is therefore proposed that star formation begun in the west, then propagated to the east along the filament axis.  Gravitational instability in the filaments initiated cloud fragmentation that hierarchically led to core collapse to form the protostars, as demonstrated by the several dense cloud cores indeed detected along the filament axis.  As seen in Figure~\ref{fig:s112_13co_vel}(a), the gas velocity is coherently high in the vicinity of BD+45\,3216, and decreases gradually further away.  Clearly the expanding \ion{H}{2} region has excavated the nearby clouds.  Toward Clump~A, gravity accelerates the gas where higher excitation temperatures ($\sim12$--13~K) are detected at velocity $< -4$~km~s$^{-1}$.  The entire cloud therefore encapsulates a complex site hosting young stellar subgroups at various evolutionary stages, with entangled filaments and fragments undergoing cloud disruption.

\section{Summary and Conclusions}
 \label{sec:s112_sum}

We have diagnosed the stellar contents in the S112 complex using multiwavelength data.  The interplay of young stellar evolution, dust clouds, and ionized gas is established to scrutinize the overall star formation history in the region. The key results are summarized as follows:

\begin{enumerate}

\item  The $K$-band extinction map in a $\sim 2\degr$ region uncovers the filamentary structure of dust along the Galactic east-west extension.  The overall extinction is moderate, averaging $A_{V} \sim 2.78$~mag, except a few clumps reaching maxima of $A_{V} \sim 17$~mag.

\item  The complex harbours more than 500 young objects identified using infrared excess, which have the best isochrone-fitting age of $\sim 1$~Myr.  Additionally, more than 350 H$\alpha$ emission-line stars are detected, standing out among even the most prominent star-forming regions, thereby signifying excessive ionization activity in the complex.  

\item  The spatial distribution of the YSOs shows a preferential alignment along the filamentary clouds, coinciding with the radio emissions, indicative of a star formation activity stretching an angular scale of $\sim2\degr$.  

\item  The molecular cloud traced by the CO gas has a median velocity of $\sim -3.65$~km~s$^{-1}$, with a gradual increase from the Galactic east ($\sim -4$~km~s$^{-1}$) to west ($\sim -2.5$~km~s$^{-1}$).  The filament, extending over $\sim 80$~pc, shows evidence of cloud fragmentation and formation of multiple compact and dense cloud cores that correlate with YSO groupings.  Overall, the filament temperature remains coherent ($\sim10$~K), except near dense cores (column density $> 10^{22}$~cm$^{-2}$) where it reaches up to $\sim32$~K.  

\item The highest concentrations of YSOs are confined at three locations, referred to as Clump~A, S112, and Clump~B, all interconnected via cloud filaments, with a  ``filament-hub'' configuration that supports the notion of mass channeling through filaments in radiant morphology to feed the massive stars or star clusters as hubs. 

\item  Analysis of the ionized gas associated with S112 and Clump~B using radio continuum observations leads to an estimate of their dynamical ages being  $\sim0.18$--1.0~Myr, consistent with the age estimation from the isochrone fitting of the young stars. 

\item Spectroscopy of luminous stars suggests the O8\,V star BD+45\,3216 to be the only viable ionizing source in the region.  Its location relative to the young population, ionized gas, and molecular cloud suggests a triggering star formation sequence in the blister-shaped S112 \ion{H}{2} region.

\end{enumerate}

\section*{Acknowledgements}

AP acknowledges the support by S. N. Bose National Centre for Basic Sciences, funded by the Department of Science and Technology, India.
We are thankful to the staff of the Indian Astronomical Observatory, Hanle and Centre for Research and Education in Science \& Technology, Hosakote for their assistance during observations at HCT, operated by the Indian Institute of Astrophysics, Bangalore.
This work has made use of data from the European Space Agency (ESA) mission {\it Gaia} (\url{https://www.cosmos.esa.int/gaia}), processed by the {\it Gaia} Data Processing and Analysis Consortium (DPAC, \url{https://www.cosmos.esa.int/web/gaia/dpac/consortium}).
This work makes use of data obtained as part of the IPHAS carried out at the INT, operated by the Isaac Newton Group.
Data products from the 2MASS, which is a joint project of the University of Massachusetts and the IPAC/Caltech are employed.
This publication has made use of data products from the WISE, which is a joint project of the University of California, Los Angeles, and the Jet Propulsion Laboratory/Caltech, funded by the National Aeronautics and Space Administration.
Part of this research is based on observations with AKARI, a Japan Aerospace Exploration Agency project with the participation of ESA.
Radio images are obtained from the NVSS archive, observed by the NRAO, a facility of the National Science Foundation operated under cooperative agreement by Associated Universities, Inc.
This work also uses observations made with the Planck mission, which is a project of the ESA with contributions provided by ESA member states.  
We are thankful to the referee for an insightful review that largely improved the scientiﬁc content of the paper.

\software{IRAF \citep{tod86}, GILDAS \citep{gil13}, Astropy Project \citep{astropy18}, APLpy \citep{robitaille19}}

\bibliography{ms_s112}{}

\begin{thebibliography}{}
\expandafter\ifx\csname natexlab\endcsname\relax\def\natexlab#1{#1}\fi
\providecommand{\url}[1]{\href{#1}{#1}}
\providecommand{\dodoi}[1]{doi:~\href{http://doi.org/#1}{\nolinkurl{#1}}}
\providecommand{\doeprint}[1]{\href{http://ascl.net/#1}{\nolinkurl{http://ascl.net/#1}}}
\providecommand{\doarXiv}[1]{\href{https://arxiv.org/abs/#1}{\nolinkurl{https://arxiv.org/abs/#1}}}

\bibitem[{{Alexander} {et~al.}(2013){Alexander}, {Kobulnicky}, {Kerton}, \&
  {Arvidsson}}]{ale13}
{Alexander}, M.~J., {Kobulnicky}, H.~A., {Kerton}, C.~R., \& {Arvidsson}, K.
  2013, \apj, 770, 1, \dodoi{10.1088/0004-637X/770/1/1}

\bibitem[{{Anderson} {et~al.}(2015){Anderson}, {Armentrout}, {Johnstone},
  {Bania}, {Balser}, {Wenger}, \& {Cunningham}}]{and15}
{Anderson}, L.~D., {Armentrout}, W.~P., {Johnstone}, B.~M., {et~al.} 2015,
  \apjs, 221, 26, \dodoi{10.1088/0067-0049/221/2/26}

\bibitem[{{Andr{\'e}} {et~al.}(2016){Andr{\'e}}, {Rev{\'e}ret}, {K{\"o}nyves},
  {Arzoumanian}, {Tig{\'e}}, {Gallais}, {Roussel}, {Le Pennec}, {Rodriguez},
  {Doumayrou}, {Dubreuil}, {Lortholary}, {Martignac}, {Talvard}, {Delisle},
  {Visticot}, {Dumaye}, {De Breuck}, {Shimajiri}, {Motte}, {Bontemps},
  {Hennemann}, {Zavagno}, {Russeil}, {Schneider}, {Palmeirim}, {Peretto},
  {Hill}, {Minier}, {Roy}, \& {Rygl}}]{andre16}
{Andr{\'e}}, P., {Rev{\'e}ret}, V., {K{\"o}nyves}, V., {et~al.} 2016, \aap,
  592, A54, \dodoi{10.1051/0004-6361/201628378}

\bibitem[{{Astropy Collaboration} {et~al.}(2018){Astropy Collaboration},
  {Price-Whelan}, {Sip{\H{o}}cz}, {G{\"u}nther}, {Lim}, {Crawford}, {Conseil},
  {Shupe}, {Craig}, {Dencheva}, {Ginsburg}, {VanderPlas}, {Bradley},
  {P{\'e}rez-Su{\'a}rez}, {de Val-Borro}, {Aldcroft}, {Cruz}, {Robitaille},
  {Tollerud}, {Ardelean}, {Babej}, {Bach}, {Bachetti}, {Bakanov}, {Bamford},
  {Barentsen}, {Barmby}, {Baumbach}, {Berry}, {Biscani}, {Boquien}, {Bostroem},
  {Bouma}, {Brammer}, {Bray}, {Breytenbach}, {Buddelmeijer}, {Burke},
  {Calderone}, {Cano Rodr{\'\i}guez}, {Cara}, {Cardoso}, {Cheedella}, {Copin},
  {Corrales}, {Crichton}, {D'Avella}, {Deil}, {Depagne}, {Dietrich}, {Donath},
  {Droettboom}, {Earl}, {Erben}, {Fabbro}, {Ferreira}, {Finethy}, {Fox},
  {Garrison}, {Gibbons}, {Goldstein}, {Gommers}, {Greco}, {Greenfield},
  {Groener}, {Grollier}, {Hagen}, {Hirst}, {Homeier}, {Horton}, {Hosseinzadeh},
  {Hu}, {Hunkeler}, {Ivezi{\'c}}, {Jain}, {Jenness}, {Kanarek}, {Kendrew},
  {Kern}, {Kerzendorf}, {Khvalko}, {King}, {Kirkby}, {Kulkarni}, {Kumar},
  {Lee}, {Lenz}, {Littlefair}, {Ma}, {Macleod}, {Mastropietro}, {McCully},
  {Montagnac}, {Morris}, {Mueller}, {Mumford}, {Muna}, {Murphy}, {Nelson},
  {Nguyen}, {Ninan}, {N{\"o}the}, {Ogaz}, {Oh}, {Parejko}, {Parley}, {Pascual},
  {Patil}, {Patil}, {Plunkett}, {Prochaska}, {Rastogi}, {Reddy Janga},
  {Sabater}, {Sakurikar}, {Seifert}, {Sherbert}, {Sherwood-Taylor}, {Shih},
  {Sick}, {Silbiger}, {Singanamalla}, {Singer}, {Sladen}, {Sooley},
  {Sornarajah}, {Streicher}, {Teuben}, {Thomas}, {Tremblay}, {Turner},
  {Terr{\'o}n}, {van Kerkwijk}, {de la Vega}, {Watkins}, {Weaver}, {Whitmore},
  {Woillez}, {Zabalza}, \& {Astropy Contributors}}]{astropy18}
{Astropy Collaboration}, {Price-Whelan}, A.~M., {Sip{\H{o}}cz}, B.~M., {et~al.}
  2018, \aj, 156, 123, \dodoi{10.3847/1538-3881/aabc4f}

\bibitem[{{Bailer-Jones} {et~al.}(2021){Bailer-Jones}, {Rybizki}, {Fouesneau},
  {Demleitner}, \& {Andrae}}]{bai21}
{Bailer-Jones}, C.~A.~L., {Rybizki}, J., {Fouesneau}, M., {Demleitner}, M., \&
  {Andrae}, R. 2021, \aj, 161, 147, \dodoi{10.3847/1538-3881/abd806}

\bibitem[{{Bailer-Jones} {et~al.}(2018){Bailer-Jones}, {Rybizki}, {Fouesneau},
  {Mantelet}, \& {Andrae}}]{bai18}
{Bailer-Jones}, C.~A.~L., {Rybizki}, J., {Fouesneau}, M., {Mantelet}, G., \&
  {Andrae}, R. 2018, \aj, 156, 58, \dodoi{10.3847/1538-3881/aacb21}

\bibitem[{{Barentsen} {et~al.}(2014){Barentsen}, {Farnhill}, {Drew},
  {Gonz{\'a}lez-Solares}, {Greimel}, {Irwin}, {Miszalski}, {Ruhland}, {Groot},
  {Mampaso}, {Sale}, {Henden}, {Aungwerojwit}, {Barlow}, {Carter}, {Corradi},
  {Drake}, {Eisl{\"o}ffel}, {Fabregat}, {G{\"a}nsicke}, {Gentile Fusillo},
  {Greiss}, {Hales}, {Hodgkin}, {Huckvale}, {Irwin}, {King}, {Knigge},
  {Kupfer}, {Lagadec}, {Lennon}, {Lewis}, {Mohr-Smith}, {Morris}, {Naylor},
  {Parker}, {Phillipps}, {Pyrzas}, {Raddi}, {Roelofs}, {Rodr{\'\i}guez-Gil},
  {Sabin}, {Scaringi}, {Steeghs}, {Suso}, {Tata}, {Unruh}, {van Roestel},
  {Viironen}, {Vink}, {Walton}, {Wright}, \& {Zijlstra}}]{bar14}
{Barentsen}, G., {Farnhill}, H.~J., {Drew}, J.~E., {et~al.} 2014, \mnras, 444,
  3230, \dodoi{10.1093/mnras/stu1651}

\bibitem[{{Bessell} \& {Brett}(1988)}]{bes88}
{Bessell}, M.~S., \& {Brett}, J.~M. 1988, \pasp, 100, 1134,
  \dodoi{10.1086/132281}

\bibitem[{{Bisbas} {et~al.}(2009){Bisbas}, {W{\"u}nsch}, {Whitworth}, \&
  {Hubber}}]{bis09}
{Bisbas}, T.~G., {W{\"u}nsch}, R., {Whitworth}, A.~P., \& {Hubber}, D.~A. 2009,
  \aap, 497, 649, \dodoi{10.1051/0004-6361/200811522}

\bibitem[{{Blitz} {et~al.}(1982){Blitz}, {Fich}, \& {Stark}}]{bli82}
{Blitz}, L., {Fich}, M., \& {Stark}, A.~A. 1982, \apjs, 49, 183,
  \dodoi{10.1086/190795}

\bibitem[{{Bolatto} {et~al.}(2013){Bolatto}, {Wolfire}, \& {Leroy}}]{bol13}
{Bolatto}, A.~D., {Wolfire}, M., \& {Leroy}, A.~K. 2013, \araa, 51, 207,
  \dodoi{10.1146/annurev-astro-082812-140944}

\bibitem[{{Bressan} {et~al.}(2012){Bressan}, {Marigo}, {Girardi}, {Salasnich},
  {Dal Cero}, {Rubele}, \& {Nanni}}]{bre12}
{Bressan}, A., {Marigo}, P., {Girardi}, L., {et~al.} 2012, \mnras, 427, 127,
  \dodoi{10.1111/j.1365-2966.2012.21948.x}

\bibitem[{{Clark} \& {Glover}(2014)}]{clark14}
{Clark}, P.~C., \& {Glover}, S. C.~O. 2014, \mnras, 444, 2396,
  \dodoi{10.1093/mnras/stu1589}

\bibitem[{{Cohen} {et~al.}(1981){Cohen}, {Frogel}, {Persson}, \&
  {Elias}}]{coh81}
{Cohen}, J.~G., {Frogel}, J.~A., {Persson}, S.~E., \& {Elias}, J.~H. 1981,
  \apj, 249, 481, \dodoi{10.1086/159308}

\bibitem[{{Condon} {et~al.}(1998){Condon}, {Cotton}, {Greisen}, {Yin},
  {Perley}, {Taylor}, \& {Broderick}}]{con98}
{Condon}, J.~J., {Cotton}, W.~D., {Greisen}, E.~W., {et~al.} 1998, \aj, 115,
  1693, \dodoi{10.1086/300337}

\bibitem[{{Danks} \& {Dennefeld}(1994)}]{dan94}
{Danks}, A.~C., \& {Dennefeld}, M. 1994, \pasp, 106, 382,
  \dodoi{10.1086/133390}

\bibitem[{{Das} {et~al.}(2021){Das}, {Jose}, {Samal}, {Zhang}, \&
  {Panwar}}]{das21}
{Das}, S.~R., {Jose}, J., {Samal}, M.~R., {Zhang}, S., \& {Panwar}, N. 2021,
  \mnras, 500, 3123, \dodoi{10.1093/mnras/staa3222}

\bibitem[{{Deharveng} {et~al.}(2005){Deharveng}, {Zavagno}, \&
  {Caplan}}]{deh05}
{Deharveng}, L., {Zavagno}, A., \& {Caplan}, J. 2005, \aap, 433, 565,
  \dodoi{10.1051/0004-6361:20041946}

\bibitem[{{Dobashi} {et~al.}(1994){Dobashi}, {Bernard}, {Yonekura}, \&
  {Fukui}}]{dob94}
{Dobashi}, K., {Bernard}, J.-P., {Yonekura}, Y., \& {Fukui}, Y. 1994, \apjs,
  95, 419, \dodoi{10.1086/192106}

\bibitem[{{Doi} {et~al.}(2015){Doi}, {Takita}, {Ootsubo}, {Arimatsu}, {Tanaka},
  {Kitamura}, {Kawada}, {Matsuura}, {Nakagawa}, {Morishima}, {Hattori},
  {Komugi}, {White}, {Ikeda}, {Kato}, {Chinone}, {Etxaluze}, \&
  {Cypriano}}]{doi15}
{Doi}, Y., {Takita}, S., {Ootsubo}, T., {et~al.} 2015, \pasj, 67, 50,
  \dodoi{10.1093/pasj/psv022}

\bibitem[{{Drew} {et~al.}(2005){Drew}, {Greimel}, {Irwin}, {Aungwerojwit},
  {Barlow}, {Corradi}, {Drake}, {G{\"a}nsicke}, {Groot}, {Hales}, {Hopewell},
  {Irwin}, {Knigge}, {Leisy}, {Lennon}, {Mampaso}, {Masheder}, {Matsuura},
  {Morales-Rueda}, {Morris}, {Parker}, {Phillipps}, {Rodriguez-Gil}, {Roelofs},
  {Skillen}, {Sokoloski}, {Steeghs}, {Unruh}, {Viironen}, {Vink}, {Walton},
  {Witham}, {Wright}, {Zijlstra}, \& {Zurita}}]{dre05}
{Drew}, J.~E., {Greimel}, R., {Irwin}, M.~J., {et~al.} 2005, \mnras, 362, 753,
  \dodoi{10.1111/j.1365-2966.2005.09330.x}

\bibitem[{{Dyson} \& {Williams}(1980)}]{dys80}
{Dyson}, J.~E., \& {Williams}, D.~A. 1980, {Physics of the interstellar medium}

\bibitem[{{Egan} {et~al.}(2003){Egan}, {Price}, {Kraemer}, {Mizuno}, {Carey},
  {Wright}, {Engelke}, {Cohen}, \& {Gugliotti}}]{ega03}
{Egan}, M.~P., {Price}, S.~D., {Kraemer}, K.~E., {et~al.} 2003, VizieR Online
  Data Catalog, V/114

\bibitem[{{Elias} {et~al.}(1982){Elias}, {Frogel}, {Matthews}, \&
  {Neugebauer}}]{eli82}
{Elias}, J.~H., {Frogel}, J.~A., {Matthews}, K., \& {Neugebauer}, G. 1982, \aj,
  87, 1029, \dodoi{10.1086/113185}

\bibitem[{{Elmegreen}(1993)}]{elm93}
{Elmegreen}, B.~G. 1993, in Protostars and Planets III, ed. E.~H. {Levy} \&
  J.~I. {Lunine}, 97

\bibitem[{{Elmegreen}(1998)}]{elm98}
{Elmegreen}, B.~G. 1998, in Astronomical Society of the Pacific Conference
  Series, Vol. 148, Origins, ed. C.~E. {Woodward}, J.~M. {Shull}, \&
  J.~{Thronson}, Harley~A., 150.
\newblock \doarXiv{astro-ph/9712352}

\bibitem[{{Elmegreen} \& {Lada}(1977)}]{elm77}
{Elmegreen}, B.~G., \& {Lada}, C.~J. 1977, \apj, 214, 725,
  \dodoi{10.1086/155302}

\bibitem[{{Esimbek} {et~al.}(2008){Esimbek}, {Wu}, \& {Wang}}]{esi08}
{Esimbek}, J., {Wu}, Y., \& {Wang}, Y. 2008, \na, 13, 144,
  \dodoi{10.1016/j.newast.2007.08.002}

\bibitem[{{Evans} {et~al.}(2018){Evans}, {Riello}, {De Angeli}, {Carrasco},
  {Montegriffo}, {Fabricius}, {Jordi}, {Palaversa}, {Diener}, {Busso},
  {Cacciari}, {van Leeuwen}, {Burgess}, {Davidson}, {Harrison}, {Hodgkin},
  {Pancino}, {Richards}, {Altavilla}, {Balaguer-N{\'u}{\~n}ez}, {Barstow},
  {Bellazzini}, {Brown}, {Castellani}, {Cocozza}, {De Luise}, {Delgado},
  {Ducourant}, {Galleti}, {Gilmore}, {Giuffrida}, {Holl}, {Kewley}, {Koposov},
  {Marinoni}, {Marrese}, {Osborne}, {Piersimoni}, {Portell}, {Pulone},
  {Ragaini}, {Sanna}, {Terrett}, {Walton}, {Wevers}, \& {Wyrzykowski}}]{eva18}
{Evans}, D.~W., {Riello}, M., {De Angeli}, F., {et~al.} 2018, \aap, 616, A4,
  \dodoi{10.1051/0004-6361/201832756}

\bibitem[{{Evans} \& {Lada}(1991)}]{evans91}
{Evans}, N.~J., I., \& {Lada}, E.~A. 1991, in Fragmentation of Molecular Clouds
  and Star Formation, ed. E.~{Falgarone}, F.~{Boulanger}, \& G.~{Duvert}, Vol.
  147, 293

\bibitem[{{Falgarone} {et~al.}(1991){Falgarone}, {Phillips}, \&
  {Walker}}]{fal91}
{Falgarone}, E., {Phillips}, T.~G., \& {Walker}, C.~K. 1991, \apj, 378, 186,
  \dodoi{10.1086/170419}

\bibitem[{{Fazio} {et~al.}(2004){Fazio}, {Hora}, {Allen}, {Ashby}, {Barmby},
  {Deutsch}, {Huang}, {Kleiner}, {Marengo}, {Megeath}, {Melnick}, {Pahre},
  {Patten}, {Polizotti}, {Smith}, {Taylor}, {Wang}, {Willner}, {Hoffmann},
  {Pipher}, {Forrest}, {McMurty}, {McCreight}, {McKelvey}, {McMurray}, {Koch},
  {Moseley}, {Arendt}, {Mentzell}, {Marx}, {Losch}, {Mayman}, {Eichhorn},
  {Krebs}, {Jhabvala}, {Gezari}, {Fixsen}, {Flores}, {Shakoorzadeh}, {Jungo},
  {Hakun}, {Workman}, {Karpati}, {Kichak}, {Whitley}, {Mann}, {Tollestrup},
  {Eisenhardt}, {Stern}, {Gorjian}, {Bhattacharya}, {Carey}, {Nelson},
  {Glaccum}, {Lacy}, {Lowrance}, {Laine}, {Reach}, {Stauffer}, {Surace},
  {Wilson}, {Wright}, {Hoffman}, {Domingo}, \& {Cohen}}]{faz04}
{Fazio}, G.~G., {Hora}, J.~L., {Allen}, L.~E., {et~al.} 2004, \apjs, 154, 10,
  \dodoi{10.1086/422843}

\bibitem[{{Flaherty} {et~al.}(2007){Flaherty}, {Pipher}, {Megeath}, {Winston},
  {Gutermuth}, {Muzerolle}, {Allen}, \& {Fazio}}]{fla07}
{Flaherty}, K.~M., {Pipher}, J.~L., {Megeath}, S.~T., {et~al.} 2007, \apj, 663,
  1069, \dodoi{10.1086/518411}

\bibitem[{{Fukuda} \& {Hanawa}(2000)}]{fuk00}
{Fukuda}, N., \& {Hanawa}, T. 2000, \apj, 533, 911, \dodoi{10.1086/308701}

\bibitem[{{Gaia Collaboration} {et~al.}(2018){Gaia Collaboration}, {Brown},
  {Vallenari}, {Prusti}, {de Bruijne}, {Babusiaux}, {Bailer-Jones}, {Biermann},
  {Evans}, {Eyer}, {Jansen}, {Jordi}, {Klioner}, {Lammers}, {Lindegren},
  {Luri}, {Mignard}, {Panem}, {Pourbaix}, {Randich}, {Sartoretti}, {Siddiqui},
  {Soubiran}, {van Leeuwen}, {Walton}, {Arenou}, {Bastian}, {Cropper},
  {Drimmel}, {Katz}, {Lattanzi}, {Bakker}, {Cacciari}, {Casta{\~n}eda},
  {Chaoul}, {Cheek}, {De Angeli}, {Fabricius}, {Guerra}, {Holl}, {Masana},
  {Messineo}, {Mowlavi}, {Nienartowicz}, {Panuzzo}, {Portell}, {Riello},
  {Seabroke}, {Tanga}, {Th{\'e}venin}, {Gracia-Abril}, {Comoretto},
  {Garcia-Reinaldos}, {Teyssier}, {Altmann}, {Andrae}, {Audard},
  {Bellas-Velidis}, {Benson}, {Berthier}, {Blomme}, {Burgess}, {Busso},
  {Carry}, {Cellino}, {Clementini}, {Clotet}, {Creevey}, {Davidson}, {De
  Ridder}, {Delchambre}, {Dell'Oro}, {Ducourant},
  {Fern{\'a}ndez-Hern{\'a}ndez}, {Fouesneau}, {Fr{\'e}mat}, {Galluccio},
  {Garc{\'\i}a-Torres}, {Gonz{\'a}lez-N{\'u}{\~n}ez}, {Gonz{\'a}lez-Vidal},
  {Gosset}, {Guy}, {Halbwachs}, {Hambly}, {Harrison}, {Hern{\'a}ndez},
  {Hestroffer}, {Hodgkin}, {Hutton}, {Jasniewicz}, {Jean-Antoine-Piccolo},
  {Jordan}, {Korn}, {Krone-Martins}, {Lanzafame}, {Lebzelter}, {L{\"o}ffler},
  {Manteiga}, {Marrese}, {Mart{\'\i}n-Fleitas}, {Moitinho}, {Mora}, {Muinonen},
  {Osinde}, {Pancino}, {Pauwels}, {Petit}, {Recio-Blanco}, {Richards},
  {Rimoldini}, {Robin}, {Sarro}, {Siopis}, {Smith}, {Sozzetti}, {S{\"u}veges},
  {Torra}, {van Reeven}, {Abbas}, {Abreu Aramburu}, {Accart}, {Aerts},
  {Altavilla}, {{\'A}lvarez}, {Alvarez}, {Alves}, {Anderson}, {Andrei},
  {Anglada Varela}, {Antiche}, {Antoja}, {Arcay}, {Astraatmadja}, {Bach},
  {Baker}, {Balaguer-N{\'u}{\~n}ez}, {Balm}, {Barache}, {Barata}, {Barbato},
  {Barblan}, {Barklem}, {Barrado}, {Barros}, {Barstow}, {Bartholom{\'e}
  Mu{\~n}oz}, {Bassilana}, {Becciani}, {Bellazzini}, {Berihuete}, {Bertone},
  {Bianchi}, {Bienaym{\'e}}, {Blanco-Cuaresma}, {Boch}, {Boeche}, {Bombrun},
  {Borrachero}, {Bossini}, {Bouquillon}, {Bourda}, {Bragaglia}, {Bramante},
  {Breddels}, {Bressan}, {Brouillet}, {Br{\"u}semeister}, {Brugaletta},
  {Bucciarelli}, {Burlacu}, {Busonero}, {Butkevich}, {Buzzi}, {Caffau},
  {Cancelliere}, {Cannizzaro}, {Cantat-Gaudin}, {Carballo}, {Carlucci},
  {Carrasco}, {Casamiquela}, {Castellani}, {Castro-Ginard}, {Charlot},
  {Chemin}, {Chiavassa}, {Cocozza}, {Costigan}, {Cowell}, {Crifo}, {Crosta},
  {Crowley}, {Cuypers}, {Dafonte}, {Damerdji}, {Dapergolas}, {David}, {David},
  {de Laverny}, {De Luise}, {De March}, {de Martino}, {de Souza}, {de Torres},
  {Debosscher}, {del Pozo}, {Delbo}, {Delgado}, {Delgado}, {Di Matteo},
  {Diakite}, {Diener}, {Distefano}, {Dolding}, {Drazinos}, {Dur{\'a}n},
  {Edvardsson}, {Enke}, {Eriksson}, {Esquej}, {Eynard Bontemps}, {Fabre},
  {Fabrizio}, {Faigler}, {Falc{\~a}o}, {Farr{\`a}s Casas}, {Federici},
  {Fedorets}, {Fernique}, {Figueras}, {Filippi}, {Findeisen}, {Fonti},
  {Fraile}, {Fraser}, {Fr{\'e}zouls}, {Gai}, {Galleti}, {Garabato},
  {Garc{\'\i}a-Sedano}, {Garofalo}, {Garralda}, {Gavel}, {Gavras}, {Gerssen},
  {Geyer}, {Giacobbe}, {Gilmore}, {Girona}, {Giuffrida}, {Glass}, {Gomes},
  {Granvik}, {Gueguen}, {Guerrier}, {Guiraud}, {Guti{\'e}rrez-S{\'a}nchez},
  {Haigron}, {Hatzidimitriou}, {Hauser}, {Haywood}, {Heiter}, {Helmi}, {Heu},
  {Hilger}, {Hobbs}, {Hofmann}, {Holland}, {Huckle}, {Hypki}, {Icardi},
  {Jan{\ss}en}, {Jevardat de Fombelle}, {Jonker}, {Juh{\'a}sz}, {Julbe},
  {Karampelas}, {Kewley}, {Klar}, {Kochoska}, {Kohley}, {Kolenberg},
  {Kontizas}, {Kontizas}, {Koposov}, {Kordopatis}, {Kostrzewa-Rutkowska},
  {Koubsky}, {Lambert}, {Lanza}, {Lasne}, {Lavigne}, {Le Fustec}, {Le
  Poncin-Lafitte}, {Lebreton}, {Leccia}, {Leclerc}, {Lecoeur-Taibi},
  {Lenhardt}, {Leroux}, {Liao}, {Licata}, {Lindstr{\o}m}, {Lister}, {Livanou},
  {Lobel}, {L{\'o}pez}, {Managau}, {Mann}, {Mantelet}, {Marchal}, {Marchant},
  {Marconi}, {Marinoni}, {Marschalk{\'o}}, {Marshall}, {Martino}, {Marton},
  {Mary}, {Massari}, {Matijevi{\v{c}}}, {Mazeh}, {McMillan}, {Messina},
  {Michalik}, {Millar}, {Molina}, {Molinaro}, {Moln{\'a}r}, {Montegriffo},
  {Mor}, {Morbidelli}, {Morel}, {Morris}, {Mulone}, {Muraveva}, {Musella},
  {Nelemans}, {Nicastro}, {Noval}, {O'Mullane}, {Ord{\'e}novic},
  {Ord{\'o}{\~n}ez-Blanco}, {Osborne}, {Pagani}, {Pagano}, {Pailler},
  {Palacin}, {Palaversa}, {Panahi}, {Pawlak}, {Piersimoni}, {Pineau}, {Plachy},
  {Plum}, {Poggio}, {Poujoulet}, {Pr{\v{s}}a}, {Pulone}, {Racero}, {Ragaini},
  {Rambaux}, {Ramos-Lerate}, {Regibo}, {Reyl{\'e}}, {Riclet}, {Ripepi}, {Riva},
  {Rivard}, {Rixon}, {Roegiers}, {Roelens}, {Romero-G{\'o}mez}, {Rowell},
  {Royer}, {Ruiz-Dern}, {Sadowski}, {Sagrist{\`a} Sell{\'e}s}, {Sahlmann},
  {Salgado}, {Salguero}, {Sanna}, {Santana-Ros}, {Sarasso}, {Savietto},
  {Schultheis}, {Sciacca}, {Segol}, {Segovia}, {S{\'e}gransan}, {Shih},
  {Siltala}, {Silva}, {Smart}, {Smith}, {Solano}, {Solitro}, {Sordo}, {Soria
  Nieto}, {Souchay}, {Spagna}, {Spoto}, {Stampa}, {Steele},
  {Steidelm{\"u}ller}, {Stephenson}, {Stoev}, {Suess}, {Surdej}, {Szabados},
  {Szegedi-Elek}, {Tapiador}, {Taris}, {Tauran}, {Taylor}, {Teixeira},
  {Terrett}, {Teyssandier}, {Thuillot}, {Titarenko}, {Torra Clotet}, {Turon},
  {Ulla}, {Utrilla}, {Uzzi}, {Vaillant}, {Valentini}, {Valette}, {van Elteren},
  {Van Hemelryck}, {van Leeuwen}, {Vaschetto}, {Vecchiato}, {Veljanoski},
  {Viala}, {Vicente}, {Vogt}, {von Essen}, {Voss}, {Votruba}, {Voutsinas},
  {Walmsley}, {Weiler}, {Wertz}, {Wevers}, {Wyrzykowski}, {Yoldas},
  {{\v{Z}}erjal}, {Ziaeepour}, {Zorec}, {Zschocke}, {Zucker}, {Zurbach}, \&
  {Zwitter}}]{gai18}
{Gaia Collaboration}, {Brown}, A.~G.~A., {Vallenari}, A., {et~al.} 2018, \aap,
  616, A1, \dodoi{10.1051/0004-6361/201833051}

\bibitem[{{Gaia Collaboration} {et~al.}(2021){Gaia Collaboration}, {Brown},
  {Vallenari}, {Prusti}, {de Bruijne}, {Babusiaux}, {Biermann}, {Creevey},
  {Evans}, {Eyer}, {Hutton}, {Jansen}, {Jordi}, {Klioner}, {Lammers},
  {Lindegren}, {Luri}, {Mignard}, {Panem}, {Pourbaix}, {Randich}, {Sartoretti},
  {Soubiran}, {Walton}, {Arenou}, {Bailer-Jones}, {Bastian}, {Cropper},
  {Drimmel}, {Katz}, {Lattanzi}, {van Leeuwen}, {Bakker}, {Cacciari},
  {Casta{\~n}eda}, {De Angeli}, {Ducourant}, {Fabricius}, {Fouesneau},
  {Fr{\'e}mat}, {Guerra}, {Guerrier}, {Guiraud}, {Jean-Antoine Piccolo},
  {Masana}, {Messineo}, {Mowlavi}, {Nicolas}, {Nienartowicz}, {Pailler},
  {Panuzzo}, {Riclet}, {Roux}, {Seabroke}, {Sordo}, {Tanga}, {Th{\'e}venin},
  {Gracia-Abril}, {Portell}, {Teyssier}, {Altmann}, {Andrae}, {Bellas-Velidis},
  {Benson}, {Berthier}, {Blomme}, {Brugaletta}, {Burgess}, {Busso}, {Carry},
  {Cellino}, {Cheek}, {Clementini}, {Damerdji}, {Davidson}, {Delchambre},
  {Dell'Oro}, {Fern{\'a}ndez-Hern{\'a}ndez}, {Galluccio}, {Garc{\'\i}a-Lario},
  {Garcia-Reinaldos}, {Gonz{\'a}lez-N{\'u}{\~n}ez}, {Gosset}, {Haigron},
  {Halbwachs}, {Hambly}, {Harrison}, {Hatzidimitriou}, {Heiter},
  {Hern{\'a}ndez}, {Hestroffer}, {Hodgkin}, {Holl}, {Jan{\ss}en}, {Jevardat de
  Fombelle}, {Jordan}, {Krone-Martins}, {Lanzafame}, {L{\"o}ffler}, {Lorca},
  {Manteiga}, {Marchal}, {Marrese}, {Moitinho}, {Mora}, {Muinonen}, {Osborne},
  {Pancino}, {Pauwels}, {Petit}, {Recio-Blanco}, {Richards}, {Riello},
  {Rimoldini}, {Robin}, {Roegiers}, {Rybizki}, {Sarro}, {Siopis}, {Smith},
  {Sozzetti}, {Ulla}, {Utrilla}, {van Leeuwen}, {van Reeven}, {Abbas}, {Abreu
  Aramburu}, {Accart}, {Aerts}, {Aguado}, {Ajaj}, {Altavilla}, {{\'A}lvarez},
  {{\'A}lvarez Cid-Fuentes}, {Alves}, {Anderson}, {Anglada Varela}, {Antoja},
  {Audard}, {Baines}, {Baker}, {Balaguer-N{\'u}{\~n}ez}, {Balbinot}, {Balog},
  {Barache}, {Barbato}, {Barros}, {Barstow}, {Bartolom{\'e}}, {Bassilana},
  {Bauchet}, {Baudesson-Stella}, {Becciani}, {Bellazzini}, {Bernet}, {Bertone},
  {Bianchi}, {Blanco-Cuaresma}, {Boch}, {Bombrun}, {Bossini}, {Bouquillon},
  {Bragaglia}, {Bramante}, {Breedt}, {Bressan}, {Brouillet}, {Bucciarelli},
  {Burlacu}, {Busonero}, {Butkevich}, {Buzzi}, {Caffau}, {Cancelliere},
  {C{\'a}novas}, {Cantat-Gaudin}, {Carballo}, {Carlucci}, {Carnerero},
  {Carrasco}, {Casamiquela}, {Castellani}, {Castro-Ginard}, {Castro Sampol},
  {Chaoul}, {Charlot}, {Chemin}, {Chiavassa}, {Cioni}, {Comoretto}, {Cooper},
  {Cornez}, {Cowell}, {Crifo}, {Crosta}, {Crowley}, {Dafonte}, {Dapergolas},
  {David}, {David}, {de Laverny}, {De Luise}, {De March}, {De Ridder}, {de
  Souza}, {de Teodoro}, {de Torres}, {del Peloso}, {del Pozo}, {Delbo},
  {Delgado}, {Delgado}, {Delisle}, {Di Matteo}, {Diakite}, {Diener},
  {Distefano}, {Dolding}, {Eappachen}, {Edvardsson}, {Enke}, {Esquej}, {Fabre},
  {Fabrizio}, {Faigler}, {Fedorets}, {Fernique}, {Fienga}, {Figueras},
  {Fouron}, {Fragkoudi}, {Fraile}, {Franke}, {Gai}, {Garabato},
  {Garcia-Gutierrez}, {Garc{\'\i}a-Torres}, {Garofalo}, {Gavras}, {Gerlach},
  {Geyer}, {Giacobbe}, {Gilmore}, {Girona}, {Giuffrida}, {Gomel}, {Gomez},
  {Gonzalez-Santamaria}, {Gonz{\'a}lez-Vidal}, {Granvik},
  {Guti{\'e}rrez-S{\'a}nchez}, {Guy}, {Hauser}, {Haywood}, {Helmi}, {Hidalgo},
  {Hilger}, {H{\l}adczuk}, {Hobbs}, {Holland}, {Huckle}, {Jasniewicz},
  {Jonker}, {Juaristi Campillo}, {Julbe}, {Karbevska}, {Kervella}, {Khanna},
  {Kochoska}, {Kontizas}, {Kordopatis}, {Korn}, {Kostrzewa-Rutkowska},
  {Kruszy{\'n}ska}, {Lambert}, {Lanza}, {Lasne}, {Le Campion}, {Le Fustec},
  {Lebreton}, {Lebzelter}, {Leccia}, {Leclerc}, {Lecoeur-Taibi}, {Liao},
  {Licata}, {Lindstr{\o}m}, {Lister}, {Livanou}, {Lobel}, {Madrero Pardo},
  {Managau}, {Mann}, {Marchant}, {Marconi}, {Marcos Santos}, {Marinoni},
  {Marocco}, {Marshall}, {Martin Polo}, {Mart{\'\i}n-Fleitas}, {Masip},
  {Massari}, {Mastrobuono-Battisti}, {Mazeh}, {McMillan}, {Messina},
  {Michalik}, {Millar}, {Mints}, {Molina}, {Molinaro}, {Moln{\'a}r},
  {Montegriffo}, {Mor}, {Morbidelli}, {Morel}, {Morris}, {Mulone}, {Munoz},
  {Muraveva}, {Murphy}, {Musella}, {Noval}, {Ord{\'e}novic}, {Orr{\`u}},
  {Osinde}, {Pagani}, {Pagano}, {Palaversa}, {Palicio}, {Panahi}, {Pawlak},
  {Pe{\~n}alosa Esteller}, {Penttil{\"a}}, {Piersimoni}, {Pineau}, {Plachy},
  {Plum}, {Poggio}, {Poretti}, {Poujoulet}, {Pr{\v{s}}a}, {Pulone}, {Racero},
  {Ragaini}, {Rainer}, {Raiteri}, {Rambaux}, {Ramos}, {Ramos-Lerate}, {Re
  Fiorentin}, {Regibo}, {Reyl{\'e}}, {Ripepi}, {Riva}, {Rixon}, {Robichon},
  {Robin}, {Roelens}, {Rohrbasser}, {Romero-G{\'o}mez}, {Rowell}, {Royer},
  {Rybicki}, {Sadowski}, {Sagrist{\`a} Sell{\'e}s}, {Sahlmann}, {Salgado},
  {Salguero}, {Samaras}, {Sanchez Gimenez}, {Sanna}, {Santove{\~n}a},
  {Sarasso}, {Schultheis}, {Sciacca}, {Segol}, {Segovia}, {S{\'e}gransan},
  {Semeux}, {Shahaf}, {Siddiqui}, {Siebert}, {Siltala}, {Slezak}, {Smart},
  {Solano}, {Solitro}, {Souami}, {Souchay}, {Spagna}, {Spoto}, {Steele},
  {Steidelm{\"u}ller}, {Stephenson}, {S{\"u}veges}, {Szabados}, {Szegedi-Elek},
  {Taris}, {Tauran}, {Taylor}, {Teixeira}, {Thuillot}, {Tonello}, {Torra},
  {Torra}, {Turon}, {Unger}, {Vaillant}, {van Dillen}, {Vanel}, {Vecchiato},
  {Viala}, {Vicente}, {Voutsinas}, {Weiler}, {Wevers}, {Wyrzykowski}, {Yoldas},
  {Yvard}, {Zhao}, {Zorec}, {Zucker}, {Zurbach}, \& {Zwitter}}]{gai21a1}
---. 2021, \aap, 649, A1, \dodoi{10.1051/0004-6361/202039657}

\bibitem[{{Garay} \& {Rodriguez}(1983)}]{gar83}
{Garay}, G., \& {Rodriguez}, L.~F. 1983, \apj, 266, 263, \dodoi{10.1086/160775}

\bibitem[{{Gildas Team}(2013)}]{gil13}
{Gildas Team}. 2013, {GILDAS: Grenoble Image and Line Data Analysis Software}.
\newblock \doeprint{1305.010}

\bibitem[{{Gutermuth} {et~al.}(2009){Gutermuth}, {Megeath}, {Myers}, {Allen},
  {Pipher}, \& {Fazio}}]{gut09}
{Gutermuth}, R.~A., {Megeath}, S.~T., {Myers}, P.~C., {et~al.} 2009, \apjs,
  184, 18, \dodoi{10.1088/0067-0049/184/1/18}

\bibitem[{{Gutermuth} {et~al.}(2005){Gutermuth}, {Megeath}, {Pipher},
  {Williams}, {Allen}, {Myers}, \& {Raines}}]{gut05}
{Gutermuth}, R.~A., {Megeath}, S.~T., {Pipher}, J.~L., {et~al.} 2005, \apj,
  632, 397, \dodoi{10.1086/432460}

\bibitem[{{Hern{\'a}ndez} {et~al.}(2004){Hern{\'a}ndez}, {Calvet},
  {Brice{\~n}o}, {Hartmann}, \& {Berlind}}]{her04}
{Hern{\'a}ndez}, J., {Calvet}, N., {Brice{\~n}o}, C., {Hartmann}, L., \&
  {Berlind}, P. 2004, \aj, 127, 1682, \dodoi{10.1086/381908}

\bibitem[{{Hunter} \& {Massey}(1990)}]{hun90}
{Hunter}, D.~A., \& {Massey}, P. 1990, \aj, 99, 846, \dodoi{10.1086/115378}

\bibitem[{{Israel}(1978)}]{isr78}
{Israel}, F.~P. 1978, \aap, 70, 769

\bibitem[{{Ivanov} {et~al.}(2005){Ivanov}, {Borissova}, {Bresolin}, \&
  {Pessev}}]{iva05}
{Ivanov}, V.~D., {Borissova}, J., {Bresolin}, F., \& {Pessev}, P. 2005, \aap,
  435, 107, \dodoi{10.1051/0004-6361:20042337}

\bibitem[{{Jacoby} {et~al.}(1984){Jacoby}, {Hunter}, \& {Christian}}]{jac84}
{Jacoby}, G.~H., {Hunter}, D.~A., \& {Christian}, C.~A. 1984, \apjs, 56, 257,
  \dodoi{10.1086/190983}

\bibitem[{{Kauffmann} {et~al.}(2013){Kauffmann}, {Pillai}, \&
  {Goldsmith}}]{kau13}
{Kauffmann}, J., {Pillai}, T., \& {Goldsmith}, P.~F. 2013, \apj, 779, 185,
  \dodoi{10.1088/0004-637X/779/2/185}

\bibitem[{{Kobulnicky} {et~al.}(2012){Kobulnicky}, {Lundquist},
  {Bhattacharjee}, \& {Kerton}}]{kob12}
{Kobulnicky}, H.~A., {Lundquist}, M.~J., {Bhattacharjee}, A., \& {Kerton},
  C.~R. 2012, \aj, 143, 71, \dodoi{10.1088/0004-6256/143/3/71}

\bibitem[{{Koenig} {et~al.}(2012){Koenig}, {Leisawitz}, {Benford}, {Rebull},
  {Padgett}, \& {Assef}}]{koe12}
{Koenig}, X.~P., {Leisawitz}, D.~T., {Benford}, D.~J., {et~al.} 2012, \apj,
  744, 130, \dodoi{10.1088/0004-637X/744/2/130}

\bibitem[{{Kumar} {et~al.}(2022){Kumar}, {Arzoumanian}, {Men'shchikov},
  {Palmeirim}, {Matsumura}, \& {Inutsuka}}]{kum22}
{Kumar}, M.~S.~N., {Arzoumanian}, D., {Men'shchikov}, A., {et~al.} 2022, \aap,
  658, A114, \dodoi{10.1051/0004-6361/202140363}

\bibitem[{{Kwan}(1997)}]{kwa97}
{Kwan}, J. 1997, \apj, 489, 284, \dodoi{10.1086/304773}

\bibitem[{{Lada} {et~al.}(1994){Lada}, {Lada}, {Clemens}, \& {Bally}}]{lad94}
{Lada}, C.~J., {Lada}, E.~A., {Clemens}, D.~P., \& {Bally}, J. 1994, \apj, 429,
  694, \dodoi{10.1086/174354}

\bibitem[{{Lahulla}(1985)}]{lah85}
{Lahulla}, J.~F. 1985, \aaps, 61, 537

\bibitem[{{Lawrence} {et~al.}(2007){Lawrence}, {Warren}, {Almaini}, {Edge},
  {Hambly}, {Jameson}, {Lucas}, {Casali}, {Adamson}, {Dye}, {Emerson},
  {Foucaud}, {Hewett}, {Hirst}, {Hodgkin}, {Irwin}, {Lodieu}, {McMahon},
  {Simpson}, {Smail}, {Mortlock}, \& {Folger}}]{law07}
{Lawrence}, A., {Warren}, S.~J., {Almaini}, O., {et~al.} 2007, \mnras, 379,
  1599, \dodoi{10.1111/j.1365-2966.2007.12040.x}

\bibitem[{{Lucas} {et~al.}(2008){Lucas}, {Hoare}, {Longmore}, {Schr{\"o}der},
  {Davis}, {Adamson}, {Bandyopadhyay}, {de Grijs}, {Smith}, {Gosling},
  {Mitchison}, {G{\'a}sp{\'a}r}, {Coe}, {Tamura}, {Parker}, {Irwin}, {Hambly},
  {Bryant}, {Collins}, {Cross}, {Evans}, {Gonzalez-Solares}, {Hodgkin},
  {Lewis}, {Read}, {Riello}, {Sutorius}, {Lawrence}, {Drew}, {Dye}, \&
  {Thompson}}]{luc08}
{Lucas}, P.~W., {Hoare}, M.~G., {Longmore}, A., {et~al.} 2008, \mnras, 391,
  136, \dodoi{10.1111/j.1365-2966.2008.13924.x}

\bibitem[{{Mart{\'\i}n-Hern{\'a}ndez}
  {et~al.}(2003){Mart{\'\i}n-Hern{\'a}ndez}, {van der Hulst}, \&
  {Tielens}}]{martin03}
{Mart{\'\i}n-Hern{\'a}ndez}, N.~L., {van der Hulst}, J.~M., \& {Tielens},
  A.~G.~G.~M. 2003, \aap, 407, 957, \dodoi{10.1051/0004-6361:20030982}

\bibitem[{{Martins} \& {Plez}(2006)}]{martins06}
{Martins}, F., \& {Plez}, B. 2006, \aap, 457, 637,
  \dodoi{10.1051/0004-6361:20065753}

\bibitem[{{Matsakis} {et~al.}(1976){Matsakis}, {Evans}, {Sato}, \&
  {Zuckerman}}]{mat76}
{Matsakis}, D.~N., {Evans}, N.~J., I., {Sato}, T., \& {Zuckerman}, B. 1976,
  \aj, 81, 172, \dodoi{10.1086/111871}

\bibitem[{{Maud} {et~al.}(2015){Maud}, {Moore}, {Lumsden}, {Mottram},
  {Urquhart}, \& {Hoare}}]{maud15b}
{Maud}, L.~T., {Moore}, T.~J.~T., {Lumsden}, S.~L., {et~al.} 2015, \mnras, 453,
  645, \dodoi{10.1093/mnras/stv1635}

\bibitem[{{Meyer} {et~al.}(1997){Meyer}, {Calvet}, \& {Hillenbrand}}]{mey97}
{Meyer}, M.~R., {Calvet}, N., \& {Hillenbrand}, L.~A. 1997, \aj, 114, 288,
  \dodoi{10.1086/118474}

\bibitem[{{Murakami} {et~al.}(2007){Murakami}, {Baba}, {Barthel}, {Clements},
  {Cohen}, {Doi}, {Enya}, {Figueredo}, {Fujishiro}, {Fujiwara}, {Fujiwara},
  {Garcia-Lario}, {Goto}, {Hasegawa}, {Hibi}, {Hirao}, {Hiromoto}, {Hong},
  {Imai}, {Ishigaki}, {Ishiguro}, {Ishihara}, {Ita}, {Jeong}, {Jeong},
  {Kaneda}, {Kataza}, {Kawada}, {Kawai}, {Kawamura}, {Kessler}, {Kester},
  {Kii}, {Kim}, {Kim}, {Kobayashi}, {Koo}, {Kwon}, {Lee}, {Lorente}, {Makiuti},
  {Matsuhara}, {Matsumoto}, {Matsuo}, {Matsuura}, {M{\"U}ller}, {Murakami},
  {Nagata}, {Nakagawa}, {Naoi}, {Narita}, {Noda}, {Oh}, {Ohnishi}, {Ohyama},
  {Okada}, {Okuda}, {Oliver}, {Onaka}, {Ootsubo}, {Oyabu}, {Pak}, {Park},
  {Pearson}, {Rowan-Robinson}, {Saito}, {Sakon}, {Salama}, {Sato}, {Savage},
  {Serjeant}, {Shibai}, {Shirahata}, {Sohn}, {Suzuki}, {Takagi}, {Takahashi},
  {Tanab{\'E}}, {Takeuchi}, {Takita}, {Thomson}, {Uemizu}, {Ueno}, {Usui},
  {Verdugo}, {Wada}, {Wang}, {Watabe}, {Watarai}, {White}, {Yamamura},
  {Yamauchi}, \& {Yasuda}}]{mur07}
{Murakami}, H., {Baba}, H., {Barthel}, P., {et~al.} 2007, \pasj, 59, S369,
  \dodoi{10.1093/pasj/59.sp2.S369}

\bibitem[{{Myers}(2009)}]{mye09}
{Myers}, P.~C. 2009, \apj, 700, 1609, \dodoi{10.1088/0004-637X/700/2/1609}

\bibitem[{{Neugebauer} {et~al.}(1984){Neugebauer}, {Habing}, {van Duinen},
  {Aumann}, {Baud}, {Beichman}, {Beintema}, {Boggess}, {Clegg}, {de Jong},
  {Emerson}, {Gautier}, {Gillett}, {Harris}, {Hauser}, {Houck}, {Jennings},
  {Low}, {Marsden}, {Miley}, {Olnon}, {Pottasch}, {Raimond}, {Rowan-Robinson},
  {Soifer}, {Walker}, {Wesselius}, \& {Young}}]{neu84}
{Neugebauer}, G., {Habing}, H.~J., {van Duinen}, R., {et~al.} 1984, \apjl, 278,
  L1, \dodoi{10.1086/184209}

\bibitem[{{Ojha} {et~al.}(2011){Ojha}, {Samal}, {Pandey}, {Bhatt}, {Ghosh},
  {Sharma}, {Tamura}, {Mohan}, \& {Zinchenko}}]{ojh11}
{Ojha}, D.~K., {Samal}, M.~R., {Pandey}, A.~K., {et~al.} 2011, \apj, 738, 156,
  \dodoi{10.1088/0004-637X/738/2/156}

\bibitem[{{Oke}(1990)}]{oke90}
{Oke}, J.~B. 1990, \aj, 99, 1621, \dodoi{10.1086/115444}

\bibitem[{{Panagia}(1973)}]{panagia73}
{Panagia}, N. 1973, \aj, 78, 929, \dodoi{10.1086/111498}

\bibitem[{{Panja} {et~al.}(2021){Panja}, {Chen}, {Dutta}, {Sun}, {Gao}, \&
  {Mondal}}]{pan21}
{Panja}, A., {Chen}, W.~P., {Dutta}, S., {et~al.} 2021, \apj, 910, 80,
  \dodoi{10.3847/1538-4357/abded4}

\bibitem[{{Panja} {et~al.}(2020){Panja}, {Mondal}, {Dutta}, {Joshi}, {Lata}, \&
  {Das}}]{pan20}
{Panja}, A., {Mondal}, S., {Dutta}, S., {et~al.} 2020, \aj, 159, 153,
  \dodoi{10.3847/1538-3881/ab737a}

\bibitem[{{Panwar} {et~al.}(2020){Panwar}, {Sharma}, {Ojha}, {Baug},
  {Dewangan}, {Bhatt}, \& {Pandey}}]{panwar20}
{Panwar}, N., {Sharma}, S., {Ojha}, D.~K., {et~al.} 2020, \apj, 905, 61,
  \dodoi{10.3847/1538-4357/abc42e}

\bibitem[{{Pecaut} \& {Mamajek}(2013)}]{pec13}
{Pecaut}, M.~J., \& {Mamajek}, E.~E. 2013, \apjs, 208, 9,
  \dodoi{10.1088/0067-0049/208/1/9}

\bibitem[{{Planck Collaboration} {et~al.}(2016){Planck Collaboration}, {Ade},
  {Aghanim}, {Arnaud}, {Ashdown}, {Aumont}, {Baccigalupi}, {Banday},
  {Barreiro}, {Bartolo}, {Battaner}, {Benabed}, {Beno{\^\i}t},
  {Benoit-L{\'e}vy}, {Bernard}, {Bersanelli}, {Bielewicz}, {Bonaldi},
  {Bonavera}, {Bond}, {Borrill}, {Bouchet}, {Boulanger}, {Bucher}, {Burigana},
  {Butler}, {Calabrese}, {Catalano}, {Chamballu}, {Chiang}, {Christensen},
  {Clements}, {Colombi}, {Colombo}, {Combet}, {Couchot}, {Coulais}, {Crill},
  {Curto}, {Cuttaia}, {Danese}, {Davies}, {Davis}, {de Bernardis}, {de Rosa},
  {de Zotti}, {Delabrouille}, {D{\'e}sert}, {Dickinson}, {Diego}, {Dole},
  {Donzelli}, {Dor{\'e}}, {Douspis}, {Ducout}, {Dupac}, {Efstathiou}, {Elsner},
  {En{\ss}lin}, {Eriksen}, {Falgarone}, {Fergusson}, {Finelli}, {Forni},
  {Frailis}, {Fraisse}, {Franceschi}, {Frejsel}, {Galeotta}, {Galli}, {Ganga},
  {Giard}, {Giraud-H{\'e}raud}, {Gjerl{\o}w}, {Gonz{\'a}lez-Nuevo},
  {G{\'o}rski}, {Gratton}, {Gregorio}, {Gruppuso}, {Gudmundsson}, {Hansen},
  {Hanson}, {Harrison}, {Helou}, {Henrot-Versill{\'e}},
  {Hern{\'a}ndez-Monteagudo}, {Herranz}, {Hildebrandt}, {Hivon}, {Hobson},
  {Holmes}, {Hornstrup}, {Hovest}, {Huffenberger}, {Hurier}, {Jaffe}, {Jaffe},
  {Jones}, {Juvela}, {Keih{\"a}nen}, {Keskitalo}, {Kisner}, {Knoche}, {Kunz},
  {Kurki-Suonio}, {Lagache}, {Lamarre}, {Lasenby}, {Lattanzi}, {Lawrence},
  {Leonardi}, {Lesgourgues}, {Levrier}, {Liguori}, {Lilje}, {Linden-V{\o}rnle},
  {L{\'o}pez-Caniego}, {Lubin}, {Mac{\'\i}as-P{\'e}rez}, {Maggio}, {Maino},
  {Mandolesi}, {Mangilli}, {Marshall}, {Martin}, {Mart{\'\i}nez-Gonz{\'a}lez},
  {Masi}, {Matarrese}, {Mazzotta}, {McGehee}, {Melchiorri}, {Mendes},
  {Mennella}, {Migliaccio}, {Mitra}, {Miville-Desch{\^e}nes}, {Moneti},
  {Montier}, {Morgante}, {Mortlock}, {Moss}, {Munshi}, {Murphy}, {Naselsky},
  {Nati}, {Natoli}, {Netterfield}, {N{\o}rgaard-Nielsen}, {Noviello},
  {Novikov}, {Novikov}, {Oxborrow}, {Paci}, {Pagano}, {Pajot}, {Paladini},
  {Paoletti}, {Pasian}, {Patanchon}, {Pearson}, {Pelkonen}, {Perdereau},
  {Perotto}, {Perrotta}, {Pettorino}, {Piacentini}, {Piat}, {Pierpaoli},
  {Pietrobon}, {Plaszczynski}, {Pointecouteau}, {Polenta}, {Pratt},
  {Pr{\'e}zeau}, {Prunet}, {Puget}, {Rachen}, {Reach}, {Rebolo}, {Reinecke},
  {Remazeilles}, {Renault}, {Renzi}, {Ristorcelli}, {Rocha}, {Rosset},
  {Rossetti}, {Roudier}, {Rubi{\~n}o-Mart{\'\i}n}, {Rusholme}, {Sandri},
  {Santos}, {Savelainen}, {Savini}, {Scott}, {Seiffert}, {Shellard}, {Spencer},
  {Stolyarov}, {Sudiwala}, {Sunyaev}, {Sutton}, {Suur-Uski}, {Sygnet},
  {Tauber}, {Terenzi}, {Toffolatti}, {Tomasi}, {Tristram}, {Tucci}, {Tuovinen},
  {Umana}, {Valenziano}, {Valiviita}, {Van Tent}, {Vielva}, {Villa}, {Wade},
  {Wandelt}, {Wehus}, {Yvon}, {Zacchei}, \& {Zonca}}]{pla16}
{Planck Collaboration}, {Ade}, P.~A.~R., {Aghanim}, N., {et~al.} 2016, \aap,
  594, A28, \dodoi{10.1051/0004-6361/201525819}

\bibitem[{{Reid} {et~al.}(2019){Reid}, {Menten}, {Brunthaler}, {Zheng}, {Dame},
  {Xu}, {Li}, {Sakai}, {Wu}, {Immer}, {Zhang}, {Sanna}, {Moscadelli}, {Rygl},
  {Bartkiewicz}, {Hu}, {Quiroga-Nu{\~n}ez}, \& {van Langevelde}}]{rei19}
{Reid}, M.~J., {Menten}, K.~M., {Brunthaler}, A., {et~al.} 2019, \apj, 885,
  131, \dodoi{10.3847/1538-4357/ab4a11}

\bibitem[{{Robitaille}(2019)}]{robitaille19}
{Robitaille}, T. 2019, {APLpy v2.0: The Astronomical Plotting Library in
  Python}, 2.0,  Zenodo, \dodoi{10.5281/zenodo.2567476}

\bibitem[{{Singh} {et~al.}(2021){Singh}, {Matzner}, {Friesen}, {Martin},
  {Pineda}, {Rosolowsky}, {Alves}, {Chac{\'o}n-Tanarro}, {Chen}, {Chen},
  {Choudhury}, {Di Francesco}, {Keown}, {Kirk}, {Punanova}, {Seo}, {Shirley},
  {Ginsburg}, {Offner}, {Arce}, {Caselli}, {Goodman}, {Myers}, {Redaelli}, \&
  {GAS Collaboration}}]{singh21}
{Singh}, A., {Matzner}, C.~D., {Friesen}, R.~K., {et~al.} 2021, \apj, 922, 87,
  \dodoi{10.3847/1538-4357/ac20d2}

\bibitem[{{Skrutskie} {et~al.}(2006){Skrutskie}, {Cutri}, {Stiening},
  {Weinberg}, {Schneider}, {Carpenter}, {Beichman}, {Capps}, {Chester},
  {Elias}, {Huchra}, {Liebert}, {Lonsdale}, {Monet}, {Price}, {Seitzer},
  {Jarrett}, {Kirkpatrick}, {Gizis}, {Howard}, {Evans}, {Fowler}, {Fullmer},
  {Hurt}, {Light}, {Kopan}, {Marsh}, {McCallon}, {Tam}, {Van Dyk}, \&
  {Wheelock}}]{skr06}
{Skrutskie}, M.~F., {Cutri}, R.~M., {Stiening}, R., {et~al.} 2006, \aj, 131,
  1163, \dodoi{10.1086/498708}

\bibitem[{{Su} {et~al.}(2019){Su}, {Yang}, {Zhang}, {Gong}, {Wang}, {Zhou},
  {Wang}, {Chen}, {Sun}, {Chen}, {Xu}, \& {Jiang}}]{su19}
{Su}, Y., {Yang}, J., {Zhang}, S., {et~al.} 2019, \apjs, 240, 9,
  \dodoi{10.3847/1538-4365/aaf1c8}

\bibitem[{{Sun} {et~al.}(2020){Sun}, {Yang}, {Xu}, {Zhang}, {Su}, {Wang},
  {Chen}, {Lu}, {Sun}, {Ju}, {Zhang}, {Zhou}, \& {Jiang}}]{sun20}
{Sun}, Y., {Yang}, J., {Xu}, Y., {et~al.} 2020, \apjs, 246, 7,
  \dodoi{10.3847/1538-4365/ab5b97}

\bibitem[{{Sun} {et~al.}(2021){Sun}, {Yang}, {Yan}, {Lin}, {Zhang}, {Su}, {Xu},
  {Chen}, {Wang}, \& {Zhou}}]{sun21}
{Sun}, Y., {Yang}, J., {Yan}, Q.-Z., {et~al.} 2021, \apjs, 256, 32,
  \dodoi{10.3847/1538-4365/ac11fe}

\bibitem[{{Tody}(1986)}]{tod86}
{Tody}, D. 1986, in Society of Photo-Optical Instrumentation Engineers (SPIE)
  Conference Series, Vol. 627, Instrumentation in astronomy VI, ed. D.~L.
  {Crawford}, 733, \dodoi{10.1117/12.968154}

\bibitem[{{Torres-Dodgen} \& {Weaver}(1993)}]{tor93}
{Torres-Dodgen}, A.~V., \& {Weaver}, W.~B. 1993, \pasp, 105, 693,
  \dodoi{10.1086/133222}

\bibitem[{{Urquhart} {et~al.}(2009){Urquhart}, {Hoare}, {Purcell}, {Lumsden},
  {Oudmaijer}, {Moore}, {Busfield}, {Mottram}, \& {Davies}}]{urq09}
{Urquhart}, J.~S., {Hoare}, M.~G., {Purcell}, C.~R., {et~al.} 2009, \aap, 501,
  539, \dodoi{10.1051/0004-6361/200912108}

\bibitem[{{Walborn} \& {Fitzpatrick}(1990)}]{wal90}
{Walborn}, N.~R., \& {Fitzpatrick}, E.~L. 1990, \pasp, 102, 379,
  \dodoi{10.1086/132646}

\bibitem[{{Wright} {et~al.}(2010){Wright}, {Eisenhardt}, {Mainzer}, {Ressler},
  {Cutri}, {Jarrett}, {Kirkpatrick}, {Padgett}, {McMillan}, {Skrutskie},
  {Stanford}, {Cohen}, {Walker}, {Mather}, {Leisawitz}, {Gautier}, {McLean},
  {Benford}, {Lonsdale}, {Blain}, {Mendez}, {Irace}, {Duval}, {Liu}, {Royer},
  {Heinrichsen}, {Howard}, {Shannon}, {Kendall}, {Walsh}, {Larsen}, {Cardon},
  {Schick}, {Schwalm}, {Abid}, {Fabinsky}, {Naes}, \& {Tsai}}]{wri10}
{Wright}, E.~L., {Eisenhardt}, P. R.~M., {Mainzer}, A.~K., {et~al.} 2010, \aj,
  140, 1868, \dodoi{10.1088/0004-6256/140/6/1868}

\end{thebibliography}
\bibliographystyle{aasjournal}

\end{document}